\documentstyle[epsfig,natbib2,natbibmnfix]{mn}

\newcommand{\be}{\begin{equation}}
\newcommand{\ee}{\end{equation}}
\newcommand{\bea}{\begin{eqnarray}}
\newcommand{\eea}{\end{eqnarray}}
\newcommand{\bc}{\begin{center}}
\newcommand{\ec}{\end{center}}

\renewcommand{\vec}[1]{ {\bmath #1} }

\newcommand{\dd}{{\rm d}}

\renewcommand{\thefootnote}{\fnsymbol{footnote}}

\title{The cosmological simulation code GADGET-2}

\author[V.~Springel]{\parbox{18cm}{Volker
Springel\footnotemark[1]\vspace{0.3cm}}\\ Max-Planck-Institut
f\"{u}r Astrophysik, Karl-Schwarzschild-Stra\ss{}e 1, 85740 Garching
bei M\"{u}nchen, Germany}

\begin{document}
\maketitle
\begin{abstract}
  We discuss the cosmological simulation code {\small GADGET-2}, a new
  massively parallel TreeSPH code, capable of following a collisionless fluid
  with the N-body method, and an ideal gas by means of smoothed particle
  hydrodynamics (SPH). Our implementation of SPH manifestly conserves energy
  and entropy in regions free of dissipation, while allowing for fully
  adaptive smoothing lengths.  Gravitational forces are computed with a
  hierarchical multipole expansion, which can optionally be applied in the
  form of a TreePM algorithm, where only short-range forces are computed with
  the `tree'-method while long-range forces are determined with Fourier
  techniques. Time integration is based on a quasi-symplectic scheme where
  long-range and short-range forces can be integrated with different
  timesteps. Individual and adaptive short-range timesteps may also be
  employed.  The domain decomposition used in the parallelisation algorithm is
  based on a space-filling curve, resulting in high flexibility and tree force
  errors that do not depend on the way the domains are cut. The code is
  efficient in terms of memory consumption and required communication
  bandwidth. It has been used to compute the first cosmological N-body
  simulation with more than $10^{10}$ dark matter particles, reaching a
  homogeneous spatial dynamic range of $10^5$ per dimension in a 3D box. It
  has also been used to carry out very large cosmological SPH simulations that
  account for radiative cooling and star formation, reaching total particle
  numbers of more than 250 million.  We present the algorithms used by the
  code and discuss their accuracy and performance using a number of test
  problems.  {\small GADGET-2} is publicly released to the research
  community.
\end{abstract}

\section{Introduction}
\renewcommand{\thefootnote}{\fnsymbol{footnote}}
\footnotetext[1]{E-mail: volker@mpa-garching.mpg.de}

Cosmological simulations play an ever more important role in theoretical
studies of the structure formation process in the Universe.  Without numerical
simulations, the $\Lambda$CDM model may arguably not have developed into the
leading theoretical paradigm for structure formation which it is today.  This
is because direct simulation is often the only available tool to compute
accurate theoretical predictions in the highly non-linear regime of
gravitational dynamics and hydrodynamics. This is particularly true for the
hierarchical structure formation process with its inherently complex geometry
and three-dimensional dynamics.

The list of important theoretical cosmological results based on simulation
work is therefore quite long, including fundamental results such as the
density profiles of dark matter halos \citep[e.g.][]{NFW}, the existence and
dynamics of dark matter substructure \citep[e.g.][]{To97}, the non-linear
clustering properties of dark matter \citep[e.g.][]{Je97}, the halo abundance
\citep[e.g.][]{Jen01}, the temperature and gas profiles of clusters of
galaxies \citep[e.g.][]{Ev90}, or the detailed properties of Lyman-$\alpha$
absorption lines in the interstellar medium \citep[e.g.][]{He96}.  Given that
many astrophysical phenomena involve a complex interplay of physical processes
on a wide range of scales, it seems clear that the importance of simulation
methods will continue to grow. This development is further fueled by the rapid
progress in computer technology, which makes an ever larger dynamic range
accessible to simulation models.  However, powerful computer hardware is only
one requirement for research with numerical simulations. The other, equally
important one, lies in the availability of suitable numerical algorithms and
simulation codes, capable of efficiently exploiting available computers to
study physical problems of interest, ideally in a highly accurate and flexible
way, so that new physics can be introduced easily.

This paper is about a novel version of the simulation code {\small
GADGET}, which was written and publicly released in its original
form four years ago \citep{SprGadget2000}, after which it found
widespread use in the research of many simulation groups. The code
discussed here has principal capabilities similar to the original
{\small GADGET} code. It can evolve all the systems (plus a number of
additional ones) that the first version could, but it does this more
accurately, and substantially faster.  It is also more flexible, more
memory efficient, and more easily extendible, making it considerably
more versatile. These improvements can be exploited for more advanced
simulations and demonstrate that progress in algorithmic methods can be
as important, or sometimes even more important, than the performance
increase offered by  new generations of computers.

The principal structure of {\small GADGET-2} is that of a TreeSPH code
\citep{He89}, where gravitational interactions are computed with a
hierarchical multipole expansion, and gas dynamics is followed with
smoothed particle hydrodynamics (SPH). Gas and collisionless dark
matter\footnote{The stars in galaxies can also be well approximated as
a collisionless fluid.} are both represented by particles in this
scheme.  Note that while there is a large variety of techniques for
computing the gravitational field, the basic N-body method for
representing a collisionless fluid is the same in all cosmological
codes, so that they ultimately only differ in the errors with which
they approximate the gravitational field.

Particle-mesh (PM) methods \citep[e.g.][]{Klypin1983,White83} are the
fastest schemes for computing the gravitational field, but for scales
below 1-2 mesh cells, the force is heavily suppressed, as a result this
technique is not well suited for work with high spatial resolution. The
spatial resolution can be greatly increased by adding short-range
direct-summation forces \citep{Hockney1981,Ef85}, or by using
additional Fourier-meshes adaptively placed on regions of interest
\citep{Cou91}. The mesh can also be adaptively refined, with the
potential found in real space using relaxation methods
\citep{Kra97,Knebe2001}.

The hierarchical tree algorithms \citep{App85,Ba86,Dehnen2000} follow
a different approach, and have no intrinsic resolution
limit. Particularly for mass distributions with low density contrast,
they can however be substantially slower than Fourier-based
methods. The recent development of TreePM hybrid methods \citep{Xu95}
tries to combine the best of both worlds by restricting the tree
algorithm to short-range scales, while computing the long-range
gravitational force by means of a particle-mesh algorithm.  {\small
GADGET-2} offers this method as well.

Compared to gravity, much larger conceptual differences exist
between the different hydrodynamical methods employed in current
cosmological codes.  Traditional `Eulerian' methods discretise space
and represent fluid variables on a mesh, while `Lagrangian' methods
discretise mass, using, for example, a set of fluid particles to model
the flow. Both methods have found widespread application in
cosmology. Mesh-based codes include algorithms with a fixed mesh
\cite[e.g.][]{Cen1992,Cen1993,Yepes1995,Pen1998}, and more recently
also with adaptive meshes
\cite[e.g.][]{Bryan1997,Norman1999,Teyssier2002,Kravtsov2002,Quilis2004}.
Lagrangian codes have almost all employed SPH thus far
\cite[e.g.][]{Evrard1988,He89,Na93,Ka96,Couchman1995,St96,Serna1996,Da97,Tissera1997,Owen1998,Serna2003,Wadsley2004},
although this is not the only possibility
\citep{Gnedin1995,Whitehurst1995}.

Mesh-codes offer superior resolving power for hydrodynamical shocks, with some
methods being able to capture shocks without artifical viscosity, and with
very low residual numerical viscosity. However, static meshes are only poorly
suited for the high dynamic range encountered in cosmology. Even for meshes as
large as $1024^3$, which is a challenge at present
\citep[e.g.][]{Kang2005,Cen2003}, individual galaxies in a cosmological volume
are poorly resolved, leaving no room for resolving internal structure such as
bulge and disk components.  A potential solution is provided by new
generations of adaptive mesh refinement codes, which begin to be more widely
used in cosmology \cite[e.g.][]{Abel2002,Kravtsov2002,Refregier2002,Motl2004}.
Some drawbacks of the mesh remain however even here. For example, the dynamics
is in general not Galilean-invariant, there are advection errors, and there
can be spurious generation of entropy due to mixing.

In contrast, Lagrangian methods like SPH are particularly well-suited to
follow the gravitational growth of structure, and to automatically increase
the resolution in the central regions of galactic halos, which are the regions
of primary interest in cosmology. The accurate treatment of self-gravity of
the gas in a fashion consistent with that of the dark matter is a further
strength of the particle-based SPH method.  Another fundamental difference
with mesh based schemes is that thermodynamic quantities advected with the
flow do not mix between different fluid elements at all, unless explicitly
modelled.  An important disadvantage of SPH is that the method has to rely on
an artificial viscosity for supplying the necessary entropy injection in
shocks. The latter are broadened over the SPH smoothing scale and not resolved
as true discontinuities.

In this paper, we give a concise description of the numerical model and the
novel algorithmic methods implemented in {\small GADGET-2}, which may also
serve as a reference for the publicly released version of this code.  In
addition we measure the code performance and accuracy for different types of
problems, and discuss the results obtained for a number  of test
problems, focusing in particular on gasdynamical simulations.

This paper is organised as follows. In Section~\ref{SecBasics}, we
summarise the set of equations the code integrates forward in
time. We then discuss in Section~\ref{SecTree} the algorithms used
to compute the `right-hand side' of these equations efficiently,
i.e.~the gravitational and hydrodynamical forces. This is followed by
a discussion of the time integration scheme in
Section~\ref{SecTimeInt}, and an explanation of the parallelisation
strategy in Section~\ref{SecParallel}. We present results for a
number of test problems in Section~\ref{SecTests}, followed by a
discussion of code performance in Section~\ref{SecPerformance}.
Finally, we summarise our findings in Section~\ref{SecDiscussion}.

\section{Basic equations} \label{SecBasics}

We here briefly summarise the basic set of equations that are studied
in cosmological simulations of structure formation. They describe the
dynamics of a collisionless component (dark matter or stars in
galaxies) and of an ideal gas (ordinary baryons, mostly hydrogen and
helium), both subject to and coupled by gravity in an expanding
background space. For brevity, we focus on the discretised forms of
the equations, noting the simplifications that apply for non-expanding
space where appropriate.

\subsection{Collisionless dynamics}

The continuum limit of non-interacting dark matter is described by the
collisionless Boltzmann equation coupled to the Poisson equation in an
expanding background Universe, the latter taken normally as a
Friedman-Lemaitre model. Due to the high-dimensionality of this
problem, these equations are best solved with the N-body method, where
phase-space density is sampled with a finite number $N$ of tracer
particles.

The dynamics of these particles is then described by the Hamiltonian
\be H= \sum_i \frac{\vec{p}_i^2}{2\,m_i\, a(t)^2} +
\frac{1}{2}\sum_{ij}\frac{m_i m_j
\,\varphi(\vec{x_i}-\vec{x}_j)}{a(t)}, \ee where
$H=H(\vec{p}_1,\ldots,\vec{p}_N,\vec{x}_1,\ldots,\vec{x}_N, t)$.  The
$\vec{x}_i$ are comoving coordinate vectors, and the corresponding
canonical momenta are given by $\vec{p}_i=a^2 m_i \dot\vec{{x}}_i$.
The explicit time dependence of the Hamiltonian arises from the
evolution $a(t)$ of the scale factor, which is given by the
Friedman-Lemaitre model.

If we assume periodic boundary conditions for a cube of size $L^3$, the
interaction potential $\varphi(\vec{x})$ is the solution of \be \nabla^2
\varphi(\vec{x}) = 4\pi G \left[ - \frac{1}{L^3} + \sum_{\vec{n}}
  \tilde\delta(\vec{x}-\vec{n}L)\right], \ee where the sum
over $\vec{n}=(n_1, n_2, n_3)$ extends over all integer triplets.  Note that
the mean density is subtracted here, so the solution corresponds to the {\em
  peculiar potential}, where the dynamics of the system is governed by
$\nabla^2 \phi(\vec{x}) = 4\pi G [\rho(\vec{x})-\overline\rho]$.  For our
discretised particle system, we define the peculiar potential as
\be
\phi(\vec{x}) = \sum_i m_i\, \varphi(\vec{x}-\vec{x}_i). \label{eqnpecpot}
\ee
The single
particle density distribution function $\tilde\delta(\vec{x})$ is the Dirac
$\delta$-function convolved with a normalised gravitational softening kernel
of comoving scale $\epsilon$. For it, we employ the spline kernel \citep{Mo85}
used in SPH and set $\tilde\delta(\vec{x}) = W(|\vec{x}|, 2.8\epsilon)$, where
$W(r)$ is given by \be W(r,h)=\frac{8}{\pi h^3} \left\{
\begin{array}{ll}
1-6\left(\frac{r}{h}\right)^2 + 6\left(\frac{r}{h}\right)^3, &
0\le\frac{r}{h}\le\frac{1}{2} ,\\
2\left(1-\frac{r}{h}\right)^3, & \frac{1}{2}<\frac{r}{h}\le 1 ,\\
0 , & \frac{r}{h}>1 .
\end{array}
\right. \label{eqkernel}
\ee
For this choice, the
Newtonian potential of a point mass at zero lag in non-periodic space is
$-G\,m/\epsilon$, the same as for a Plummer `sphere' of size $\epsilon$.

If desired, we can simplify to Newtonian space by setting $a(t)=1$, so
that the explicit time dependence of the Hamiltonian vanishes. For
vacuum boundaries, the interaction potential simplifies to the usual
Newtonian form, i.e. for point masses it is given by $\varphi(\vec{x})
= -G/|\vec{x}|$ modified by the softening for small separations.

Note that independent of the type of boundary conditions, a complete
force computation involves a double sum, resulting in a $N^2$-scaling
of the computational cost. This reflects the long-range nature of
gravity, where each particle interacts with every other particle,
making high-accuracy solutions for the gravitational forces {\em very
expensive} for large $N$.  Fortunately, the force accuracy needed for
collisionless dynamics is comparatively modest.  Force errors up to
$\sim 1$ per cent tend to only slightly increase the numerical
relaxation rate without compromising results \citep{He93c}, provided
the force errors are {\em random}.  This allows the use of
approximative force computations by methods such as those discussed in
Section~\ref{SecTree}. We note however that the situation is different
for collisional N-body systems, such as star clusters. Here direct
summation can be necessary to deliver the required force accuracy, a
task that triggered the development of powerful custom-made computers
such as {\small GRAPE} \cite[e.g][]{Mak90,Makino2003}. These systems
can then also be applied to collisionless dynamics using a
direct-summation approach \citep[e.g.][]{St96,Ma97}, or by combining
them with tree- or treePM-methods \citep{Fukushige2005}.

\subsection{Hydrodynamics}

Smoothed particle hydrodynamics (SPH) uses a set of discrete tracer particles
to describe the state of a fluid, with continuous fluid quantities being
defined by a kernel interpolation technique \citep{Lu77,Gi77,Mo92}. The
particles with coordinates $\vec{r}_i$, velocities $\vec{v}_i$, and masses
$m_i$ are best thought of as fluid elements that sample the gas in a
Lagrangian sense. The thermodynamic state of each fluid element may either be
defined in terms of its thermal energy per unit mass, $u_i$, or in terms of
the entropy per unit mass, $s_i$.  We prefer to use the latter as the
independent thermodynamic variable evolved in SPH, for reasons discussed in
full detail in \citet{SH02}.  Our formulation of SPH manifestly conserves both
energy and entropy even when fully adaptive smoothing lengths are used.
Traditional formulations of SPH on the other hand can violate entropy
conservation in certain situations.

We begin by noting that it is more convenient to work with an entropic
function defined by $A \equiv P/\rho^\gamma$, instead of directly using the
entropy $s$ per unit mass.  Because $A=A(s)$ is only a function of $s$ for an
ideal gas, we will often refer to $A$ as `entropy'.

Of fundamental importance for any SPH formulation is the density estimate,
which {\small GADGET-2} does in the form \be \rho_i = \sum_{j=1}^N m_j
W(|\vec{r}_{ij}|,h_i),
\label{eqndens}\ee where $\vec{r}_{ij}\equiv \vec{r}_i - \vec{r}_j$, and
$W(r,h)$ is the SPH smoothing kernel defined in equation~(\ref{eqkernel})
\footnote{We note that most of the literature on SPH defines the smoothing
length such that the kernel drops to zero at a distance $2h$, 
and not at $h$ as we have chosen here
for consistency with \citet{SprGadget2000}. This is only a 
difference in notation without bearing on the results.}.
In our `entropy formulation' of SPH, the adaptive smoothing lengths
$h_i$ of each particle are defined such that their kernel volumes contain a
constant mass for the estimated density, i.e.~the smoothing lengths and the
estimated densities obey the (implicit) equations \be \frac{4\pi}{3} h_i^3
\rho_i = N_{\rm sph}\overline{m},
\label{eqhsml}\ee where $N_{\rm sph}$ is the typical number of smoothing
neighbours, and $\overline{m}$ is an average particle mass. Note that in many
other formulations of SPH, smoothing lengths are typically chosen such that
the {\em number} of particles inside the smoothing radius $h_i$ is nearly
equal to a constant value $N_{\rm sph}$.

Starting from a discretised version of the fluid Lagrangian, one can show
\citep{SH02} that the equations of motion for the SPH particles are given by
\be \frac{\dd \vec{v}_i}{\dd t} = - \sum_{j=1}^N m_j \left[ f_i
  \frac{P_i}{\rho_i^2} \nabla_i W_{ij}(h_i) + f_j \frac{P_j}{\rho_j^2}
  \nabla_i W_{ij}(h_j) \right],
\label{eqnmot} 
\ee where the coefficients $f_i$ are defined by \be f_i = \left[ 1 +
  \frac{h_i}{3\rho_i}\frac{\partial \rho_i}{\partial h_i} \right]^{-1} \, ,
\ee and the abbreviation $W_{ij}(h)= W(|\vec{r}_{i}-\vec{r}_{j}|, h)$ has been
used. The particle pressures are given by $P_i=A_i \rho_i^{\gamma}$. Provided
there are no shocks and no external sources of heat, the equations above
already fully define reversible fluid dynamics in SPH. The entropy $A_i$ of
each particle stays  constant in such a flow.

However, flows of ideal gases can easily develop discontinuities,
where entropy is generated by microphysics. Such shocks need to
be captured by an artificial viscosity in SPH. To this end,
{\small GADGET-2} uses a viscous force \be \left. \frac{\dd
\vec{v}_i}{\dd t}\right|_{\rm visc}  = -\sum_{j=1}^N m_j \Pi_{ij}
\nabla_i\overline{W}_{ij} \, ,
\label{eqnvisc}
\ee where $\Pi_{ij}\ge 0$ is non-zero only when particles approach
each other in physical space.  The viscosity generates entropy at a
rate \be \frac{\dd A_i}{\dd t} =
\frac{1}{2}\frac{\gamma-1}{\rho_i^{\gamma-1}}\sum_{j=1}^N m_j \Pi_{ij}
\vec{v}_{ij}\cdot\nabla_i \overline{W}_{ij} \,, \ee transforming
kinetic energy of gas motion irreversibly into heat. The symbol
$\overline{W}_{ij}$ is here the arithmetic average of the two kernels
$W_{ij}(h_i)$ and $W_{ij}(h_j)$.

The Monaghan-Balsara form of the artificial viscosity
 \citep{Mo83,Balsara1995} is probably the most widely employed
 parameterisation of the viscosity in SPH codes. It takes the form \be
\label{eqvisc}
\Pi_{ij}=\left\{
\begin{tabular}{cl}
${\left[-\alpha c_{ij} \mu_{ij} +\beta \mu_{ij}^2\right]}/{\rho_{ij}}$ & \mbox{if
$\vec{v}_{ij}\cdot\vec{r}_{ij}<0$} \\
0 & \mbox{otherwise}, \\
\end{tabular}
\right. \label{eqnMoBals} \ee with 
\be
\mu_{ij}=\frac{h_{ij}\,\vec{v}_{ij}\cdot\vec{r}_{ij} }
{\left|\vec{r}_{ij}\right|^2}.\label{egnMu} \ee
Here $h_{ij}$ and $\rho_{ij}$ denote
arithmetic means of the corresponding quantities for the two particles $i$
and $j$, with $c_{ij}$ giving the mean sound speed.   The strength of the viscosity
is regulated by the parameters $\alpha$ and $\beta$, with typical values in
the range $\alpha\simeq 0.5-1.0$ and the frequent choice of
$\beta=2\,\alpha$.

Based on an analogy with the Riemann problem and using the notion of a signal
velocity $v_{ij}^{\rm sig}$ between two particles, \citet{Monaghan1997}
derived a slightly modified parameterisation of the viscosity, namely the
ansatz $\Pi_{ij} = - \frac{\alpha}{2} w_{ij} v_{ij}^{\rm sig}/ \rho_{ij}$.  In
the simplest form, the signal velocity can be estimated as \be v_{ij}^{\rm
  sig} = c_i + c_j - 3w_{ij}, \ee where
$w_{ij}={\vec{v}_{ij}\cdot\vec{r}_{ij}} / {\left|\vec{r}_{ij}\right|}$ is the
relative velocity projected onto the separation vector, provided the particles
approach each other, i.e.~for $\vec{v}_{ij}\cdot\vec{r}_{ij}<0$, otherwise we
set $w_{ij}=0$.  This gives a viscosity of the form \be \Pi_{ij} =
-\frac{\alpha}{2} \frac{\left[ c_{i} + c_{j} - 3 w_{ij} \right] w_{ij}}
{\rho_{ij}} , \label{eqnViscNew} \ee which is identical to (\ref{eqnMoBals})
if one sets $\beta=3/2\, \alpha$ and replaces $w_{ij}$ with $\mu_{ij}$. The
main difference between the two viscosities lies therefore in the additional
factor of $h_{ij}/r_{ij}$ that $\mu_{ij}$ carries with respect to $w_{ij}$. In
equations (\ref{eqnMoBals}) and (\ref{egnMu}), this factor weights the viscous
force towards particle pairs with small separations. In fact, after
multiplying with the kernel derivative, this weighting is strong enough to
make the viscous force diverge as $\propto 1/r_{ij}$ for small pair
separations, unless $\mu_{ij}$ in equation (\ref{egnMu}) is softened at small
separations by adding some fraction of $h_{ij}^2$ in the denominator, as it is
often done in practice.

In the equation of motion, the viscosity acts like an excess pressure
 $P_{\rm visc} \simeq \frac{1}{2} \rho_{ij}^2 \Pi_{ij}$ assigned to
the particles.  For the new form (\ref{eqnViscNew}) of the viscosity,
this is given by \be P_{\rm visc} \simeq \frac{\alpha}{2} \gamma
\left[ \frac{w_{ij}}{c_{ij}} +
\frac{3}{2}\left(\frac{w_{ij}}{c_{ij}}\right)^2 \right] P_{\rm therm},
\ee assuming roughly equal sound speeds and densities of the two
particles for the moment.  This viscous pressure depends only on a
Mach-number like quantity $w /c$, and not explicitly on the particle
separation or smoothing length.  We found that the modified viscosity
(\ref{eqnViscNew}) gives equivalent or improved results in our tests
compared to the standard formulation of equation (\ref{eqnMoBals}). In simulations with dissipation
it has the advantage that the occurrence of very large viscous
accelerations is reduced, so that a more efficient and stable time
integration results.  For these reason, we usually adopt the viscosity
(\ref{eqnViscNew}) in {\small GADGET-2}.

The signal-velocity approach naturally leads to a Courant-like
hydrodynamical timestep of the form \be \Delta t^{(\rm hyd)}_i =
\frac{C_{\rm courant}\, h_i}{\max_j(c_i+c_j-3w_{ij})} \ee which is
adopted by {\small GADGET-2}.  The maximum is here determined with
respect to all neighbours $j$ of particle $i$.

Following \citet{Balsara1995} and \citet{St96}, {\small GADGET-2} also
uses an additional viscosity-limiter to alleviate spurious angular
momentum transport in the presence of shear flows. This is done by
multiplying the viscous tensor with $(f_i+f_j)/2$, where \be f_i=
\frac{|\nabla \times \vec{v}|_i}{|\nabla \cdot \vec{v}|_i + |\nabla
\times \vec{v}|_i } \ee is a simple measure for the relative amount of
shear in the flow around particle $i$, based on standard SPH estimates
for divergence and curl \citep{Mo92}.

The above equations for the hydrodynamics were all expressed using physical
coordinates and velocities. In the actual simulation code, we use comoving
coordinates $\vec{x}$, comoving momenta $\vec{p}$ and comoving densities as
internal computational variables, which are related to physical variables in
the usual way. Since we continue to use the physical entropy, adiabatic
cooling due to expansion of the universe is automatically treated accurately.

\subsection{Additional physics} \label{SecAddPhys}

A number of further physical processes have already been implemented in
{\small GADGET-2}, and were applied to study structure formation problems.  A
full discussion of this physics (which is not included in the public release
of the code) is beyond the scope of this paper.  However, we here give a brief
overview of what has been done so far and refer the reader to the cited papers
for physical results and technical details.

Radiative cooling and heating by photoionisation has been implemented in
{\small GADGET-2} in a similar way as in \citet{Ka96}, i.e.~the ionisation
balance of helium and hydrogen is computed in the presence of an externally
specified time-dependent UV background under the assumption of collisional
ionisation equilibrium.  \cite{Yoshida2003} recently added a network for the
nonequilibrium treatment of the primordial chemistry of nine species, allowing
cooling by molecular hydrogen to be properly followed.

Star formation and associated feedback processes have been modelled
with {\small GADGET} by a number of authors using different physical
approximations. \citet{Spr99} considered a feedback model based on a
simple turbulent pressure term, while \citet{Kay2004} studied thermal
feedback with delayed cooling. A related model was also implemented by
\citet{Cox2004}.  \citet{SprHerMultiPhase,SprHerSFR} implemented a
subresolution multiphase model for the treatment of a star-forming
interstellar medium. Their model also accounts for energetic feedback
by galactic winds, and includes a basic treatment of metal enrichment.
More detailed metal enrichment models that allow for separate
treatments of type-II and type-I supernovae while also properly
accounting for the lifetimes of different stellar populations have
been independently implemented by \citet{Tornatore2004} and
\citet{Scannapieco2005}.  A different, more explicit approach to
describe a multiphase ISM has been followed by \citet{Marri2003}, who
introduced a hydrodynamic decoupling of cold gas and the ambient hot
phase.  A number of studies also used more ad-hoc models of feedback
in the form of preheating prescriptions
\citep{Springel2001,vandenBosch2003,Tornatore2003}.

A treatment of thermal conduction in hot ionised gas has been
implemented by \citet{Jubelgas2004} and was used to study modifications
of the intracluster medium of rich clusters of galaxies
\citep{DolagJub2004} caused by conduction.  An SPH approximation of
ideal magneto-hydrodynamics has been added to {\small GADGET-2} and
was used to study deflections of ultra-high energy cosmic rays the
Local Universe \citet{DolagGrasso2004,Dolag2005}.

\cite{DiMatteo2005} and \cite{Springel2005a} introduced a model for
the growth of supermassive black holes at the centres of galaxies, and
studied how energy feedback from gas accretion onto a supermassive
black hole regulates quasar activity and nuclear star
formation. \cite{Cuadra2005} added the ability to model stellar winds
and studied the feeding of Sgr A* by the stars orbiting in the
vicinity of the centre of the Galaxy.

Finally, non-standard dark matter dynamics has also been investigated with
{\small GADGET}.  \citet{Lindner2003} and \citet{DolagBart2004}
independently studied dark energy cosmologies.  Also, both \citet{Yos2000b}
and \citet{DaveSpergel2001} studied halo formation with self-interacting
dark matter, modelled by explicitly introducing collisional terms for the
dark matter particles.

\section{Gravitational algorithms}  \label{SecTree}

Gravity is the driving force of structure formation. Its computation
thus forms the core of any cosmological code. Unfortunately, its
long-range nature, and the high-dynamic range posed by the structure
formation problem, make it particularly challenging to compute
the gravitational forces accurately {\em and} efficiently.  In the
{\small GADGET-2} code, both the collisionless dark matter and the
gaseous fluid are represented as particles, allowing the self-gravity
of both components to be computed with gravitational N-body
methods, which we discuss next.

\subsection{The tree algorithm}

The primary method that {\small GADGET-2} uses to achieve the required spatial
adaptivity is a hierarchical multipole expansion, commonly called a tree
algorithm.  These methods group distant particles into ever larger cells,
allowing their gravity to be accounted for by means of a single multipole
force. Instead of requiring $N-1$ partial forces per particle as needed in a
direct-summation approach, the gravitational force on a single particle can
then be computed with just ${\cal O}({\rm log} N)$ interactions.

In practice, the hierarchical grouping that forms the basis of the multipole
expansion is most commonly obtained by a recursive subdivision of space.  In
the approach of \citet[][BH hereafter]{Ba86}, a cubical root node is used to
encompass the full mass distribution, which is repeatedly subdivided into
eight daughter nodes of half the side-length each, until one ends up with
`leaf' nodes containing single particles.  Forces are then obtained by
``walking'' the tree, i.e.~starting at the root node, a decision is made
whether or not the multipole expansion of the node is considered to provide an
accurate enough partial force (which will in general be the case for nodes
that are small and distant enough). If the answer is `yes', the multipole
force is used and the walk along this branch of the tree can be terminated, if
it is `no', the node is ``opened'', i.e.~its daughter nodes are considered in
turn.

It should be noted that the final result of the tree algorithm will in
general only represent an approximation to the true force. However, the
error can be controlled conveniently by modifying the opening criterion for
tree nodes, because higher accuracy is obtained by walking the tree to
lower levels. Provided sufficient computational resources are invested, the
tree force can then be made arbitrarily close to the well-specified correct
force.

\subsubsection{Details of the tree code}

There are three important characteristics of a gravitational tree code: the
type of grouping employed, the order chosen for the multipole expansion, and
the opening criterion used.  As a grouping algorithm, we prefer the
geometrical BH oct-tree instead of alternatives such as those based on
nearest-neighbour pairings \citep{Je89} or a binary kD-tree \cite{Stadel2001}.
The oct-tree is `shallower' than the binary tree, i.e.~fewer internal nodes
are needed for a given number $N$ of particles. In fact, for a nearly
homogeneous mass distribution, only $\approx 0.3\,N$ internal nodes are
needed, while for a heavily clustered mass distribution in a cosmological
simulation, this number tends to increase to about $\approx 0.65\,N$, which is
still considerably smaller than the number of $\approx N$ required in the
binary tree. This has advantages in terms of memory consumption.  Also, the
oct-tree avoids problems due to large aspect ratios of nodes, which helps to
keep the magnitude of higher order multipoles small. The clean geometric
properties of the oct-tree make it  ideal for use as a range-searching
tool, a further application of the tree we need for finding SPH neighbours.
Finally, the geometry of the oct-tree has  a close correspondence with a
space-filling Peano-Hilbert curve, a fact we exploit for our parallelisation
algorithm.

With respect to the multipole order, we follow a different approach
than used in {\small GADGET-1}, where an expansion including octopole
moments was employed.  Studies by \citet{He87} and \citet{BaH89}
indicate that the use of quadrupole moments may increase the
efficiency of the tree algorithm in some situations, and
\citet{Wadsley2004} even advocate hexadecopole order as an optimum
choice.  Higher order typically allows larger cell-opening angles,
i.e.~for a desired accuracy fewer interactions need to be
evaluated. This advantage is partially compensated by the increased
complexity per evaluation and the higher tree construction and tree
storage overhead, such that the performance as a function of multipole
order forms a broad maximum, where the precise location of the optimum
may depend sensitively on fine details of the software implementation
of the tree algorithm.

In {\small GADGET-2}, we deliberately went back to monopole moments,
because they feature a number of distinct advantages which make them
very attractive compared to schemes that carry the expansions to
higher order.  First of all, gravitational oct-trees with monopole
moments can be constructed in an extremely memory efficient way.  In
the first stage of our tree construction, particles are inserted one
by one into the tree, with each internal node holding storage for
indices of 8 daughter nodes or particles.  Note that for leaves
themselves, no node needs to be stored.  In a second step, we compute
the multipole moments recursively by walking the tree once in full. It
is interesting to note that these 8 indices will not be needed anymore
in the actual tree walk -- all that is needed for each internal node
is the information which node would be the next one to look at in case
the node needs to be opened, or alternatively, which is the next node
in line in case the multipole moment of the node can be used. We can
hence reuse the memory used for the 8 indices and store in it the two
indices needed for the tree walk, plus the multipole moment, which in
the case of monopole moments is the mass and and the centre-of-mass
coordinate vector. We additionally need the node side length, which
adds up to 7 variables, leaving  one variable still free, which we
use however in our parallelisation strategy. In any case, this method
of constructing the tree at no time requires more than $\sim 0.65
\times 8\times 4 \simeq 21$ bytes per particle (assuming 4 bytes per
variable), for a fully threaded tree.  This compares favourably with
memory consumptions quoted by other authors, even compared with the
storage optimised tree construction schemes of \cite{Dubinski2004},
where the tree is only constructed for part of the volume at a given
time, or with the method of \citet{Wadsley2004}, where particles are
bunched into groups, reducing the number of internal tree nodes by
collapsing ends of trees into nodes. Note also that the memory
consumption of our tree is lower than required for just storing the
phase-space variables of particles, leaving aside additional variables
that are typically required to control time-stepping, or to label the
particles with an identifying number.  In the standard version of
{\small GADGET-2}, we do not quite realize this optimum because we
also store the geometric centre of each tree in order to simplify the
SPH neighbour search.  This can in principle be omitted for purely
collisionless simulations.

Very compact tree nodes as obtained above are also highly advantageous given
the architecture of modern processors, which typically feature several layers
of fast `cache'-memory as work-space. Computations that involve data that is
already in cache can be carried out with close to maximum performance, but
access to the comparatively slow main memory imposes large numbers of wait
cycles. Small tree nodes thus help to make better use of the available memory
bandwidth, which is often a primary factor limiting the sustained performance
of a processor. By ordering the nodes in main memory in a special way (see
Section~\ref{SecPeano}), we can in addition help the processor and optimise
its cache utilisation.

Finally, a further important advantage of monopole moments is that they allow
simple dynamical tree updates that are consistent with the time integration
scheme discussed in detail in Section~\ref{SecTimeInt}. {\small GADGET-1}
already allowed dynamic tree updates, but it neglected the time variation of
the quadrupole moments.  This introduced a time asymmetry, which had the
potential to introduce secular integration errors in certain situations. Note
that particularly in simulations with radiative cooling, the dynamic range of
timesteps can  easily become so large that the tree construction overhead would
become dominant unless such dynamic tree update methods can be used.

\begin{figure}
\bc
\resizebox{8cm}{!}{\includegraphics{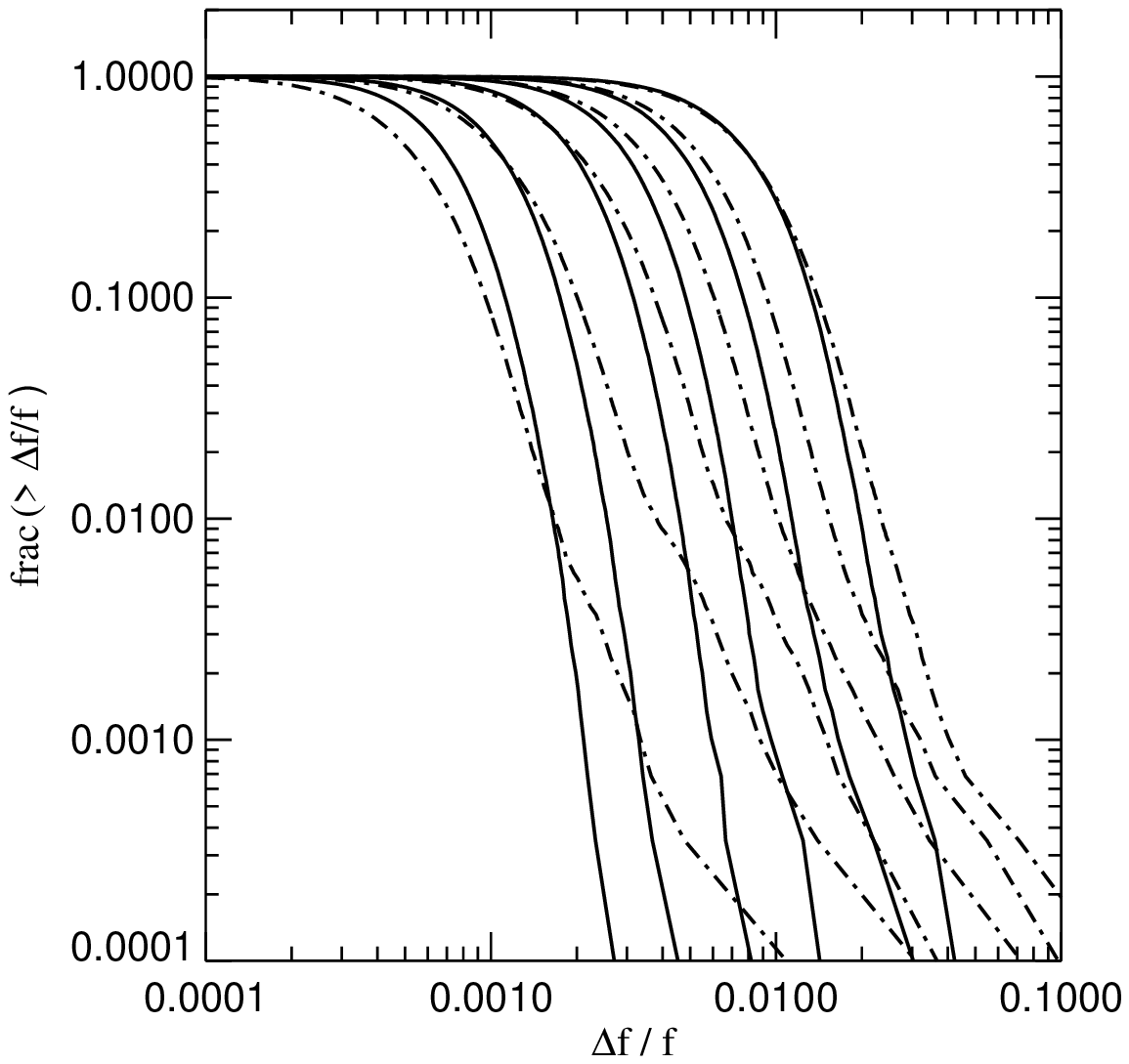}}\\
\resizebox{8cm}{!}{\includegraphics{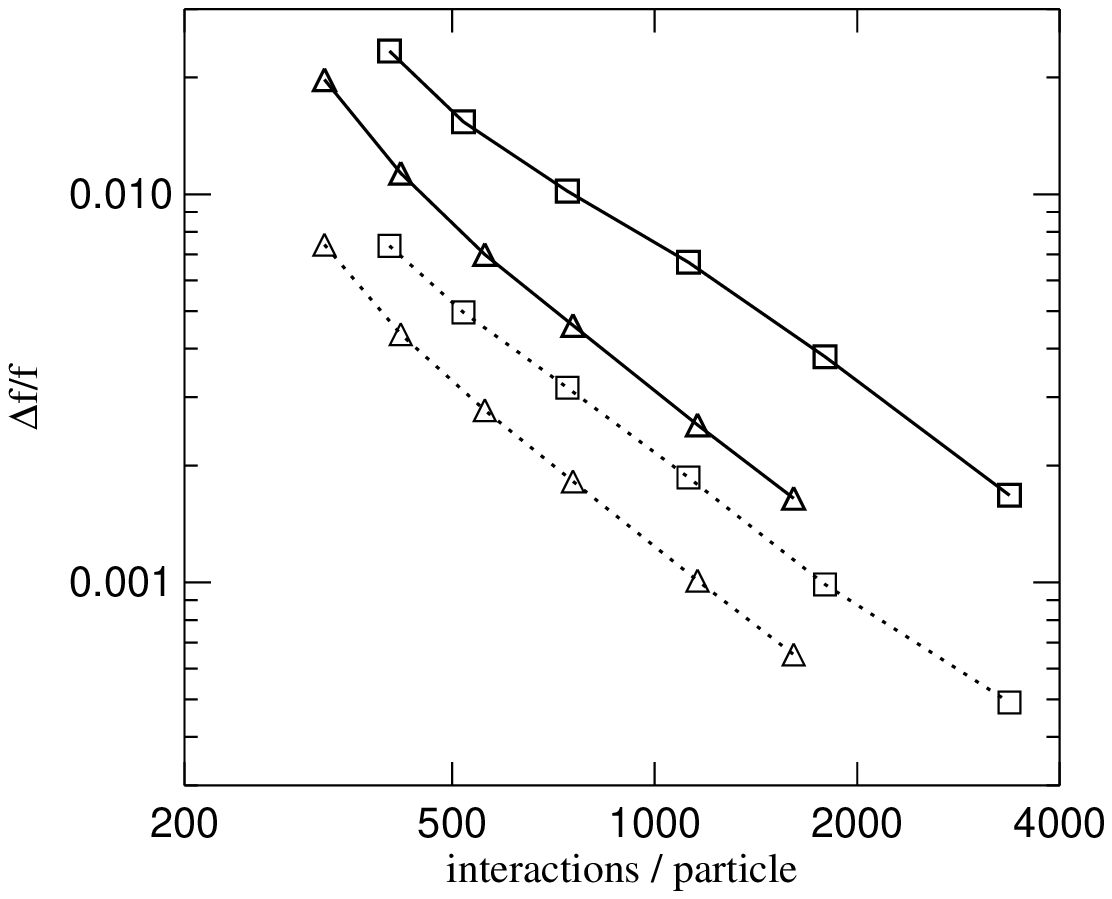}}%
\caption{Force-errors of the tree code for an isolated galaxy,
  consisting of a dark halo and a stellar disk.  In the top panel, each line
  shows the fraction of particles with force errors larger than a given value.
  The different line-styles are for different cell-opening criteria: the
  relative criterion is shown as solid lines and the standard BH criterion as
  dot-dashed lines. Both are shown for different values of the corresponding
  tolerance parameters, taken from the set $\{0.0005, 0.001, 0.0025, 0.005,
  0.01, 0.02\}$ for $\alpha$ in the case of the relative criterion, and from
  $\{0.3, 0.4, 0.5, 0.6, 0.7, 0.8\}$ in the case of the opening angle $\theta$
  used in the BH-criterion. In the lower panel, we compare the computational
  cost as a function of force accuracy. Solid lines compare the force accuracy
  of the 99.9\% percentile as a function of computational cost for the
  relative criterion (triangles) and the BH criterion (boxes).  At the same
  computational cost, the relative criterion always delivers somewhat more
  accurate forces. The dotted lines show the corresponding comparison for the
  50\% percentile of the force error distribution.
\label{figForceAccGalaxy}}
\ec
\end{figure}

With respect to the cell-opening criterion, we usually employ a
relative opening criterion similar to that used in {\small
GADGET-1}, but adjusted to our use of monopole moments. Specifically,
we consider a node of mass $M$ and extension $l$ at distance $r$ for
usage if \be \frac{GM}{r^2}\, \left(\frac{l}{r}\right)^2 \le \alpha\,
|\vec{a}|, \label{EqRelOpen} \ee where $|\vec{a}|$ is the size of the
total acceleration obtained in the last timestep, and $\alpha$ is a
tolerance parameter.  This criterion tries to limit the absolute force
error introduced in each particle-node interaction by comparing a rough
estimate of the truncation error with the size of the total expected
force. As a result, the typical {\em relative} force error is kept
roughly constant, and if needed, the opening criterion adjusts to the
dynamical state of the simulation to achieve this goal; at high
redshift, where peculiar accelerations largely cancel out, the average
opening angles are very small, while they can become larger once
matter clusters. Also, the opening angle varies with the distance of
the node. The net result is an opening criterion that typically
delivers higher force accuracy at a given computational cost compared
to a purely geometrical criterion such as that of BH.  In
Figure~\ref{figForceAccGalaxy}, we demonstrate this explicitly with
measurements of the force accuracy in a galaxy collision
simulation. Note that for the first force computation, an estimate of
the size of the force from the previous timestep is not yet
available. We then use the ordinary BH opening criterion to obtain
such estimates, followed by a recomputation of the forces in order to
have consistent accuracy for all steps.

\citet{Sal94} pointed out that tree codes can produce rather large worst-case
force errors when standard opening criteria with commonly employed opening
angles are used. These errors originate in situations where the distance to
the nearest particle in the node becomes very small.  When getting very close
or within the node, the error can even become unbounded. Our relative opening
criterion (\ref{EqRelOpen}) may suffer such errors since one may in principle
encounter a situation where a particle falls inside a node while still
satisfying the cell-acceptance criterion. To protect against this possibility,
we impose an additional opening criterion, viz.  \be |{r}_k - {c}_k|
\le 0.6 \,l . \ee Here $\vec{c}$ is the geometric centre of the node, $\vec{r}$
is the particle coordinate, and the inequality applies for each coordinate
axis $k$ separately. We hence require that the particle lies outside a box about 20\%
larger than the tree node.  Tests have shown that this robustly protects against
the occurrence of pathologically large force errors while incurring an acceptably
small increase in the average cost of the tree walk.

\subsubsection{Neighbour search using the tree}

We also use the BH oct-tree for the search of SPH neighbours, following the
range-search method of \cite{He89}. For a given spherical search region of
radius $h_i$ around a target location $\vec{r}_i$, we walk the tree with an
opening criterion that examines whether there is any geometric overlap between
the current tree node and the search region. If yes, the daughter nodes of the
node are considered in turn. Otherwise the walk along this branch of the tree
is immediately discarded. The tree walk is hence restricted to the region
local to the target location, allowing an efficient retrieval of the
desired neighbours. This use of the tree as a hierarchical search grid makes
the method extremely flexible and insensitive in performance to particle
clustering.

A difficulty arises for the SPH force loop, where the neighbour search depends
not only on $h_i$, but also on properties of the target particles.  We here
need to find all pairs with distances $|\vec{r}_{i}-\vec{r}_j|<\max(h_i,h_j)$,
including those where the distance is smaller than $h_j$ but not smaller than
$h_i$.  We solve this issue by storing in each tree node the maximum SPH
smoothing length occuring among all particles represented by the node. Note
that we update these values consistently when the SPH smoothing lengths are
redetermined in the first part of the SPH computation (i.e.~the density loop).
Using this information, it is straightforward to modify the opening criterion
such that all interacting pairs in the SPH force computation are always
correctly found.

Finally, a few notes on how we solve the implicit equation (\ref{eqhsml}) for
determining the desired SPH smoothing lengths of each particle in the first
place. For simplicity and for allowing a straightforward integration into our
parallel communication strategy, we find the root of this equation with a
binary bisection method.  Convergence is significantly accelerated by choosing
a Newton-Raphson value as the next guess instead of the the mid-point of the
current interval. Given that we compute $\partial \rho_i/\partial h_i$ anyway
for our SPH formulation, this comes at no additional cost. Likewise, for each
new timestep, we start the iteration with a new guess for $h_i$ based on the
expected change from the velocity divergence of the flow. Because we
usually only require that equation (\ref{eqhsml}) is solved to a few per cent
accuracy, finding and adjusting the SPH smoothing lengths is a subdominant
task in the CPU time consumption of our SPH code.

\subsubsection{Periodic boundaries in the tree code}

The summation over the infinite grid of particle images required for
simulations with periodic boundary conditions can also be treated in the tree
algorithm. {\small GADGET-2} uses the technique proposed by \citet{He91} for
this purpose.  The global BH-tree is only constructed for the primary mass
distribution, but it is walked such that each node is periodically mapped to
the closest image as seen from the coordinate under consideration. This
accounts for the dominant forces of the nearest images.  For each of the
partial forces, the Ewald summation method can be used to complement the force
exerted by the nearest image with the contribution of all other images of the
fiducial infinite grid of nodes.  In practice, {\small GADGET-2} uses a 3D
lookup table (in one octant of the simulation box) for the Ewald correction,
as proposed by \citet{He91}.

In the first version of our code, we carried out the Ewald correction for each
of the nodes visited in the primary tree walk over nearest node images,
leading to roughly a doubling of the computational cost. However, the sizes of
Ewald force correction terms have a very different distance dependence than
the ordinary Newtonian forces of tree nodes. For nodes in the vicinity of a
target particle, i.e.~for separations small against the boxsize, the
correction forces are negligibly small, while for separations approaching half
the boxsize they become large, eventually even cancelling the Newtonian force.
In principle, therefore, the Ewald correction only needs to be evaluated for
distant nodes with the same opening criterion as the ordinary Newtonian force,
while for nearby ones, a coarser opening angle can be chosen.  In {\small
  GADGET-2} we take advantage of this and carry out the Ewald corrections in a
separate tree walk, taking the above considerations into account. This leads
to a significant reduction of the overhead incurred by the periodic
boundaries.

\begin{figure}
\bc
\resizebox{8cm}{!}{\includegraphics{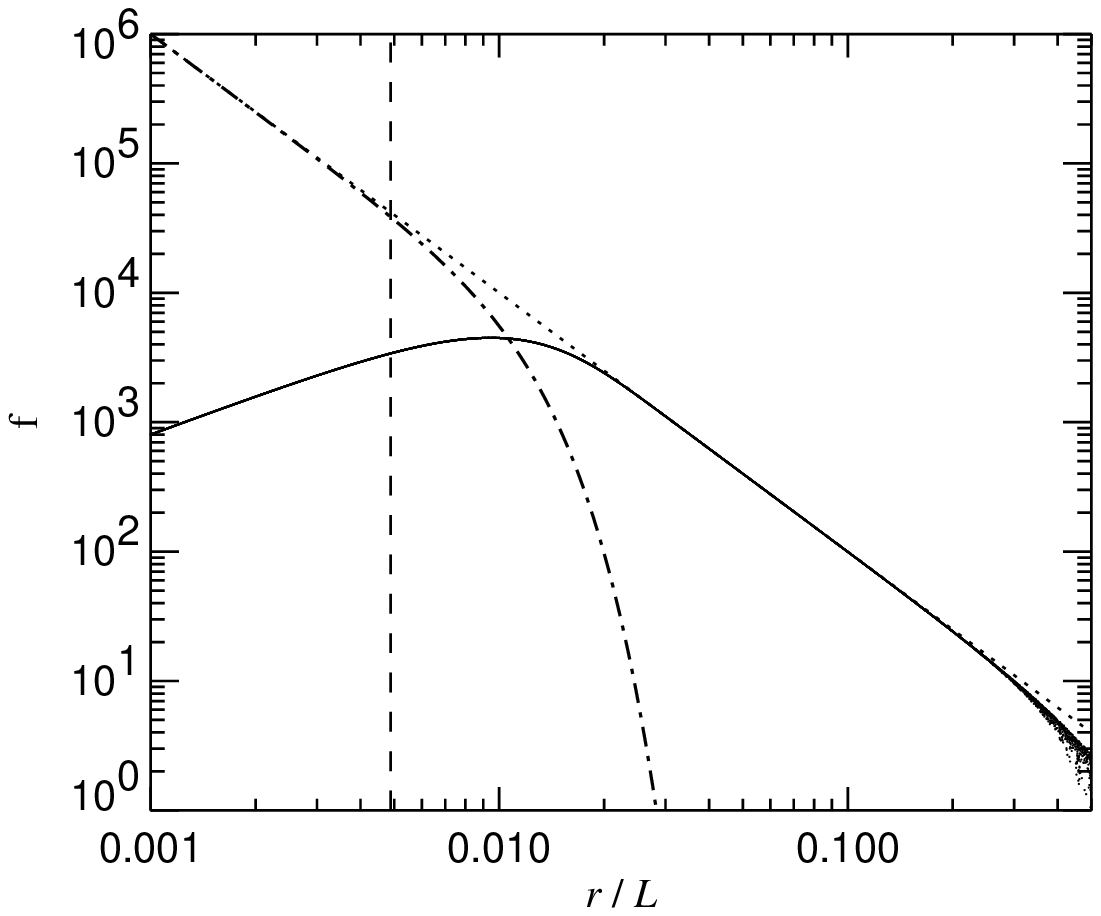}}\\
\resizebox{8cm}{!}{\includegraphics{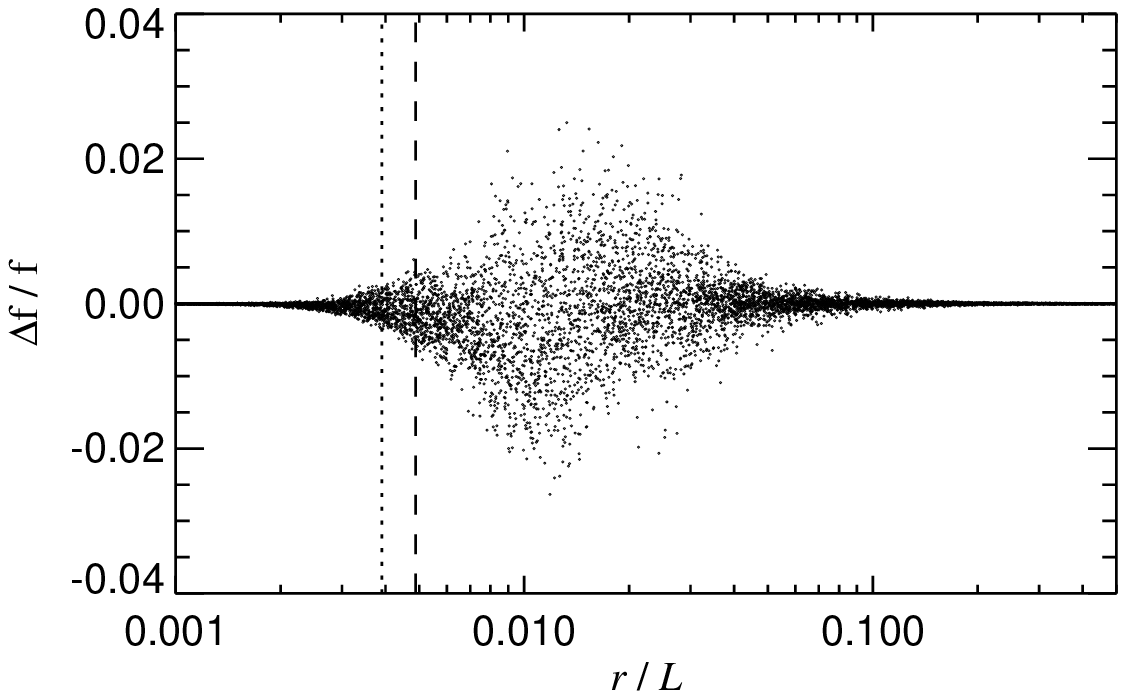}}%
\caption{Force decomposition and force error of the TreePM scheme. The top
  panel illustrates the size of the short-range (dot-dashed) and long-range
  force (solid) as a function of distance in a periodic box. The spatial scale
  $r_s$ of the split is marked with a vertical dashed line. The bottom panel
  compares the TreePM force with the exact force expected in a periodic box.
  For separations of order the mesh scale (marked by a vertical dotted line),
  maximum force errors of $1-2$ per cent due to the mesh anisotropy arise, but
  the rms force error is well below 1 per cent even in this range, and the
  mean force tracks the correct result accurately. If a larger force-split
  scale is chosen, the residual force anisotropies can be further reduced, if
  desired.
\label{figPMaccuracy}}
\ec
\end{figure}

\subsection{The TreePM method}

The new version of {\small GADGET-2} used in this study optionally
allows the pure tree algorithm to be replaced by a hybrid method consisting
of a synthesis of the particle-mesh method and the tree algorithm.
{\small GADGET-2}'s mathematical implementation of this so-called TreePM method
\citep{Xu95,Bode2000,Bagla02a} is similar to that of \citet{Bagla02b}.  The
potential of Eqn.~(\ref{eqnpecpot}) is explicitly split in Fourier space
into a long-range and a short-range part according to $\phi_{\vec{k}} =
\phi^{{\rm long}}_{\vec{k}} + \phi^{{\rm short}}_{\vec{k}}$, where \be
\phi^{{\rm long}}_{\vec{k}} = \phi_{\vec{k}} \exp(-\vec{k}^2 r_s^2), \ee
with $r_s$ describing the spatial scale of the force-split.  This long
range potential can be computed very efficiently with mesh-based Fourier
methods. Note that if $r_s$ is chosen slightly larger than the mesh scale,
force anisotropies that exist in plain PM methods can be suppressed to
essentially arbitrarily small levels.

The short range part of the potential can be solved in real space by noting
that for $r_s \ll L$ the short-range part of the real-space solution of
Eqn.~(\ref{eqnpecpot}) is given by \be \phi^{{\rm short}}(\vec{x}) = - G \sum_i
\frac{m_i}{r_i} {\rm erfc}\left(\frac{r_i} {2 r_s}\right).
\label{eqshortfrc}\ee Here $r_i = \min(|\vec{x}-\vec{r}_i - \vec{n}L|)$ is
defined as the smallest distance of any of the images of particle $i$ to the
point $\vec{x}$.  Because the complementary error function rapidly suppresses
the force for distances large compared to $r_s$ (the force drops to about 1\%
of its Newtonian value for $r\simeq 4.5\, r_s$), only this nearest image has
any chance to contribute to the short-range force.

The short-range force corresponding to Equation~(\ref{eqshortfrc}) can
now be computed by the tree algorithm, except that the force law is
modified by a short-range cut-off factor. However, the tree only needs
to be walked in a small spatial region around each target particle,
and no corrections for periodic boundary conditions are
required. Together this can result in a very substantial performance
improvement, and in addition one typically gains accuracy in the long
range force, which is now basically exact, and not an approximation as
in the tree method. Note that the TreePM approach maintains all of the
most important advantages of the tree algorithm, namely its
insensitivity to clustering, its essentially unlimited dynamic range,
and its precise control about the softening scale of the gravitational
force.

\subsubsection{Details of the TreePM algorithm}

To compute the PM part of the force, we use clouds-in-cells (CIC)
assignment \citep{Hockney1981} to construct the mass density field on
to the mesh. We carry out a discrete Fourier transform of the mesh,
and multiply it with the Greens function for the potential in periodic
boundaries, $-4\pi G/k^2$, modified with the exponential truncation of
the short-range force.  We then deconvolve for the clouds-in-cell
kernel by dividing twice with ${\rm sinc}^2(k_xL/2N_g)\,{\rm
sinc}^2(k_yL/2N_g)\,{\rm sinc}^2(k_zL/2N_g)$.  One deconvolution
corrects for the smoothing effect of the CIC in the mass assignment,
the other for the force interpolation. After performing an inverse
Fourier transform, we then obtain the gravitational potential on the
mesh.

We approximate the forces on the mesh by finite differencing the
potential, using the four-point differencing rule \bea
\left.\frac{\partial\phi}{\partial x}\right|_{ijk} & = & \frac{1}{\Delta
x}\left( \frac{2}{3}\left[ \phi_{i+1,j,k} - \phi_{i-1,j,k}\right] -
\right. \\
\nonumber & & \;\;\;\;\;\;\;\;\left.
\frac{1}{12}\left[ \phi_{i+2,j,k} - \phi_{i-2,j,k}\right] \right) \eea
which offers order ${\cal O}(\Delta x^4)$ accuracy, where $\Delta
x=L/N_{\rm mesh}$ is the mesh spacing. It would also be possible to
carry out the differentiation in Fourier space, by pulling down a
factor $-i\,\vec{k}$ and obtaining the forces directly instead of the
potential. However, this would require an inverse Fourier transform
separately for each coordinate, i.e.~three instead of one, with little
(if any) gain in accuracy compared to the four-point formula.

Finally, we interpolate the forces to the particle position using
again a CIC, for consistency. Note that the FFTs required here can be
efficiently carried out using real-to-complex transforms and their
inverse, which saves memory and execution time compared to fully
complex transforms.

In Figure~\ref{figPMaccuracy}, we illustrate the spatial decomposition
of the force and show the force error of the PM scheme.  This has been
computed by randomly placing a particle of unit mass in a periodic
box, and then measuring the forces obtained by the simulation code for
a set of randomly placed test particles. We compare the force to the
theoretically expected exact force, which can be computed by Ewald
summation over all periodic images, and then by multiplying with the
pre-factor \be f _l = 1 - {\rm erfc}\left(\frac{r}{2\,r_s}\right) -
\frac{r}{\sqrt{\pi} r_s} \exp\left(- \frac{r^2}{4\,r_s^2}\right), \ee
which takes out the short-range force, exactly the part that will be
supplied by the short-range tree walk. The force errors of the PM
force are mainly due to mesh anisotropy, which shows up on scales
around the mesh size. However, thanks to the smoothing of the
short-range force and the deconvolution of the CIC kernel, the mean
force is very accurate, and the rms-force error due to mesh anisotropy
is well below one percent. Note that these force errors compare
favourably to the ones reported by P$^3$M codes
\citep[e.g.]{Ef85}. Also, note that in the above formalism, the force
anisotropy can be reduced further to essentially arbitrarily small
levels by simply increasing $r_s$, at the expense of slowing down the
tree part of the algorithm.  Finally we remark that while
Fig~\ref{figPMaccuracy} characterises the magnitude of PM force errors
due to a single particle, it is not yet a realistic error distribution
for a real mass distribution. Here the PM force errors on the
mesh-scale can partially average out, while there can be additional
force errors from the tree algorithm on short-scales.


\subsubsection{TreePM for `zoom' simulations}

{\small GADGET-2} is capable of applying the PM algorithm also for
non-periodic boundary conditions. Here, a sufficiently large mesh needs to be
used such that the mass distribution completely fits in one octant of the
mesh. The potential can then be obtained by a real-space convolution with the
usual $1/r$ kernel of the potential, and this convolution can be efficiently
carried out using FFTs in Fourier space. For simplicity, {\small GADGET}
obtains the necessary Green's function by simply Fourier transforming $1/r$
once, and storing the result in memory.

However, it should be noted that the 4-point differencing of the
potential requires that there are at least 2 correct potential values
on either side of the region that is actually occupied by
particles. Since the CIC assignment/interpolation involves two cells,
one therefore has the following requirement for the minimum dimension
$N_{\rm mesh}$ of the employed mesh: \be  (N_{\rm mesh} -5)\, d \ge L, \label{eqcond1} \ee where
$L$ is the spatial extension of the region occupied by particles and
$d$ is the size of a mesh cell. Recall that due to the necessary
zero-padding, the actual dimension of the FFT that will be carried out
is $(2N_{\rm mesh})^3$.

The code is also able to use a two-level hierarchy of FFT meshes. This
was designed for `zoom-simulations', which focus on a small region
embedded in a much larger cosmological volume. Some of these
simulations can feature a rather large dynamic range, being as extreme
as putting much more than 99\% of the particles in less than
$10^{-10}$ of the volume \citep{Liang2005}. Here, the standard TreePM
algorithm is of little help since a mesh covering the full volume
would have a mesh-size still so large that the high-resolution region
would fall into one or  a few cells, so that the tree algorithm would
effectively degenerate to an ordinary tree method within the
high-resolution volume.

One possibility to improve upon this situation is to use a second
FFT-mesh that covers the high resolution region, such that the
long-range force is effectively computed in two steps. Adaptive
mesh-placement in the ${\rm AP^3M}$ code \citep{Couchman1995} follows
a similar scheme. {\small GADGET-2} allows the use of such a
secondary mesh-level and places it automatically, based on a
specification of which of the particles are `high-resolution
particles'. However, there are a number of additional technical
constraints in using this method.  The intermediate-scale FFT works
with vacuum boundaries, i.e.~the code will use zero-padding and a FFT
of size $(2N_{\rm mesh})^3$ to compute it. If $L_{\rm HR}$ is the
maximum extension of the high-resolution zone (which may not overlap
with the boundaries of the box in case the base simulation is
periodic), then condition (\ref{eqcond1}) for the minimum
high-resolution cell size applies. But in addition, this intermediate
scale FFT must properly account for the part of the short-range force
that complements the long-range FFT of the whole box, i.e.~it must be
able to properly account for all mass in a sphere of size $R_{\rm
cut}$ around each of the high-resolution particles.  There must hence
be at least a padding region of size $R_{\rm cut}$ still covered by
the mesh-octant used for the high-resolution zone. Because of the CIC
assignment, this implies the constraint $L_{\rm HR} + 2R_{\rm cut} \le
d_{\rm HR} (N_{\rm mesh}-1)$.  This limits the dynamic range one can
achieve with a single additional mesh-level.  In fact, the
high-resolution cell size must satisfy \be d_{\rm HR} \ge \max
\left(\frac{L_{\rm HR} + 2 R_{\rm cut}}{N_{\rm mesh} -1 } \; , \;
\frac{L_{\rm HR}}{N_{\rm mesh}-5}\right). \ee For our typical choice of
$R_{\rm cut} = 4.5 \times r_s = 1.25\times 4.5 \times d_{\rm LR}$,
this means that the high-resolution mesh-size cannot be made smaller
than $d_{\rm HR} \simeq 10\, d_{\rm LR} /(N_{\rm mesh}-1)$, i.e.~at
least slightly more than 10 low-resolution mesh-cells must be covered
by the high-resolution mesh.  Nevertheless, provided one has a very
large number of particles in a quite small high-resolution region, the
resulting reduction of the tree walk time can outweight the additional
cost of doing a large, zero-padded FFT for the high-resolution region.

\begin{figure}
\bc
\resizebox{8cm}{!}{\includegraphics{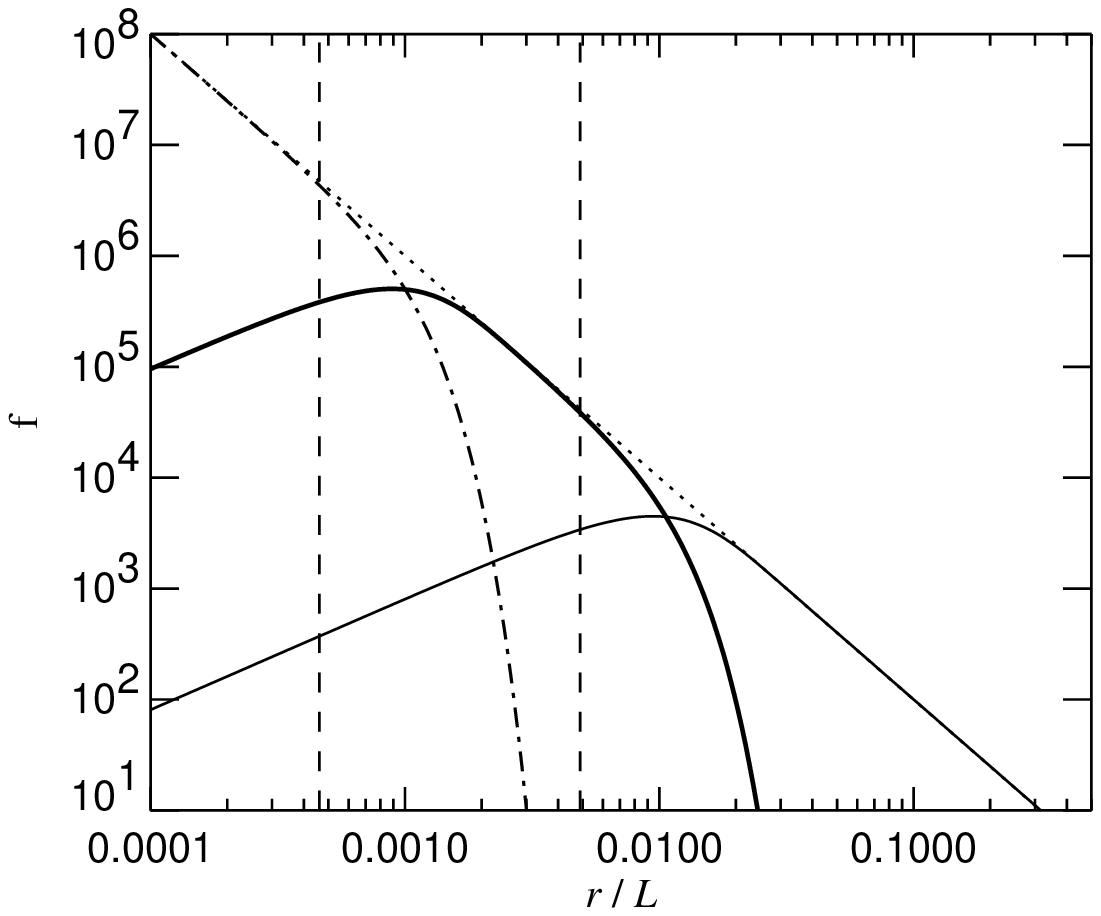}}\\
\resizebox{8cm}{!}{\includegraphics{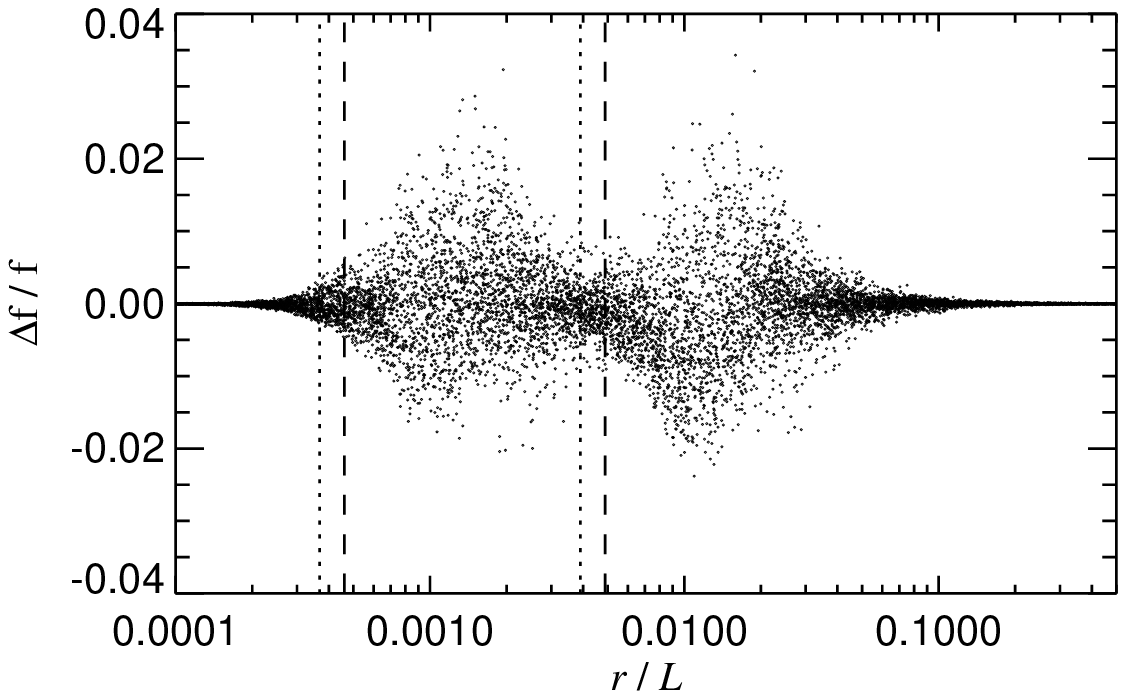}}%
\caption{Force decomposition and force error of the TreePM scheme in the case
  when two meshes are used (`zoom-simulations'). The top panel illustrates the
  strength of the short-range (dot-dashed), intermediate-range (thick solid),
  and long-range force (solid) as a function of distance in a periodic box.
  The spatial scales of the two splits are marked with vertical dashed lines.
  The bottom panel shows the error distribution of the PM force. The outer
  matching region exhibits a very similar error characteristic as the inner
  match of tree- and PM-force. In both cases, for separations of order the
  fine or coarse mesh scale (dotted lines), respectively, force errors of up
  to $1-2$ per cent arise, but the rms force error stays well below 1 per
  cent, and the mean force tracks the correct result accurately.
\label{figPMaccuracyZoom}}
\ec
\end{figure}

In Figure~\ref{figPMaccuracyZoom}, we show the PM force error
resulting for such a two-level decomposition of the PM force. We here
placed a particle of unit mass randomly inside a high-resolution
region of side-length 1/20 of a periodic box. We then measured the PM
force accuracy of {\small GADGET-2} by randomly placing test
particles. Particles that were falling inside the high-resolution
region were treated as high-resolution particles such that their
PM-force consists of two FFT contributions, while particles outside
the box receive only the long-range FFT force.  In real simulations,
the long-range forces are decomposed in an analogous way. With respect
to the short-range force, the tree is walked with different values for
the short-range cut-off, depending on whether a particle is
characterised as belonging to the high-res zone or not. Note however
that only one global tree is constructed containing all the mass.  The
top panel of Figure~\ref{figPMaccuracyZoom} shows the contributions of
the different force components as a function of scale, while the
bottom panel gives the distribution of the PM force errors. The
largest errors occur at the matching regions of the forces.  For
realistic particle distributions, a large number of force components
contribute, further reducing the typical error due to averaging.

\begin{figure}
\bc 
\vspace*{-0.5cm}\resizebox{7.2cm}{!}{\includegraphics{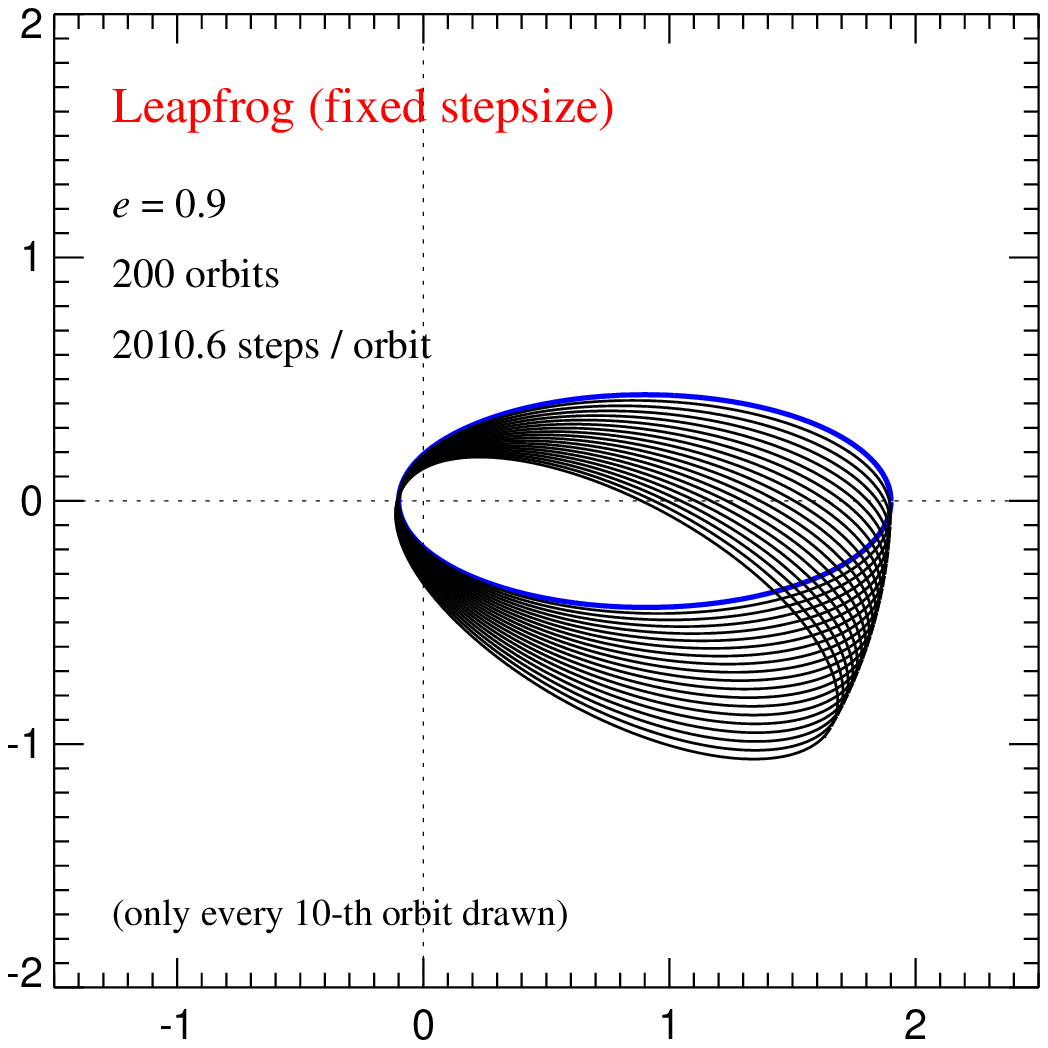}}\\%
\vspace*{-0.5cm}\resizebox{7.2cm}{!}{\includegraphics{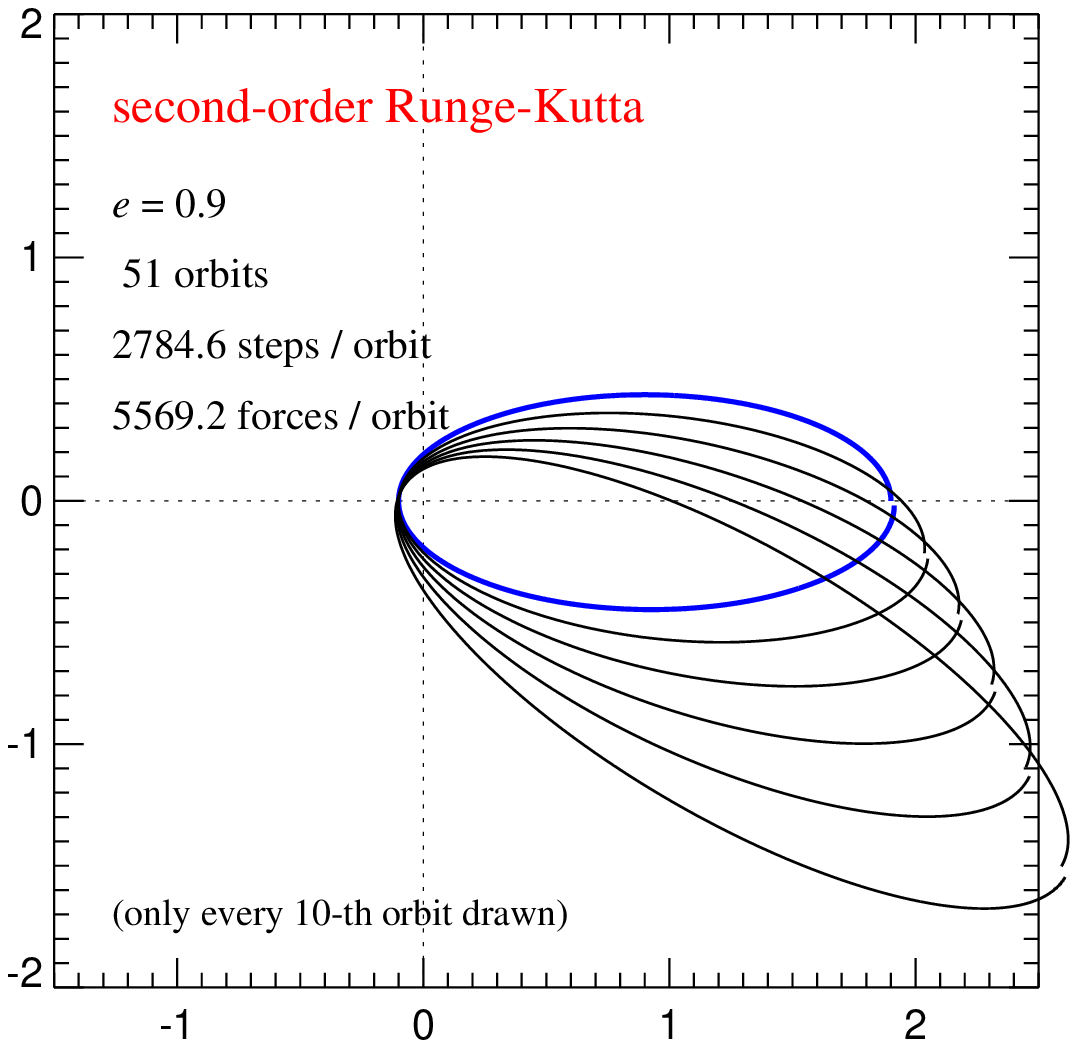}}\\%
\vspace*{-0.5cm}\resizebox{7.2cm}{!}{\includegraphics{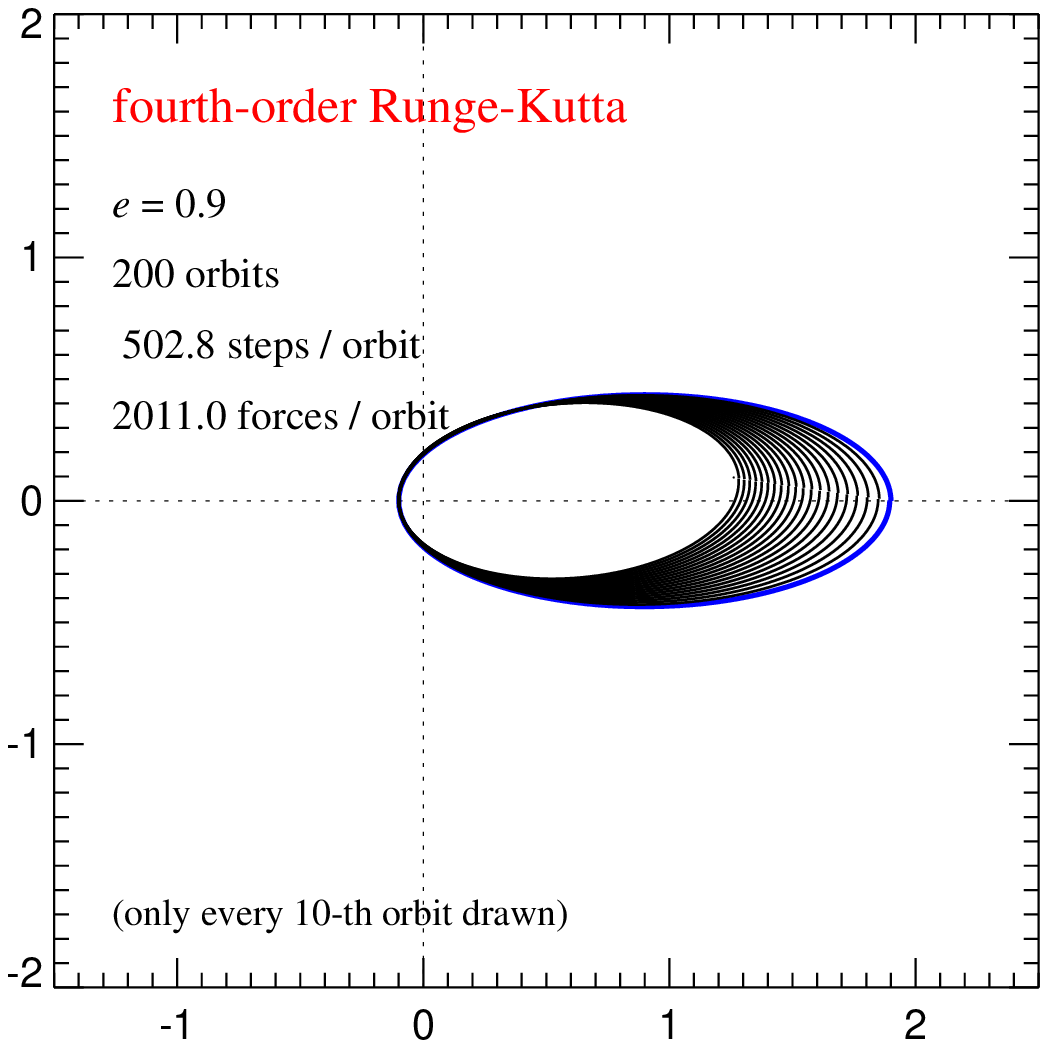}}\\%
\vspace*{-0.25cm}\caption{A Kepler problem of high eccentricity
evolved with different simple time integration schemes, using an equal
timestep in all cases. Even though the leapfrog and the 2nd order
Runge-Kutta produce comparable errors in a single step, the long term
stability of the integration is very different. Even a computationally
much more expensive 4th order Runge-Kutta scheme, with a smaller error
per step, performs dramatically worse than the leapfrog in this
problem.
\label{figKepler}}
\ec
\end{figure}

\section{Time integration}  \label{SecTimeInt}

\subsection{The symplectic nature of the leapfrog}

Hamiltonian systems are not robust in the sense that they are not structurally
stable against non-Hamiltonian perturbations. Numerical approximations for the
temporal evolution of a Hamiltonian system obtained from an ordinary numerical
integration method (e.g. Runge-Kutta) in general introduce non-Hamiltonian
perturbations, which can completely change the long-term behaviour. Unlike
dissipative systems, Hamiltonian systems do not have attractors.

The Hamiltonian structure of the system can be preserved during the
time integration if each step of it is formulated as a canonical
transformation.  Since canonical transformations leave the
symplectic two-form invariant (equivalent to preserving Poincare's
integral invariants, or stated differently, to preserving
phase-space), such an integration scheme is called symplectic
\citep[e.g.]{Hairer2002}.  Note that the time-evolution of a system can be
viewed as a continuous canonical transformation generated by the
Hamiltonian.  If an integration step is the exact solution of a
(partial) Hamiltonian, it represents the result of a phase-space
conserving canonical transformation and is hence symplectic.

We now note that the Hamiltonian of the usual N-body problem is separable in
the form \be H = H_{\rm kin} + H_{\rm pot}. \ee In this simple case, the
time-evolution operators for each of the parts $H_{\rm kin}$ and $H_{\rm pot}$
can be computed {\rm exactly}.  This gives rise to the following {\em drift}
and {\em kick} operators \citep{Qu97}:
\be D_t(\Delta t) :\left\{
\begin{array}{ccl} 
\vec{p}_i & \mapsto & \vec{p}_i   \\
\vec{x}_i & \mapsto & \vec{x}_i +  \frac{\vec{p}_i}{m_i}\int_t^{t+\Delta
  t}\frac{{\rm d}t}{a^2}  \\
\end{array}
\right.
\ee
\be
K_t(\Delta t): \left\{
\begin{array}{ccl} 
\vec{x}_i & \mapsto & \vec{x}_i   \\
\vec{p}_i & \mapsto & \vec{p}_i + \vec{f}_i \int_t^{t+\Delta
  t}\frac{{\rm d}t}{a}  \\
\end{array}
\right.
\ee
where $\vec{f}_i = -  \sum_j m_i m_j
\frac{\partial \phi(\vec{x}_{ij})}{\partial \vec{x}_i}$ is the
force on particle $i$.

Note that both $D_t$ and $K_t$ are symplectic operators because they are exact
solutions for arbitrary $\Delta t$ for the canonical transformations generated
by the corresponding Hamiltonians.  A time integration scheme can now be
derived by the idea of operator splitting.  For example, one can try to
approximate the time evolution operator $U(\Delta t)$ for an interval $\Delta
t$ by \be \tilde U(\Delta t) = D\left(\frac{\Delta t}{2}\right) K(\Delta t)
\,D\left(\frac{\Delta t}{2}\right), \ee or \be \tilde U(\Delta t) =
K\left(\frac{\Delta t}{2}\right) D(\Delta t) \,K\left(\frac{\Delta
    t}{2}\right), \ee which correspond to the well-known {\em
  drift-kick-drift} (DKD) and {\em kick-drift-kick} (KDK) leapfrog integrators. Both of
these integration schemes are symplectic, because they are a succession of
symplectic phase-space transformations. In fact, $\tilde U$ generates the
exact time evolution of a modified Hamiltonian $\tilde H$.  Using the
Baker-Campbell-Hausdorff identity for expanding $U$ and $\tilde U$, one can
investigate the relation between $\tilde H$ and $H$.  Writing $\tilde H = H+
H_{\rm err}$, one finds \citep{Saha1992} \be H_{\rm err} = \frac{\Delta
  t^2}{12} \left\{ \left\{ H_{\rm kin}, H_{\rm pot}\right\} , H_{\rm kin} +
  \frac{1}{2} H_{\rm pot} \right\} + {\cal O} (\Delta t^4) \label{EqnerrHam} \ee for the
kick-drift-kick leapfrog.  Provided $H_{\rm err} \ll H$, the evolution under
$\tilde H$ will be typically close to that under $H$. In particular, most of
the Poincare integral invariants of $\tilde H$ can be expected to be close to
those of $H$, so that the long-term evolution of $\tilde H$ will remain
qualitatively similar to that of $H$. If $H$ is time-invariant and conserves
energy, then $\tilde H$ will be {\rm conserved} as well. For a periodic
system, this will then usually mean that the energy in the numerical solution
oscillates around the true energy, but there cannot be a long-term secular
trend.

We illustrate these surprising properties of the leapfrog in
Figure~\ref{figKepler}. We show the numerical integration of a Kepler
problem of high eccentricity $e=0.9$, using second order accurate
leapfrog and Runga-Kutta schemes with fixed timestep. There is no-long
term drift in the orbital energy for the leapfrog result (top panel),
only a small residual precession of the elliptical orbit is observed.
On the other hand, the Runge-Kutta integrator, which has formally the
same error per step, catastrophically fails for an equally large
timestep (middle panel).  Already after 50 orbits the binding energy
has increased by $\sim 30\%$. If we instead employ a fourth order
Runge-Kutta scheme using the same timestep (bottom panel), the
integration is only marginally more stable, giving now a decline of
the binding energy by $\sim 40\%$ over 200 orbits. Note however that
such a higher order integration scheme requires several force
computations per timestep, making it computationally much more
expensive for a single step than the leapfrog, which requires only one
force evaluation per step. The underlying mathematical reason for the
remarkable stability of the leapfrog integrator observed here lies in
its symplectic properties.

\subsection{Individual and adaptive timesteps}

In cosmological simulations, we are confronted with a large dynamic range in
timescales. In high-density regions, like at the centres of galaxies, orders
of magnitude smaller timesteps are required than in low-density regions of the
intergalactic medium, where a large fraction of the mass resides. Evolving all
particles with the smallest required timestep hence implies a substantial
waste of computational resources. An integration scheme with individual
timesteps tries to cope with this situation more efficiently. The principal
idea is to compute forces only for a certain group of particles in a given
kick operation, with the other particles being evolved on larger timesteps and
being `kicked' more rarely.

\begin{figure}
\bc 
\vspace*{-0.5cm}\resizebox{7.2cm}{!}{\includegraphics{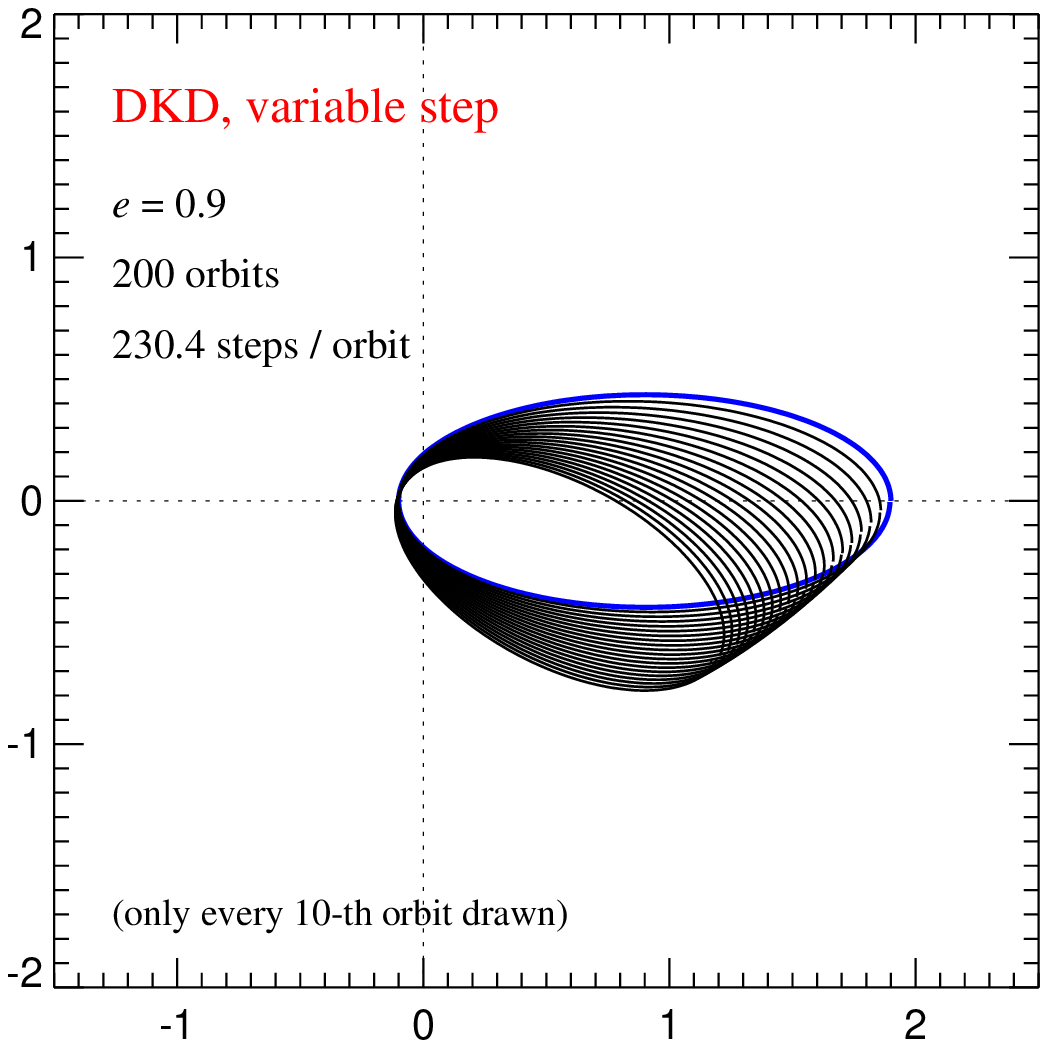}}\\%
\vspace*{-0.5cm}\resizebox{7.2cm}{!}{\includegraphics{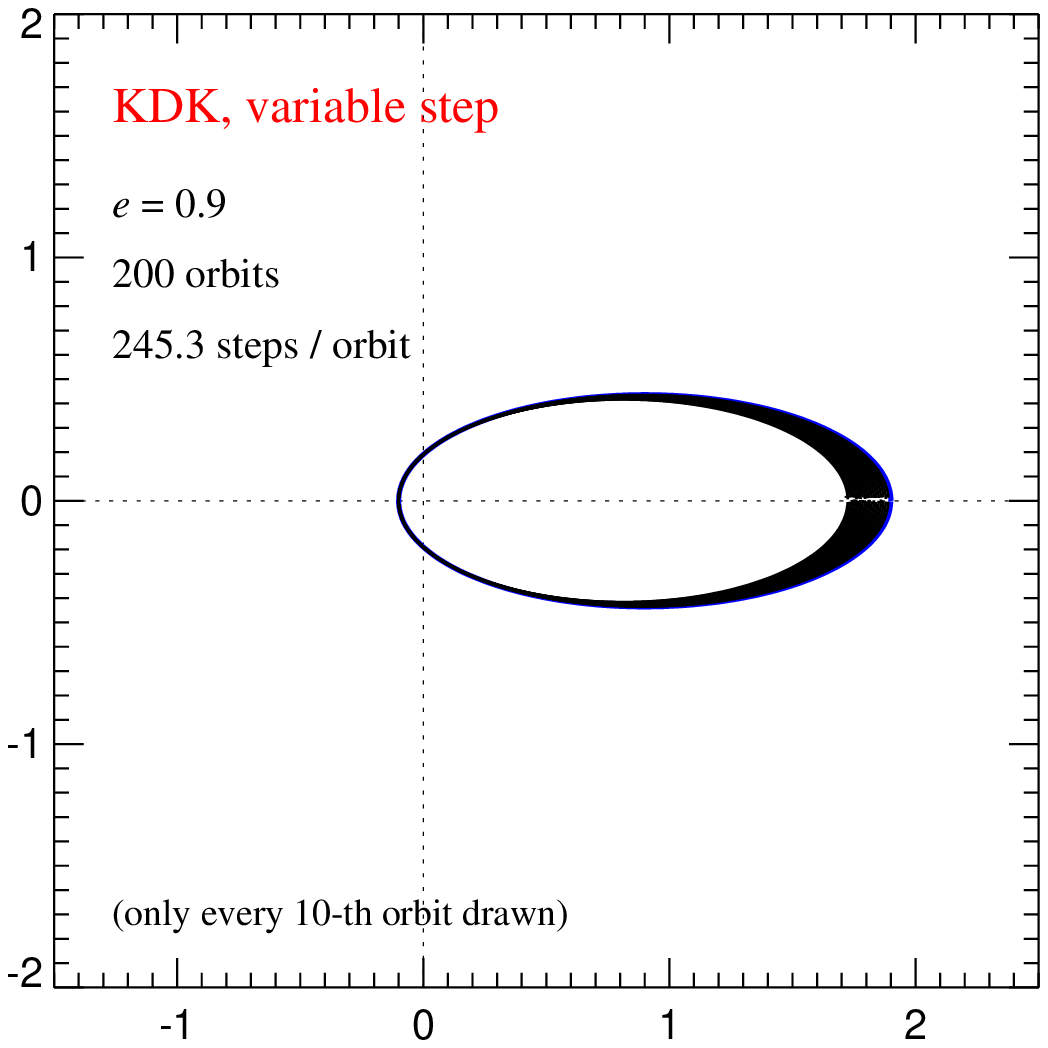}}\\%
\vspace*{-0.25cm}\caption{A Kepler problem of high eccentricity
integrated with leapfrog schemes using a variable timestep from step
to step, based on the $\Delta t \propto 1/\sqrt{|\vec{a}|}$ criterion
commonly employed in cosmological simulations. As a result of the
variable timesteps, the integration is no longer manifestly time
reversible, and long term secular errors develop. Interestingly, the
error in the KDK variant grows four times slower than in the DKD
variant, despite being of equal computational cost.
\label{figKeplerDKDvsKDK}}
\ec
\end{figure}

Unfortunately, due to the pairwise coupling of particles, a formally
symplectic integration scheme with individual timesteps is not
possible, simply because the potential part of the Hamiltonian is not
separable.  However, one can partition the potential between two
particles into a long-range and a short-range part, as we have done in
the TreePM algorithm.  This leads to a separation of the Hamiltonian
into \be H = H_{\rm kin} + H_{\rm sr} + H_{\rm lr}.\ee We can now
easily obtain symplectic integrators as a generalisation of the
ordinary leapfrog schemes by ``subcycling'' the evolution under
$H_{\rm kin} + H_{\rm sr}$ \citep{Duncan1998}. For example, we can
consider \bea \tilde U(\Delta t) = & &  \label{eqnIntschem} \\&\hspace*{-1cm} K_{\rm
lr}\left(\frac{\Delta t}{2}\right) \left[ K_{\rm sr}\left(\frac{\Delta
t}{2m}\right) D\left(\frac{\Delta t}{m}\right) K_{\rm
sr}\left(\frac{\Delta t}{2m}\right) \right]^{m} \,K_{\rm
lr}\left(\frac{\Delta t}{2}\right)  & \nonumber
\eea with $m$ being a positive integer. This is the scheme {\small
GADGET-2} uses for integrating simulations run with the TreePM
algorithm. The long-range PM force has a comparatively large timestep,
which is sufficient for the slow time-variation of this force.  Also,
we always evaluate this force for all particles. The evolution under
the short-range force, however, which varies on shorter timescales, is
done on a power of 2 subdivided timescale. Here, we optionally also
allow particles to have individual timesteps, even though this
perturbs the symplectic nature of the integration (see below). Note
that unlike the PM algorithm, tree forces can be easily computed for a
small fraction of the particles, at a computational cost that is to
first order strictly proportional to the number of particles
considered. This is true as long as the subfraction is not so small
that tree construction overhead becomes significant. PM-forces on the
other hand are either ``all'' or ``nothing''.  The above decomposition
is hence ideally adjusted to these properties.

\begin{figure}
\bc
\resizebox{8.5cm}{!}{\includegraphics{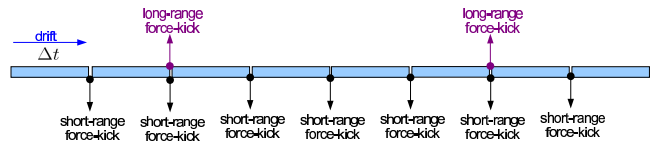}}%
\caption{Schematic illustration of the short- and long-range
  timestepping used by {\small GADGET-2}. The code always drifts the
  whole particle system to the next time when a force computation is
  required. At that time `kicks', i.e.~changes of the particle
  momenta, are applied based on short-range or long-range forces, or
  on both.
\label{FigIntSketch}}
\ec
\end{figure}

Note that despite the somewhat complicated appearance of equation
(\ref{eqnIntschem}), the integration scheme is still a simple
alternation of drift and kick operators. In practice, the simulation
code simply needs to drift the whole particle system to the next
synchronisation point where a force computation is necessary. There, a
fraction of the particles receive a force computation and their
momenta are updated accordingly, as illustrated in the sketch of
Figure~\ref{FigIntSketch}. Then the system is drifted to the next
synchronisation point.

As we have discussed, the integration is no longer symplectic in a formal
sense when individual short-range timesteps are chosen for different
particles.  However, in the limit of collisionless dynamics, we can argue that
the particle number is so large that particles effectively move in a
collective potential, where we assume that any force between two particles is
always much smaller than the total force. In this desired limit, two-body
collisions become unimportant, and the motion of particles is to good
approximation collisionless. We can then approximate the particles as moving
quasi-independently in their collective potential, which we may describe by a
global potential $\Phi(\vec{x},t)$. Obviously, in this approximation the
Hamiltonian separates into a sum of single particle Hamiltonians, where we
have now hidden their coupling in the collective potential $\Phi(\vec{x},t)$.
Provided we follow the evolution of each particle accurately in this fiducial
collective potential $\Phi(\vec{x},t)$, the evolution of the potential itself
will also be faithful, justifying the integration of particles with individual
timesteps in a N-body system that behaves collisionlessly.  While not formally
being symplectic, the evolution can then be expected to reach comparable
accuracy to a phase-space conserving symplectic integration.


Treating the potential as constant for the duration of a kick, each
particle can be integrated by a sequence of KDK leapfrogs, which may
have different timestep from step to step. Note that changing the
timestep in the leapfrog from step to step does not destroy the
simplecticity of the integrator, because the implied transformation is
constructed from steps which are simplectic individually. However,
what one finds in practice is that the superior long-term stability of
periodic motion is typically lost.  This is because each time the
timestep is changed, the error Hamiltonian appearing in equation
(\ref{EqnerrHam}) is modified.  This introduces an artificial temporal
dependence into the numerical Hamiltonian which is not in phase with
the orbit itself because the timestep criterion usually involves
information from the previous timestep. The associated time-asymmetry
destroys the formal time reversibility of the integration, and the
phase-lag of the timestep cycle in each orbit produces a secular
evolution. We illustrate this behaviour in
Figure~\ref{figKeplerDKDvsKDK} for an integration of the Kepler
problem considered earlier, but this time using a leapfrog with an
adaptive timestep according to $\Delta t\propto 1/\sqrt{|\vec{a}|}$,
where $\vec{a}$ is the acceleration of the last timestep.
Interestingly, while being equivalent for a fixed timestep, the DKD
and KDK leapfrogs behave quite differently in this test. For the same
computational effort, the energy error grows four times as fast in the
DKD scheme compared with the KDK scheme. This is simply because the
effective time asymmetry in the DKD scheme is effectively twice as
large. To see this, consider what determines the size of a given
timestep when integrating forward or backwards in time. In the DKD
scheme, the relevant acceleration that enters the timestep criterion
stems from a moment that lies half a timestep before or behind the
given step. As a result, there is temporal lapse of two timesteps
between forward and backwards integration. For the KDK, the same
consideration leads only to a temporal asymmetry of one timestep, half
as large.

The KDK scheme is hence clearly superior once one allows for individual
timesteps. It is also possible to try to recover time reversibility more
precisely. \citet{Hut1995} discuss an implicit timestep criterion that depends
both on the beginning and end of the timestep, and similarly, \citet{Qu97}
discuss a binary hierarchy of trial steps that serves a similar purpose.
However, these schemes are computationally impractical for large collisionless
systems.  But fortunately, here the danger to build up large errors by
systematic accumulation over many periodic orbits is much smaller, because the
gravitational potential is highly time-dependent and the particles tend to
make comparatively few orbits over a Hubble time.

In the normal integration mode of {\small GADGET-2}, we discretise the
timesteps in a power of 2 hierarchy, where all timesteps are a power
of 2 subdivision of a global timestep. Particles may always move to a
smaller timestep, but to a larger one only every second step, when
this leads to synchronisation with the higher timestep hierarchy. The
level of synchronisation achieved by this is beneficial for minimising
the required number of particle drifts and tree constructions.
Alternatively, the code also allows a more flexible way to populate
timesteps, where timesteps are discretised as integer multiples of the
minimum timestep occuring among the particle set. This has the
advantage of producing a more homogenous distribution of particles
across the timeline, which can simplify work-load balancing.

\subsection{Time integration scheme of SPH particles}

For gas particles, similar considerations apply in principle, since in the
absence of viscosity, SPH can also be formulated as a Hamiltonian system.
However, because shocks occur in any non-trivial flow, hydrodynamics will in
practice always be irreversible, hence the long-term integration aspects of
Hamiltonian systems do not apply as prominently here. Also, in systems in
hydrodynamic equilibrium the gas particles do not move, and hence do not tend
to accumulate errors over many orbits as in dynamical equilibrium.  However,
if SPH particles are cold and rotate in a centrifugally supported disk,
long-term integration aspects can become important again. So it is desirable
to treat the kinematics of SPH particles in close analogy to that of the
collisionless particles.

The reversible part of hydrodynamics can be described by adding the thermal
energy to the Hamiltonian, viz.  \be H_{\rm therm} = \frac{1}{\gamma -1
}\sum_{i} m_i A_i \rho_i^{\gamma -1} .  \ee Note that the SPH smoothing
lengths are implicitly given by equations (\ref{eqhsml}), i.e.~the thermal
energy depends only on the entropy per unit mass, and the particle
coordinates. Hence the same considerations as for the collisionless leapfrog
apply, and as long as there is no entropy production included, time
integration is fully time reversible.  This is actually a considerable
difference to mesh codes which in non-trivial flows always produce some
entropy due to mixing, even when the fluid motion should in principle be fully
adiabatic. These errors arise from the advection over the mesh, and are absent
in the above formulation of SPH.

\section{Parallelisation strategies}  \label{SecParallel}

There are a number of different design philosophies for constructing powerful
supercomputers. So-called {\em vector}-machines employ particularly potent
CPUs which can simultaneously carry out computational operations on whole
arrays of floating point numbers.  However, not all algorithms can easily
exploit the full capabilities of such vector processors. It is easier to use
{\em scalar} architectures, but here large computational throughput is only
achieved by the simultaneous use of a large number of processors. The goal is
to let these CPUs work together on the same problem, thereby reducing the time
to solution and allowing larger problem sizes. Unfortunately, the required
{\em parallelisation} of the application program is not an easy task in
general.

On symmetric multiprocessing (SMP) computers, several scalar CPUs share the
same main memory, so that time-intensive loops of a computation can be
distributed easily for parallel execution on several CPUs using a technique
called threading. The code for creation and destruction of threads can
be generated automatically by sophisticated modern compilers, guided by hints
inserted into the code in the form of compiler directives (e.g. based on the
OpenMP standard).  The primary advantage of this method lies in its ease of
use, requiring few (if any) algorithmic changes in existing serial code. A
disadvantage is that the compiler-assisted parallelisation may not always
produce an optimum result, and depending on the code, sizable serial parts may
remain.  A more serious limitation is that this technique prevents one from
using processor numbers and memory larger than available on a particular SMP
computer. Also, such shared-memory SMP computers tend to be {\em
  substantially} more expensive than a set of single computers with comparable
performance, with the price-tag quickly rising the more CPUs are contained
within one SMP computer.

A more radical approach to parallelisation is to treat different scalar CPUs
as independent computers, each of them having their own separate physical
memory, and each of them running a separate instance of the application
code. This approach requires extension of the program with instructions that
explicitly deal with the necessary communication between the CPUs to split up the
computational work and to exchange partial results.  Memory is {\em
  distributed} in this method. In order to allow a scaling of the problem size
with the total available memory, each CPU should only store  a fraction of
the total data of the problem in its own memory.  Successful implementation of
this paradigm therefore requires substantial algorithmic changes compared to
serial programs, and depending on the problem, a considerably higher
complexity than in corresponding serial codes may result. However, such {\em
  massively parallel} programs have the potential to be scalable up to very
large processor number, and to exploit the combined performance of the CPUs in
a close to optimum fashion.  Also, such codes can be run on computers of
comparatively low cost, like clusters of ordinary PCs.

{\small GADGET-2} follows this paradigm of a massively parallel simulation
code. It contains instructions for communication using the standardised
`Message Passing Interface' (MPI). The code itself was deliberately written
using the language C (following the ANSI-standard) and the open-source
libraries GSL and FFTW.  This results in a very high degree of portability to
the full family of UNIX systems, without any reliance on special features of
proprietary compilers. The parallelisation algorithms of the code are flexible
enough to allow its use on an arbitrary number of processors, including just
one. As a result {\small GADGET-2} can be run on large variety of machines,
ranging from a laptop to clusters of the most powerful SMP computers presently
available. In the following, we describe in more detail the parallelisation
algorithms employed by the code.

\subsection{Domain decomposition and Peano-Hilbert order} \label{SecPeano}

Since large cosmological simulations are often memory-bound, it is essential
to decompose the full problem into parts that are suitable for distribution to
individual processors.  A commonly taken approach in the gravitational
N-body/SPH problem is to decompose the computational volume into a set of
domains, each assigned to one processor.  This has often been realised with a
hierarchical orthogonal bisection, with cuts chosen to approximately balance
the estimated work for each domain \cite[e.g.][]{Du96b}.  However, a
disadvantage of some existing implementations of this method is that the geometry
of the tree eventually constructed for each domain depends on the geometry of
the domains themselves. Because the tree force is only an approximation, this
implies that individual particles may experience a different force error when
the number of CPUs is changed, simply because this in general modifies the way
the underlying domains are cut.  Of course, provided the typical size of force
errors is sufficiently small, this should not pose a severe problem for the
final results of collisionless simulations.  However, it complicates code
validation, because individual particle orbits will then depend on the number
of processors employed. Also, there is the possibility of subtle correlations
of force errors with domain boundaries, which could especially in the very
high redshift regime show up as systematic effects.

\begin{figure*}
\bc
\resizebox{5.3cm}{!}{\includegraphics{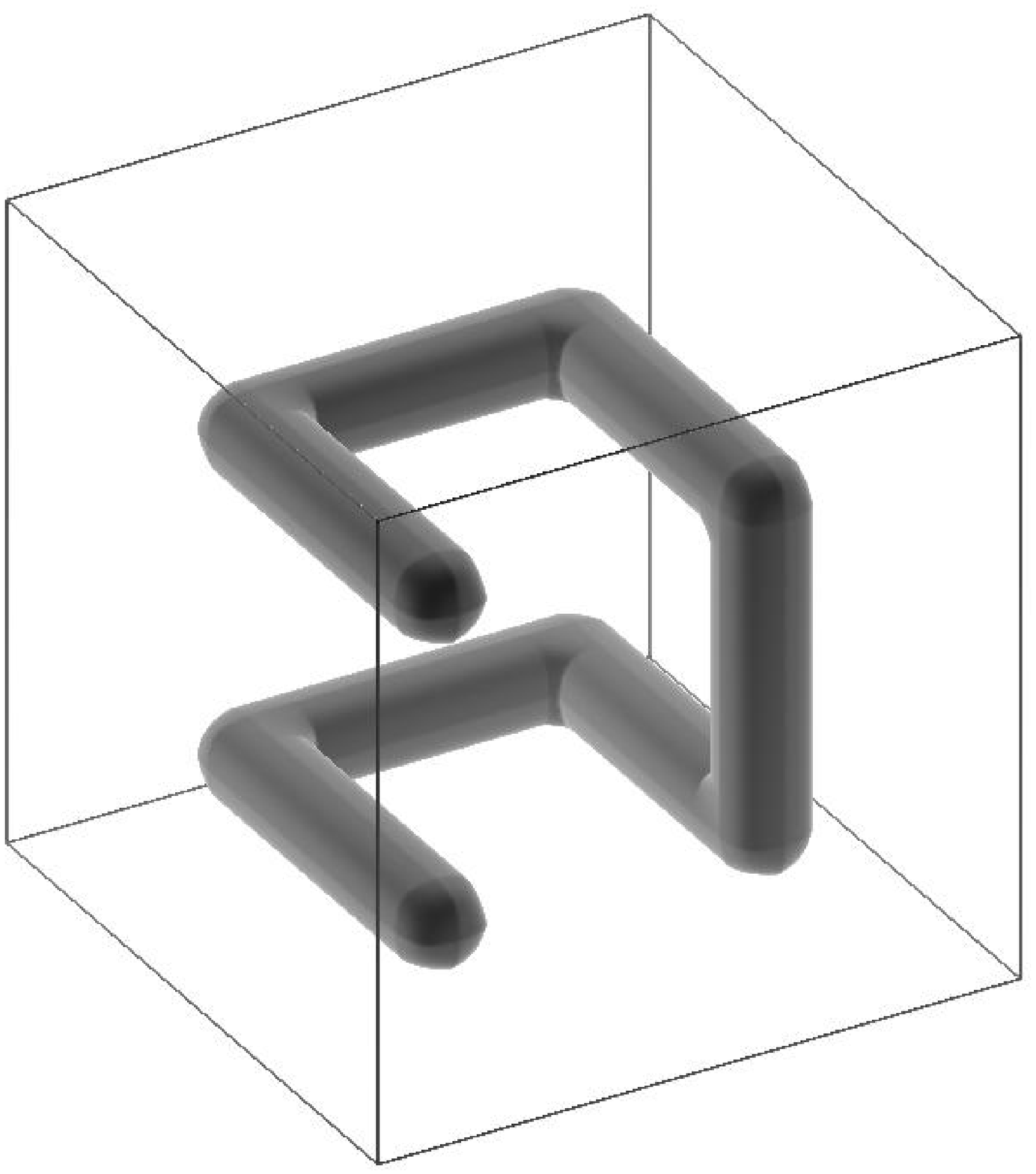}}%
\resizebox{5.3cm}{!}{\includegraphics{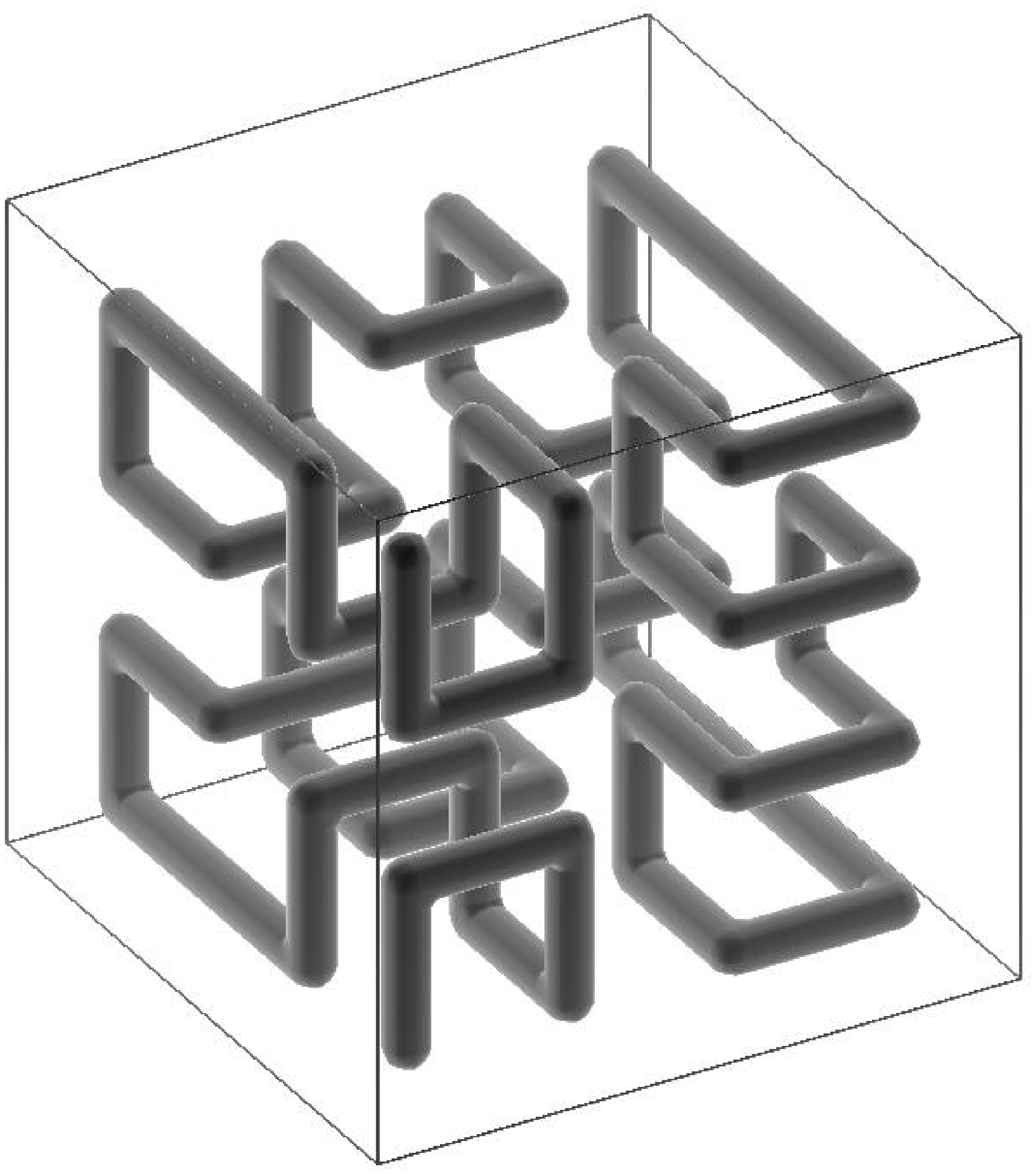}}%
\resizebox{5.3cm}{!}{\includegraphics{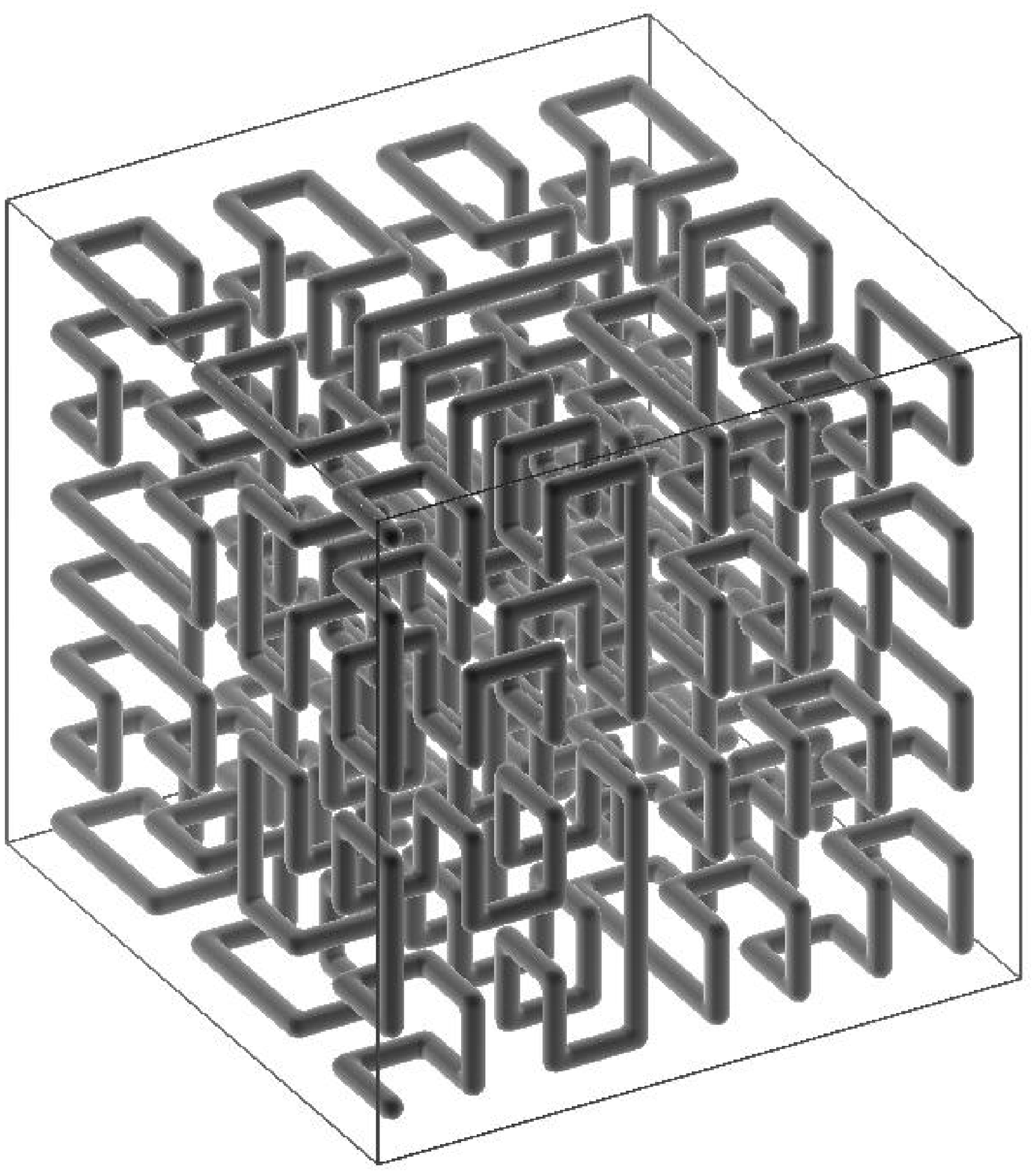}}\\
\ \\
\resizebox{11.0cm}{!}{\includegraphics{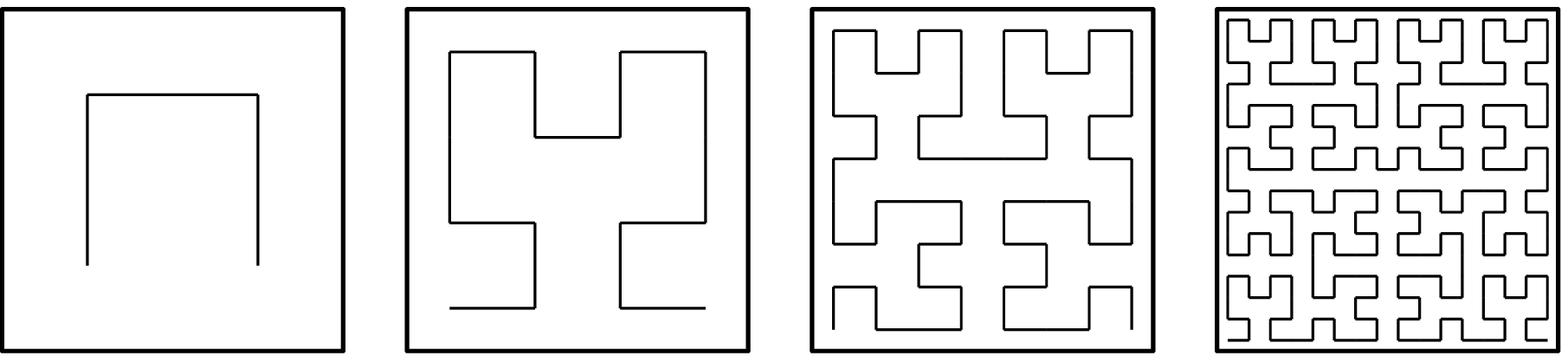}}\\
\caption{Space-filling Peano-Hilbert curve in two (bottom) and three
  (top) dimensions.
\label{figPeanoHilbert}}
\ec
\end{figure*}

Here we propose a new scheme for domain decomposition that guarantees
a force that is independent of the processor number. It also avoids
other shortcomings of the orthogonal bisection, such as high
aspect-ratios of domains. Our method uses a space-filling fractal, the
Peano-Hilbert curve, to map 3D space onto a 1D curve. The latter is
then simply chopped off into pieces that define the individual
domains. The idea of using a space-filling curve for the domain
decomposition of a tree code has first been proposed by
\citet{Warren1993,Warren1995}. They however used Morton ordering
for the underlying curve, which produces irregularly shaped domains.

In Figure~\ref{figPeanoHilbert}, we show examples of the Peano-Hilbert curve
in two and three dimensions.  The Peano-curve in 2D can be constructed
recursively from its basic `U'-shaped form that fills a $2\times2$ grid,
together with the rules that determine the extension of this curve onto a
$4\times 4$ grid. As can be seen in Fig~\ref{figPeanoHilbert}, these rules
mean that the bar of the ``U'' has to be replaced with two smaller copies of
the underlying ``U'', while at the two ends, rotated and mirrored copies have
to be placed. By repeated application of these rules one can construct an
area-filling curve for arbitrarily large grids of size $2^n\times 2^n$.  In
three dimensions, a basic curve defined on a $2\times2\times 2$ grid can be
extended in an analogous way, albeit with somewhat more complicated mapping
rules, to the three dimensional space-filling curve shown in
Fig.~\ref{figPeanoHilbert}.

An interesting property of these space-filling curves is their
self-similarity. Suppose we describe the Peano-Hilbert curve that fills a $2^n
\times 2^n \times 2^n$ grid with a one-to-one mapping $p_n(i,j,k)$, where the
value $p_n \in [{0,\ldots,n^3-1}]$ of the function is the position of the cell
$(i,j,k)$ along the curve.  Then we have $p_{n/2}(i/2,j/2,k/2) =
p_n(i,j,k)/8$, where all divisions are to be understood as integer divisions.
We can hence easily ``contract'' a given Peano-Hilbert curve and again obtain
one of lower order. This is a property we exploit in the code.

\begin{figure*}
\bc
\resizebox{16.0cm}{!}{\includegraphics{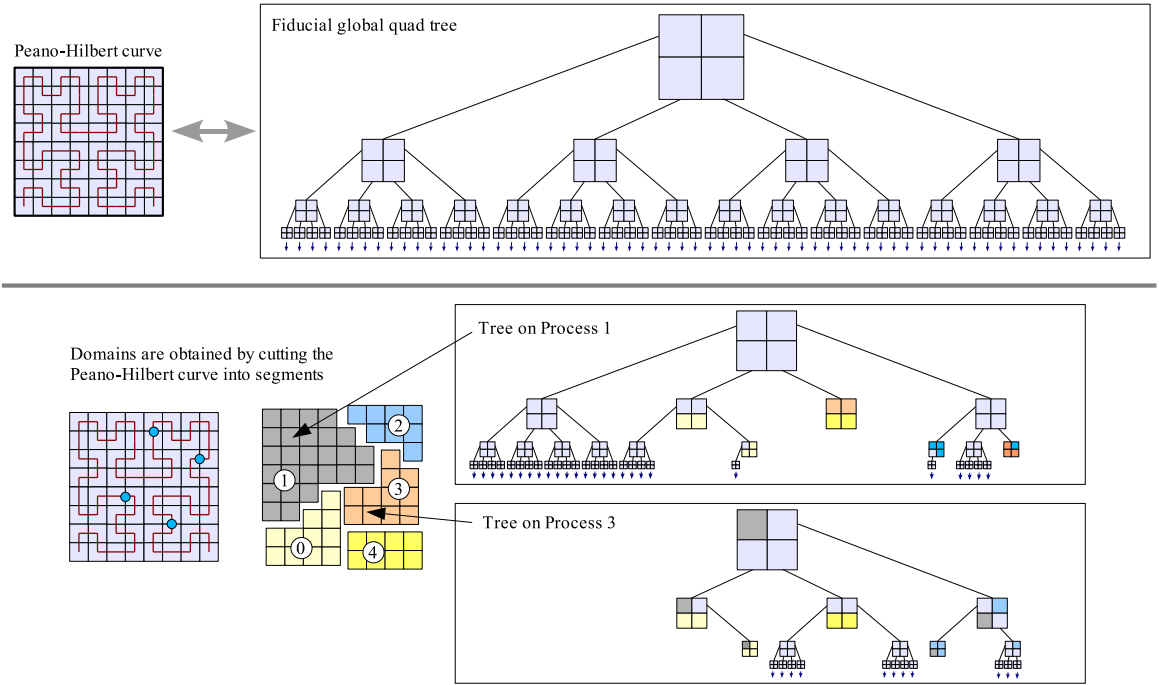}}\vspace*{-1cm}%
\caption{Illustration of the relation between the BH oct-tree and a
  domain decomposition based on a Peano-Hilbert curve. For clarity,
  the sketch is drawn in two dimensions. The fiducial Peano curve
  associated with the simulation volume visits each cell of a regular
  mesh exactly once. The simulation volume is cut into domains by
  segmenting this curve at arbitrary intermediate points on cell
  boundaries. This generates a rule for distributing the particle set
  onto individual processors. Since the geometric structure of the BH
  tree is commensurable with the mesh, each mesh-cell corresponds to a
  certain branch of a fiducial global BH tree. These branches then
  reside entirely on single processors. In addition, each processor
  constructs a `top-level tree' where all nodes at higher level are
  represented. The missing data on other processors is marked using
  `pseudo-particles' in this tree.
\label{figDomainDecomp}}
\ec
\end{figure*}

A second important property is that points that are close along the
one-dimensional Peano-Hilbert curve are in general also close in three
dimensional space, i.e.~the mapping preserves locality. If we simply cut a
space-filling Peano-curve into segments of a certain length, we obtain a
domain decomposition which has the property that the spatial domains are
simply connected and quite `compact', i.e.~they tend to have small
surface-to-volume ratios and low aspect ratios, a highly desirable property
for reducing communication costs with neighbouring domains.

Third, we note that there is a close correspondence between the
spatial decomposition obtained by a hierarchical BH oct-tree, and that
obtained from segmenting a Peano-Hilbert curve. For example, consider
a fiducial Peano-Hilbert curve that fills a box (the root node),
encompassing the whole particle set.  Cutting this curve into 8
equally long pieces, and then recursively cutting each segment into 8
pieces again, we regenerate the spatial oct-tree structure of the
corresponding BH tree. If we hence assign an arbitrary segment of the
Peano-Hilbert curve to a processor, the corresponding volume is
compatible with the node structure of a fiducial global BH tree
covering the full volume, i.e.~we effectively assign a collection of
branches of this tree to each processor. Because of this property, we
obtain a tree whose geometry is not affected by the parallelisation
method, and the results for the tree force become strictly independent
of the number of processors used.

We illustrate these concepts in Figure~\ref{figDomainDecomp}, where we show a
sketch of a global BH tree and its decomposition into domains by a
Peano-Hilbert curve. For simplicity, we show the situation in 2D.  Note that
the sizes of the largest nodes assigned to each processor in this way need not
all be of the same size. Instead, the method can quite flexible adjust to
highly clustered particle distributions, if required.

In order to carry out the domain decomposition in practice, we first
compute a Peano-Hilbert {\em key} for each particle. This is simply
the integer returned by the function $p$, where the coordinates of
particles are mapped onto integers in the range $[0,2^n-1]$. The
construction of the Peano-Hilbert key can be carried out with a number
of fast bit-shift operations, and short look-up tables that deal with
the different orientations of the fundamental figure. We typically use
$n=20$, such that the key fits into a 64-bit long integer, giving a
dynamic range of the Peano-Hilbert curve of $\sim 10^6$ per
dimension. This is enough for all present applications but
could be  easily extended if needed.

In principle, one would then like to sort these keys and divide the
sorted list into segments of approximately constant
work-load. However, since the particle data (including the keys) is
distributed, a global sort is a non-trivial operation. We solve this
problem using an adaptive hashing method.  Each processor first
considers only its locally sorted list of keys and uses it to
recursively construct a set of segments (by chopping segments into 8
pieces of equal length) until each holds at most $s N/N_{\rm cpu}$
particles, where we usually take $s\simeq 0.1$. This operation
partitions the load on each processor into a set of reasonably fine
pieces, but the total number of these segments remains small,
independent of the clustering state of matter.  Next, a global list of
all these segments is established, and segments that overlap are
joined and split as needed, so that a global list of segments
results. It corresponds to a BH tree where the leaf nodes hold of
order $s N/N_{\rm cpu}$ particles. We can now assign one or several
consecutive segments to each processor, with the divisions chosen such
that an approximate work-load balance is obtained, subject to the
constraint of a maximum allowed memory imbalance. The net result of
this procedure is that a range of keys is assigned to each processor,
which defines the domain decomposition and is now used to move the
particles to their target processors, as needed. Note that unlike a
global sort, the above method requires little communication.

For the particles of each individual processor, we then construct a
BH-tree in the usual fashion, using the full extent of the particle
set as the root grid size. In addition, we insert ``pseudo-particles''
into the tree, which represent the mass on all other processors. Each
of the segments in the global domain-list which was not assigned to
the local processor is represented by a pseudo-particle.  In the tree,
these serve as placeholders for branches of the tree that reside
completely on a different processor.  We can inquire the multipole
moments of such a branch from the corresponding remote processor, and
give the pseudo particle these properties. Having inserted the
pseudo-particles into each local tree therefore results in a
``top-level tree'' that complements the tree branches generated by
local particles. The local tree is complete in the sense that all
internal nodes of the top-level tree have correct multipole-moments,
and they are independent of the domain decomposition resulting for a
given processor number. However, the local tree has some nodes that
consist of pseudo-particles. These nodes cannot be opened since the
corresponding particle data resides on a different processor, but when
encountered in the tree walk, we know precisely on which processor
this information resides.

The parallel tree-force computation proceeds therefore as follows. For
each of its (active) local particles, a processor walks its tree in
the usual way, collecting force contributions from a set of nodes,
which may include top-level tree nodes and pseudo particles.  If the
node represented by a pseudo-particle needs to opened, the walk along
the corresponding branch of the tree cannot be continued. In this
case, the particle is flagged for export to the processor the
pseudo-particle came from, and its coordinates are written into a
buffer-list, after which the tree walk is continued. If needed, the
particle can be put several times into the buffer list, but at most
once for each target processor. After all local particles have been
processed, the particles in the buffer are sorted by the rank of the
processor they need to be sent to.  This collects all the data that
needs to be sent to a certain processor in a contiguous block, which
can then be communicated in one operation based on a collective hypercube
communication model. The result is a list of imported particles for
which the local tree is walked yet again. Unlike in the normal tree
walk for local particles, all branches of the tree that do not
exclusively contain local mass can be immediately discarded, since the
corresponding force contributions have already been accounted for by
the processor that sent the particle.  Once the partial forces for all
imported particles have been computed, the results are communicated
back to the sending processors, using a second hypercube
communication. A processor that sent out particles receives in this
way force contributions for nodes that it could not open locally.
Adding these contributions to the local force computed in the first
step, the full force for each local particle is then obtained.  The
forces are independent of the number of processors used and the domain
cuts that where made.  In practice, numerical round-off can still
introduce differences however, since the sequence of arithmetic
operations that leads to a given force changes when the number of CPUs
is modified.

Unlike in {\small GADGET-1}, particles are not automatically exported to other
processors, and if they are, then only to those processors that hold
information that is directly needed in the tree walk. Particularly in the
TreePM scheme and in SPH, this leads to a drastic reduction in the required
communication during the parallel force computations, an effect that is
particularly important when the number of CPUs is large. Since the domains are
locally compact and the tree-walk is restricted to a small short-range region
in SPH and TreePM, most particles will lie completely inside the local domain,
requiring no information from other processors at all. And if they have to be
exported, then typically only to one or a few other processors.  We also
remark that the above communication scheme tends to hide communication
latency, because the processors can work independently on (long) lists of
particles before they meet for an exchange of particles or results.

Finally, we note that we apply the Peano-Hilbert curve for a second purpose as
well. Within each local domain, we order the particles in memory according to
a finely resolved Peano-Hilbert curve. This is done as a pure optimisation
measure, designed to increase the computational speed. Because particles that
are adjacent in memory after Peano-Hilbert ordering will have close spatial
coordinates, they also tend to have very similar interaction lists.  If the
microprocessor works on them consecutively, it will hence in many cases find
the required data for tree-nodes already in local cache-memory, which reduces
wait cycles for the slower main memory. Our test results show that the
Peano-Hilbert ordered particle set can result in nearly twice the performance
compared to random order, even though the actual tree code that is executed is the
same in both cases. The exact speed-up obtained by this trick is architecture-
and problem-dependent, however.

\subsection{Parallel Fourier transforms}

In the TreePM algorithm, we not only need to parallelise the tree algorithm,
but also the PM computations. For the Fourier transforms themselves we employ
the massively parallel version of the FFTW library developed at MIT. The
decomposition of the data is here based on slabs along one coordinate axis.
The Fourier transform can then be carried out locally for the coordinate axes
parallel to the slabs, but the third dimension requires a global transpose of
the data cube, a very communication intensive step which tends to be quite
restrictive for the scalability of massively parallel FFTs, unless the
communication bandwidth of the computer is very high. Fortunately, in most
applications of interest, the cost of the FFTs is so subdominant that even a
poor scaling remains unproblematic up to relatively large processor numbers.

A more important problem lies in the slab data layout required by the FFT,
which is quite different from the, to first order, `cubical' domain
decomposition that is ideal for the tree algorithm.  \citet{Dubinski2004} and
\citet{WhiteM2002} approached this problem by choosing a slab decomposition
also for the tree algorithm. While being simple, this poses severe
restrictions on the combinations of mesh size and processor number that can
be run efficiently. In particular, in the limit of large processor number, the
slabs become very thin, so that work-load balancing can become poor. In
addition, due to the large surface to volume ratio of the thin slabs, the
memory cost of ghost layers required for the CIC assignment and interpolation
schemes can become quite sizable. In fact, in the extreme case of slabs that
are one mesh cell wide, one would have to store three ghost layer zones, which
would then have to come from more than one processor on the `left' and
`right'.

An obvious alternative is to use different decompositions for the tree
algorithm and the PM part. This is the approach {\small GADGET-2} uses. One
possibility would be to swap the data between the Peano-Hilbert decomposition,
and the slab decomposition whenever a PM force computation is necessary.
However, this approach has a number of drawbacks. First of all, it would
require the exchange of a substantial data volume, because almost all
particles and their associated data would have to be moved in the general
case. Second, since the slab decomposition essentially enforces an equal
volume decomposition, this may give rise to large particle-load imbalance in
highly clustered simulations, for example in ``zoom'' simulations. An extreme
case of this problem would be encountered when FFTs with vacuum boundaries are
used. Here at least half of the slabs, and hence processors, would be
completely devoid of particles if the particle set was actually swapped to the
slab decomposition.

We therefore implemented a second possibility, where the particle data
remains in place, i.e.~in the order established for the tree
algorithm. For the FFT, each processor determines by itself with which
slab its local particle data overlaps. For the corresponding patch,
the local particle data is then CIC-binned, and this patch is
transmitted to the processor that holds the slab in the parallel
FFT. In this way, the required density field for each slab is
constructed from the contributions of several processors. In this
scheme only the scalar density values are transmitted, which is a
substantially smaller data volume than in the alternative scheme, even
when the PM grid is chosen somewhat larger than the effective particle
grid. After the gravitational potential has been computed, we collect
in the same way the potential for a mesh that covers the local
particle set. We can here pull the corresponding parts from the slabs
of individual processors, including the ghost layers required around
the local patch for finite differencing of the potential. Since the
local domains are compact, they have a much smaller surface-to-volume
ratio than the slabs, so that the memory cost of the ghost layers
remains quite small. After the local patch of the potential has been
assembled, it can be finite differenced and interpolated to the
particle coordinates without requiring any additional
communication. This method hence inlines the PM computation in a quite
flexible way with the tree algorithm, without putting any restriction
on the allowed processor number, and avoiding, in particular, the
memory- and work-load balancing issues mentioned above.

\subsection{Parallel I/O}

Current cosmological simulations have reached a substantial size, with
particle numbers well in excess of $10^7$ done quite routinely. Time
slices of such simulations can reach up to a few GByte in size, at
which point it becomes very time-consuming to write or read this data
sequentially on a single processor. Also, it can be impractical to
store the data in a single file. {\small GADGET-2} therefore allows
simulation data to be split across several files. Each file is written
or read by one processor only, with data sent to or received by  a
group of processors. Several of these file can be processed in
parallel. This number can be either equal to the total number of files
requested, or restricted to a smaller value in order to prevent a
``flooding'' of the I/O subsystem of the operating system, which can
be counterproductive. Unlike in previous versions of the code, 
{\small GADGET-2} does not pose restrictions on the number of files
and the number of simultaneously processed files in relation to the
number of processors that is used.

In the largest simulation carried out with {\small GADGET-2} thus far,
a simulation with $2160^3$ particles \citep{SpringelMS2005}, the total
size of a snapshot slice was more than 300 GB. Using parallel I/O on
the high-performance IBM p690 system of the MPG computing centre in
Garching, these time slices could be written in slightly less than 300
seconds, translating in an effective disk bandwidth of $\sim{\rm 1\,
GB/sec}$. Without parallel I/O, this would have taken a factor $\simeq
50-60$ longer.

\subsection{Miscellaneous features}

We note that unlike previous versions, {\small GADGET-2} can be run on
an arbitrary number of processors, including a single processor. There
is hence no need any more for separate serial and parallel
versions. Lifting the restriction for the processor numbers to be
powers of two can be quite useful, particularly for loosely coupled
clusters of workstations, where windows of opportunity for simulations
may arise that offer `odd' processor numbers for production runs.

This flexibility is achieved despite the code's use of a communication
model that operates with synchronous communication exclusively. The
principal model for communication in the force computations follows a
hypercube strategy.  If the processor number is a power of two,
say~$2^p$, then a full all-to-all communication cycle can be realized
by $2^p-1$ cycles, where in each cycle $2^{p-1}$ disjoint processor
pairs are formed that exchange messages.  If the processor number is
not a power of two, this scheme can still be used, but the processors
need to be embedded in the hypercube scheme corresponding to the next
higher power of two. As a result, some of the processors will be
unpaired in a subfraction of the communication cycle, lowering the
overall efficiency somewhat.

\begin{figure}
\bc
\resizebox{8.4cm}{!}{\includegraphics{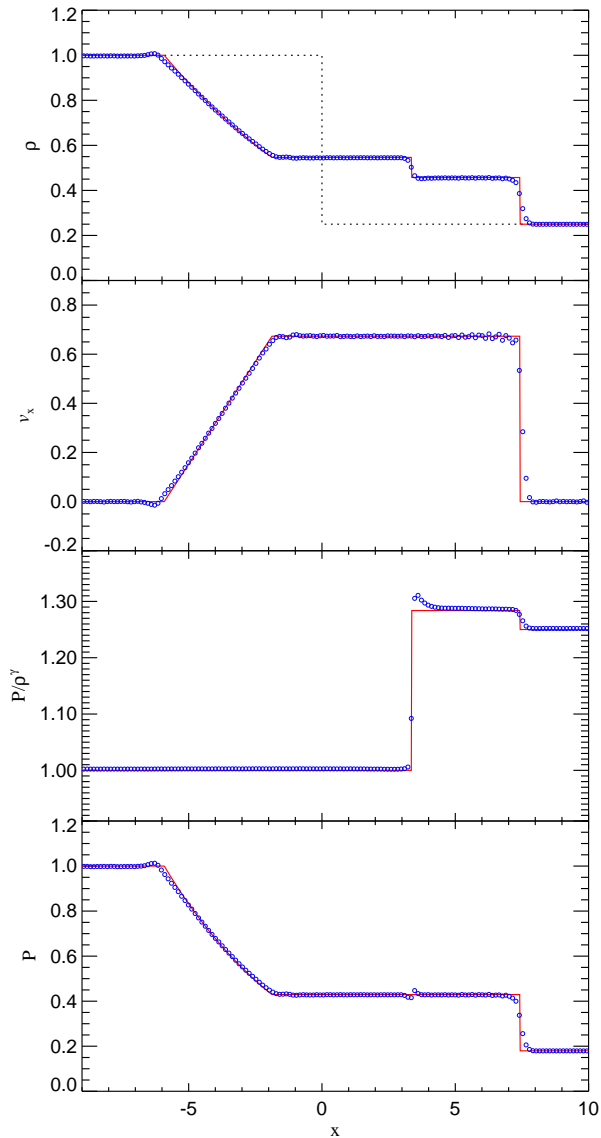}}
\caption{Sod shock test carried out in three dimensions. The gas is initially
  at rest with $\rho_1=1.0$, $P_1=1.0$ for $x<0$, and $\rho_2=0.25$,
  $P_2=9.1795$ for $x>0$. The numerical result is shown with circles (with a
  spacing equal to the mean particle spacing in the low-density region) and
  compared with the analytic result at $t=5.0$ . A shock of Mach-number
  $M=1.48$ develops.
\label{figSod}}
\ec
\end{figure}

{\small GADGET-2} can also be used to set-up `glass' initial conditions, as
suggested by \citet{WhiteSDM1996}. Such a particle distribution arises when a
Poisson sample in an expanding periodic box is evolved with the sign of
gravity reversed until residual forces have dropped to negligible values. The
glass distribution then provides an alternative to a regular grid for use as
an unperturbed initial mass distribution in cosmological simulations of
structure formation. To speed up convergence, the code uses an ``inverse
Zel'dovich'' approximation based on the measured forces to move the particles
to their estimated Lagrangian positions.

We have also added the ability to simulte gas-dynamical simulations in two
dimensions, both with and without periodic boundary conditions.
A further new feature in {\small GADGET-2} is the optional use of the {\em
  Hierarchical Data Format} (HDF5), developed by NCSA. This allows storage of
snapshot files produced by {\small GADGET-2} in a platform independent form,
simplifying data exchange with a variety of analysis software.

\section{Test problems}   \label{SecTests}

Unfortunately, it is not possible to formally demonstrate the
correctness of complex simulation codes such as {\small
GADGET-2}. However, the reliability of a code can be studied
empirically by applying it to a wide range of problems, under a broad
range of values of nuisance code parameters.  By comparing with known
analytic solutions and other independent numerical methods, an
assessment of the numerical reliability of the method can be
established, which is essential for trusting the results of
simulations where no analytic solutions are known (which is of course the
reason to perform simulations to begin with).

We begin with a simple shock-tube test for the SPH component of
{\small GADGET-2}, which has known analytic solutions.  We then
consider the more elaborate problem of the collapse of a cold sphere
of gas under self-gravity. This three-dimensional problem couples
self-gravity and gas dynamics over a dynamic range similar to that
encountered in structure formation simulations. There are no analytic
solutions, but highly accurate results from 1D shock-capturing codes
exist for comparison. We then move on and consider the highly
dissipative collapse of an isothermal cloud of gas, the ``standard
isothermal test case'' of \citet{Boss1979}, where we carry out a
resolution study that examines the reliability of the onset of
fragmentation.

As a test of the accuracy of the dark matter dynamics, we consider the
dark matter halo mass function and the two-point correlation function
obtained for two $256^3$ simulations of cosmological structure
formation. Our initial conditions are the same as those used recently
by \citet{Heitmann2004} in a comparison of several cosmological
codes. We also use their results obtained for these different codes to
compare with {\small GADGET-2}.

We then consider the formation of the `Santa Barbara cluster'
\citep{Frenk99}, a realistic hydrodynamical simulation of the
formation of a rich cluster of galaxies.  The correct solution for
this complex problem, which is directly tied to our theoretical
understanding of the intracluster medium, is not known. However,
results for {\small GADGET-2} can be compared to the 12 codes examined
in \citet{Frenk99}, which can serve as a broad consistency check.

Finally, we briefly consider a further hydrodynamical test problem,
which involves strong shocks and vorticity generation. This is the
interaction of a blast wave with a cold cloud of gas embedded at
pressure equilibrium in ambient gas. This forms an advanced test of
the capabilities of the SPH solver and has physical relevance for
models of the interstellar medium, for example.

\subsection{Shock tube}

We begin by considering a standard Sod shock-tube test, which provides
a useful validation of the code's ability to follow basic
hydrodynamical phenomena. We consider an ideal gas with $\gamma=1.4$,
initially at rest, where the half-space $x<0$ is filled with gas at
unit pressure and unit density ($\rho_1=1$, $P_1=1$), while $x>0$ is
filled with high pressure gas ($P_2=0.1795$) of lower density
($\rho_2=0.25$). These initial conditions have been frequently used as
a test for SPH codes \citep[e.g.][]{He89,Rasio91,Wadsley2004}. We
realize the initial conditions in 3D using an irregular glass-like
distribution of particles of equal mass, embedded in a periodic box
that is longer in the $x$-direction than in the $y$- and
$z$-directions.

\begin{figure*}
\bc
\resizebox{17cm}{!}{\includegraphics{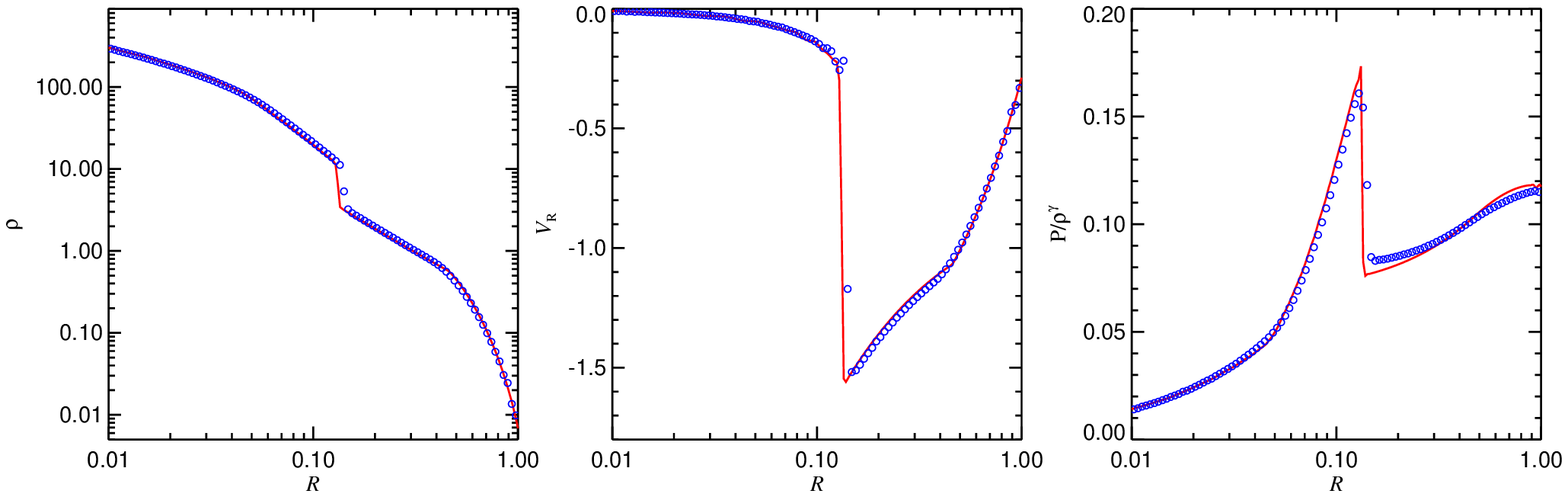}}\\%
\resizebox{17cm}{!}{\includegraphics{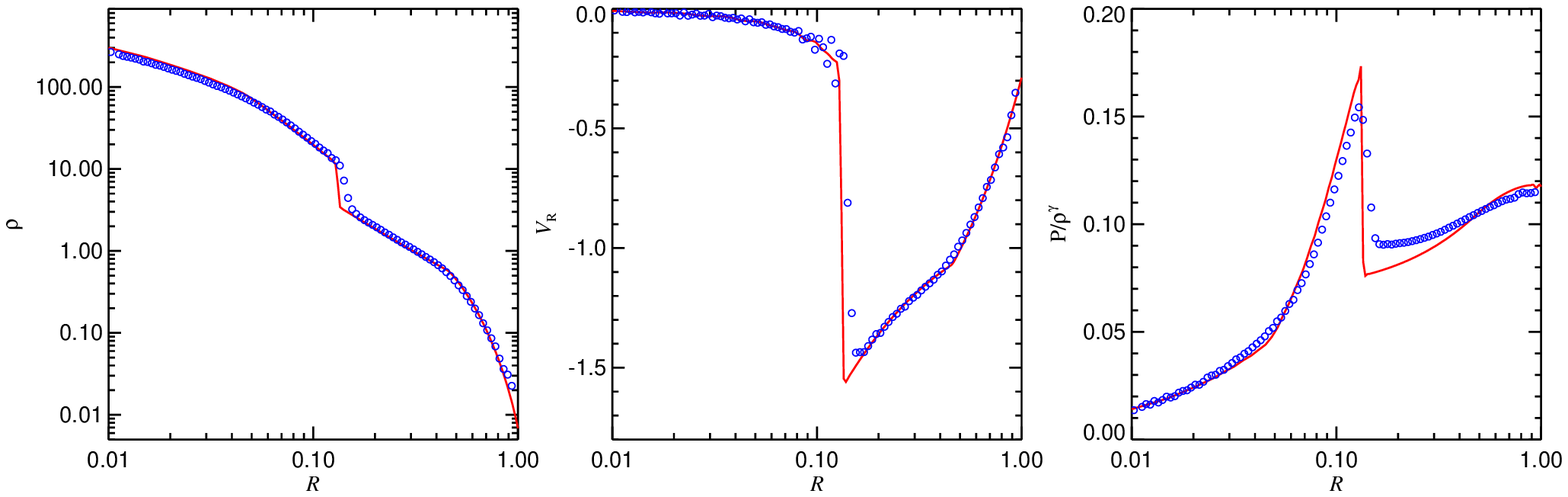}}\\%
\caption{Adiabatic collapse of a gas sphere (`Evrard'-test). At time
  $t=0.8$, we show radial profiles of density, velocity, and entropy
  for two different resolutions, in the top row for $1.56\times 10^6$
  particles, in the bottom row for $1.95\times 10^5$ particles. The
  solid lines mark the result of a 1D PPM calculation
  \citep{Steinmetz1993}, which can be taken as a quasi exact result in
  this case. The 3D SPH calculations reproduce the principal features
  of this solution generally quite well. But as expected, the shock is
  broadened, and also shows some pre-shock entropy generation. The
  latter effect is particularly strong in this spherically symmetric
  problem because of the rapid convergence of the flow in the infall
  region in front of the shock, which triggers the artificial
  viscosity. However, the post-shock properties of the flow are
  correct.
\label{figEvrard}}
\ec
\end{figure*}

In Figure~\ref{figSod}, we show the result obtained with {\small
GADGET-2} at time $t=5.0$. The agreement with the analytic solution is
good, with discontinuities resolved in about $\sim 3$ interparticle
separations, or equivalently $2-3$ SPH smoothing lengths. At the
contact discontinuity, a characteristic pressure blip is observed, and
some excess entropy has been produced there as a result of the sharp
discontinuity in the initial conditions, which has not been smoothed
out and therefore is not represented well by SPH at $t=0$. Note that
while the shock is broadened, the post-shock temperature and density
are computed very accurately.

\subsection{Collapse of an adiabatic gas sphere}

A considerably more demanding test problem is the adiabatic collapse of
an initially cold gas cloud under its own self-gravity. Originally
proposed by \citet{Ev88}, this problem has been considered by many
authors \citep[e.g.][]{He89,Da97,Wadsley2004} as a test of
cosmological codes. The initial conditions in natural units ($G=1$)
take the form of a spherical $\gamma=5/3$ cloud of unit mass and unit
radius, with a $\rho \propto 1/r$ density profile, and with an initial
thermal energy per unit mass of $u=0.05$. When evolved forward in
time, the cloud collapses gravitationally until a central bounce
develops with a strong shock moving outward. 

In Figure~\ref{figEvrard} we show spherically averaged profiles of density,
radial velocity and entropy of the system at time $t=0.8$, and compare it to a
1D high-precision calculation carried out with a PPM scheme by
\citet{Steinmetz1993}. An analytic solution is not available for this problem.
We show results for two different resolutions, $1.56\times 10^6$ and
$1.95\times 10^5$ particles; lower resolution runs are still able to reproduce
the overall solution well, although the shock becomes increasingly more
broadened. We see that for sufficiently high resolution, the 3D SPH
calculation reproduces the 1D PPM result reasonably well. In the region just
outside the shock, we see appreciable pre-shock entropy generation. As pointed
out by \citet{Wadsley2004}, this arises due to the artificial viscosity which
is here already triggered at some level by the strong convergence of the flow
in the pre-shock region. However, this does not lead to any apparent problems
for the post-shock flow. Note that thanks to our entropy formulation, the
entropy profile is also well reproduced at the outer edge of the flow, unlike
the test calculation by \citet{Wadsley2004} using a traditional SPH
formulation.

\begin{figure*}
\bc
\resizebox{8.0cm}{!}{\includegraphics{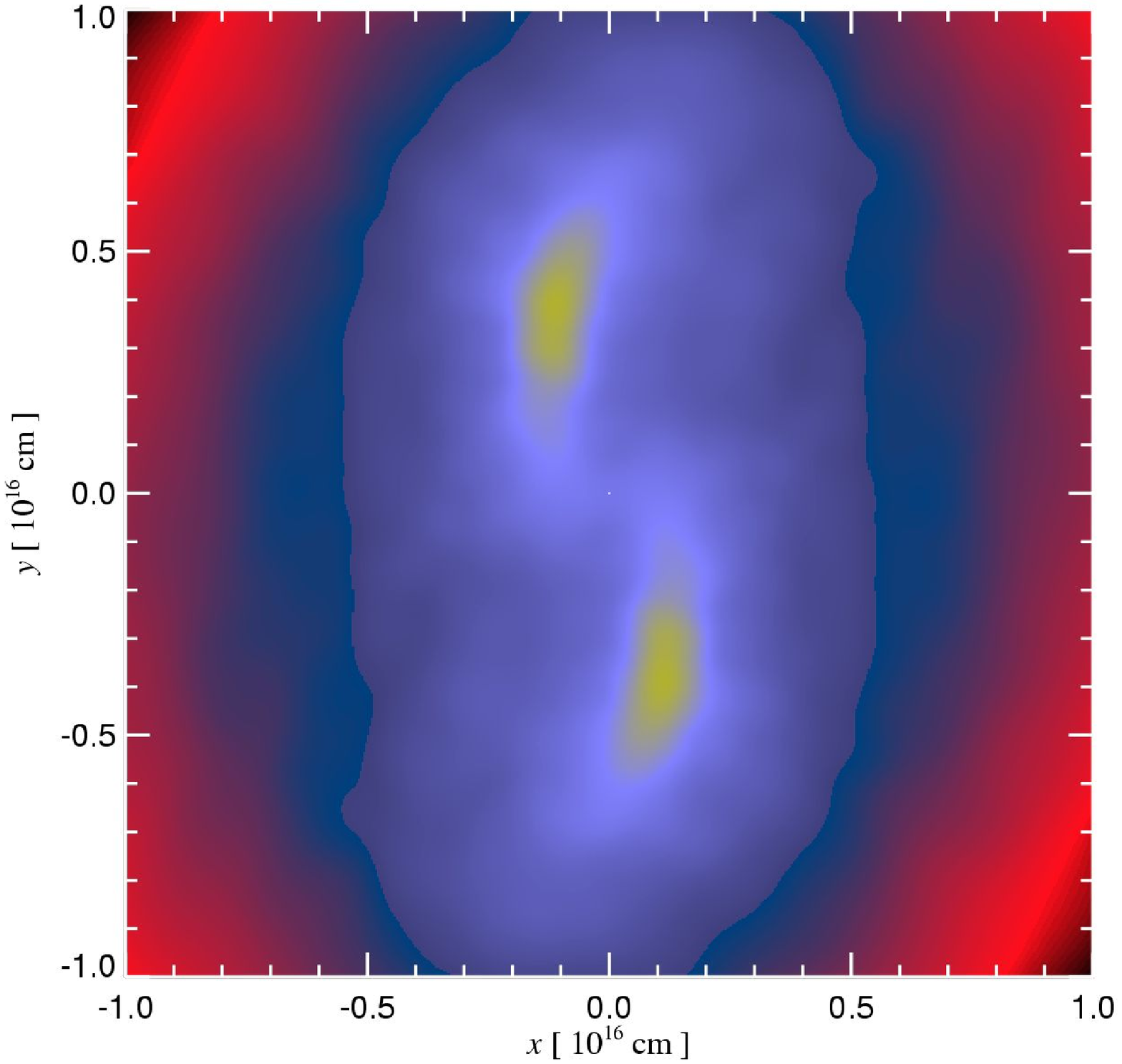}}
\resizebox{8.0cm}{!}{\includegraphics{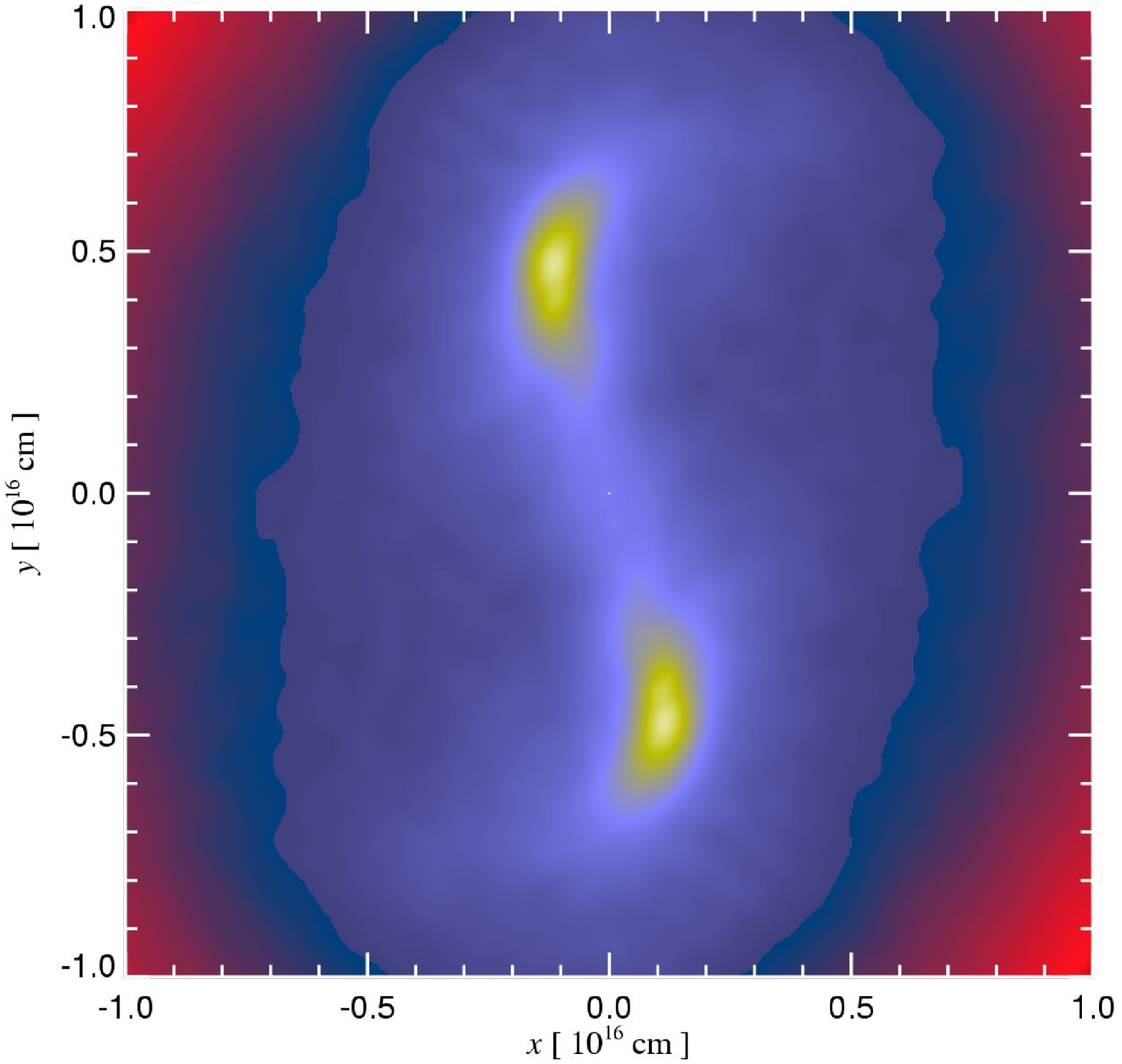}}\\
\resizebox{8.0cm}{!}{\includegraphics{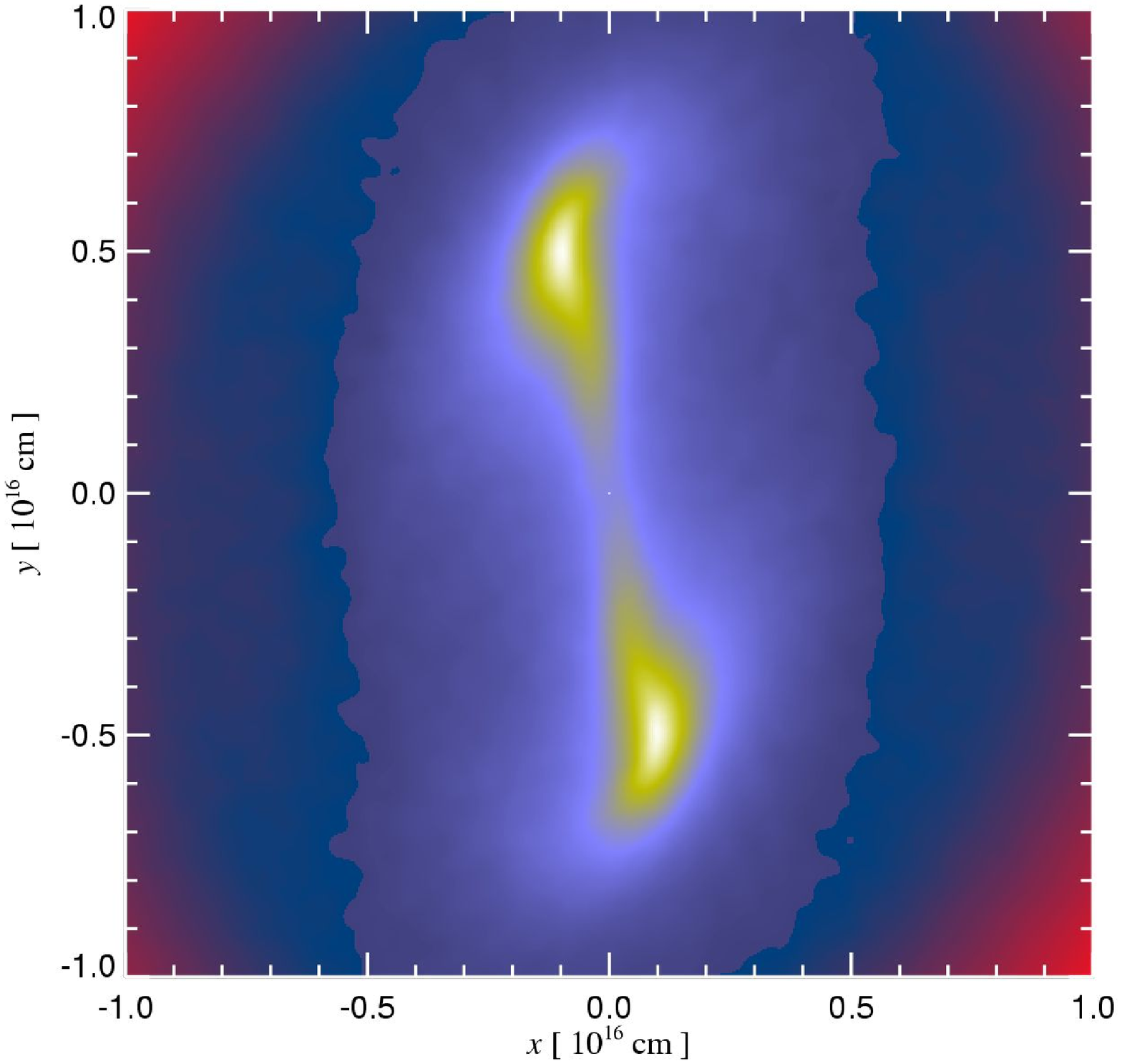}}
\resizebox{8.0cm}{!}{\includegraphics{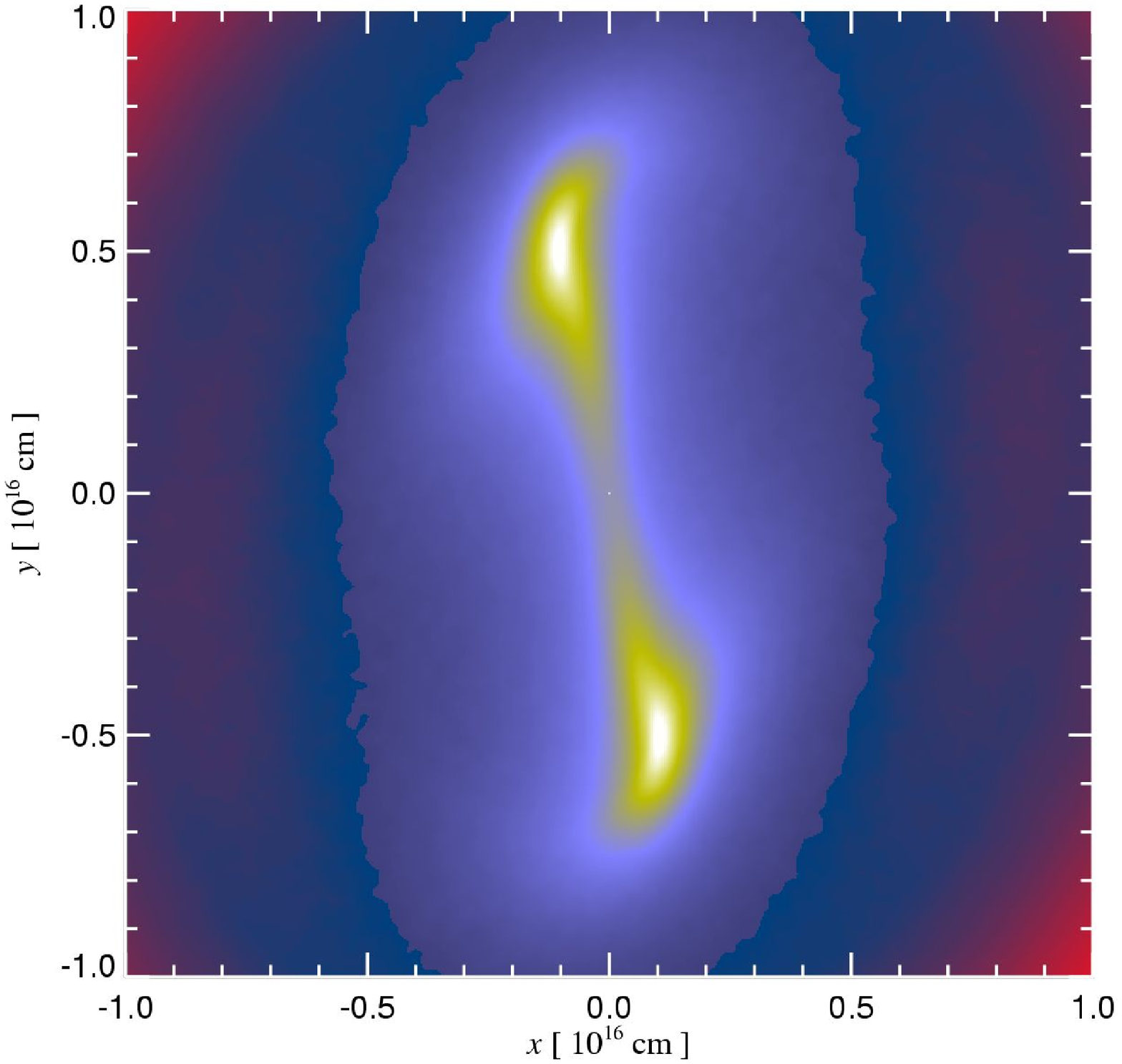}}\\
\caption{Resolution study for the `standard isothermal collapse'
simulation. We show the gas density in a slice trough the centre of
the simulated volume at $1.24$ free fall times, roughly when the two
perturbations at the ends of the bar-like structure become
self-gravitating and undergo gravitational collapse. From the top left
to the bottom row, the particle number increases from $3.3\times 10^4$
to $1.71\times 10^7$ by factors of 8.
\label{figCompIsothermalCollapse}}
\ec
\end{figure*}

\subsection{Isothermal collapse}

Another demanding test problem that couples the evolution under
self-gravity and hydrodynamics is the `standard isothermal test case'
introduced by \citet{Boss1979}. We consider this fragmentation
calculation in the variant proposed by \citet{Burkert1993}, where a
smaller initial non-axisymmetric perturbation is employed; this form
of the initial conditions has been used in numerous test calculations
since then.  The initial state consists of a spherical cloud with
sound speed $c_s = 1.66 \times 10^4 \,{\rm cm\, s^{-1}}$ and an
isothermal equation of state, $P= c_s^2 \rho$. The cloud radius is
$R=5 \times 10^{16} {\rm cm}$, its mass is $M= 1\,{\rm M}_\odot$, and
it is in solid body rotation with an angular velocity of $\omega =
7.2\times 10^{-13}\,{\rm rad \, s^{-1}}$. The underlying constant
density distribution ($\rho_0 = 3.82\times 10^{-18}{\rm g\, cm^{-3}}$)
is modulated with a $m=2$ density perturbation, \be \rho(\phi) =
\rho_0 \left[ 1+ 0.1 \cos(2\phi)\right], \ee where $\phi$ is the
azimuthal angle around the rotation axis. We implement the initial
conditions with a sphere of particles carved out of a regular grid,
where the $10\%$ density perturbation is achieved with a mass
perturbation in the otherwise equal-mass particles.

This simultaneous collapse and fragmentation problem requires high
spatial resolution and accuracy both in the treatment of self-gravity
and in the hydrodynamics. A particular difficulty is that only a small
fraction of the simulated mass eventually becomes sufficiently
self-gravitating to form fragments. As \cite{Bate1997} discuss,
numerical results are only trustworthy if the Jeans mass is resolved
during the calculation. Also, if the gravitational softening is too
large, collapse may be inhibited and the forming clumps may have too
large mass. In fact, \citet{SommerLarsen1998} show that for a finite
choice of softening length an arbitrarily large mass of gas in
pressure equilibrium can be deposited in a nonsingular isothermal
density distribution with radius of order the softening length.  On
the other hand, a gravitational softening much smaller than the SPH
smoothing length can lead to artificial clumping of particles. The
best strategy for this type of fragmentation calculations therefore
appear to be to make the gravitational softening equal to the SPH
softening length, an approach we use in this test calculation. While a
varying gravitational softening formally changes the potential energy
of the system, this energy perturbation can be neglected in the highly
dissipative isothermal case we consider here. Note that once
fragmentation occurs, the density rises rapidly on a free fall
timescale, and the smallest resolved spatial scale as well as the
timestep drop rapidly. This quickly causes the simulation to stall,
unless the dense gas is eliminated somehow, for example by modelling
star formation with sink particles \citep{Bonnell1997}.

\begin{figure}
\bc
\resizebox{8.0cm}{!}{\includegraphics{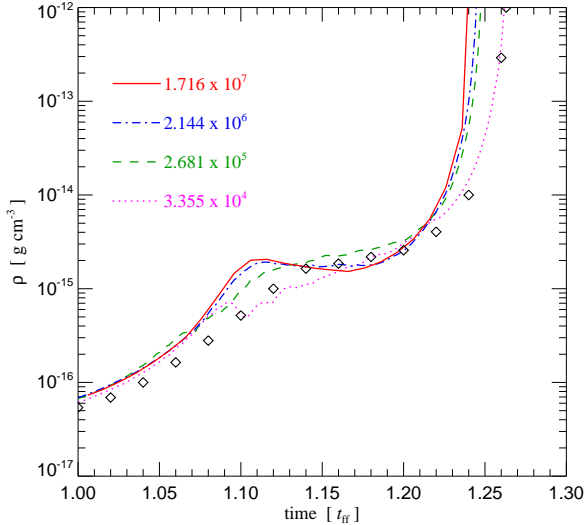}}
\caption{Resolution study for the `standard isothermal collapse'
simulation. We here compare the temporal evolution of the maximum
density reached in simulations of different particle number, as
indicated in the legend. Symbols give the SPH result ($8\times 10^4$
particles) of \citet{Bate1997}, which agrees quite well with our
result at comparable resolution. The small residual differences are
plausibly due to differences in the employed SPH density estimator or
the neighbour number.
\label{figCompMaxDens}}
\ec
\end{figure}

In Figure~\ref{figCompIsothermalCollapse}, we compare the density
fields at $t=1.24$ free fall times in the $z=0$ plane, orthogonal to
the rotation axis, for four different numerical resolutions, ranging
from $3.3\times 10^4$ to $1.71\times 10^7$. At this time, an elongated
bar-like structure has formed with two high-density regions at its
ends. Due to a converging gas-flow onto these ends, they become
eventually self-gravitating and collapse to form two fragments. The
onset of this collapse can be studied in Figure~\ref{figCompMaxDens},
where we plot the maximum density reached in the simulation volume as
a function of time. It can be seen that the three high-resolution
computations converge reasonably well, with a small residual trend
towards slightly earlier collapse times with higher resolution,
something that is probably to be expected. The low-resolution run
behaves qualitatively very similarly, but shows some small oscillations
in the maximum density in the early phases of the collapse. Overall,
our results compare favourably with those of \citet{Bate1997}, but we
are here able to reach higher resolution and are also able to
reproduce more cleanly a first density maximum at $t\simeq 1.1$, which
is also seen in the mesh calculations considered by \citet{Bate1997}.

\begin{figure}
\bc
\resizebox{8.0cm}{!}{\includegraphics{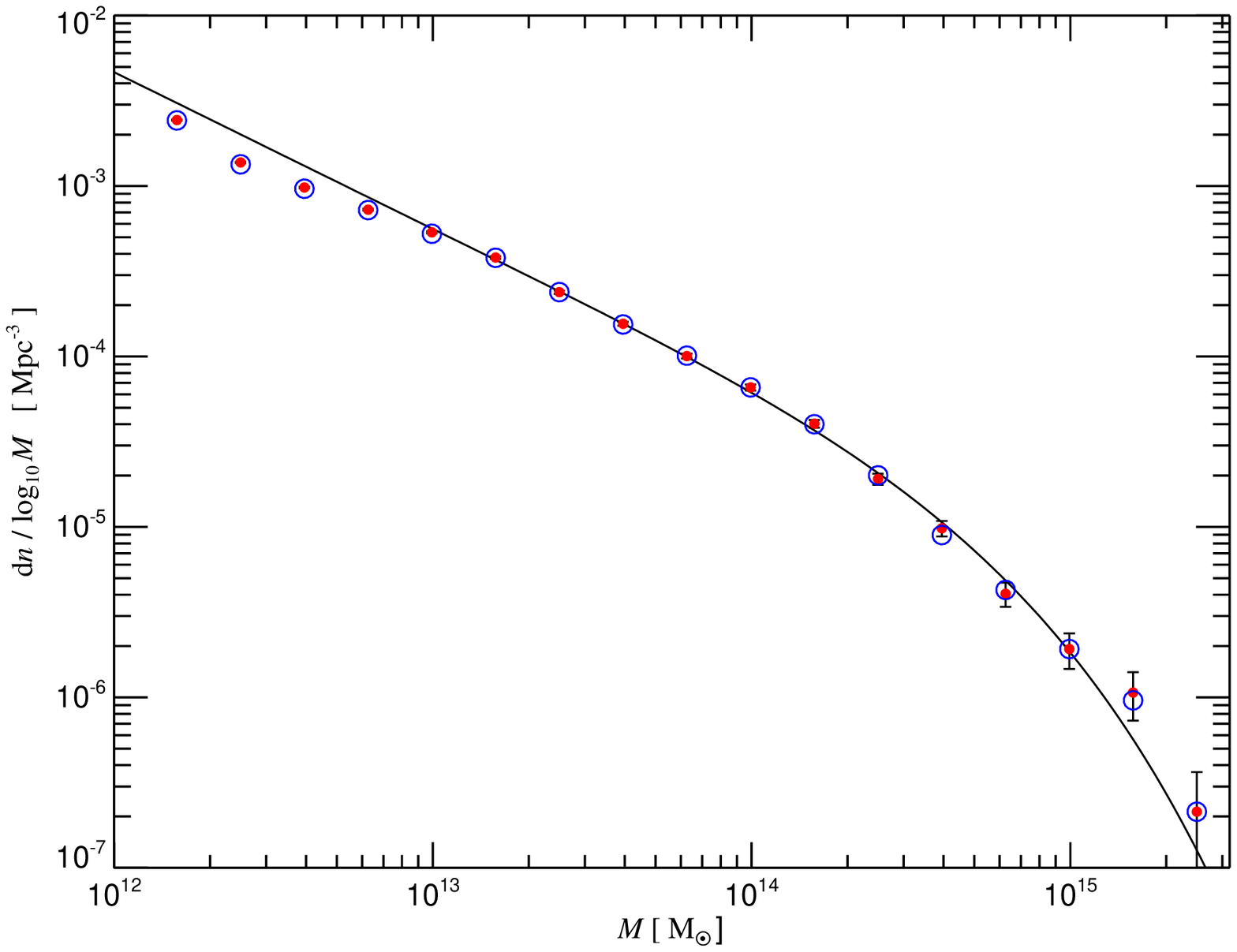}}
\resizebox{8.0cm}{!}{\includegraphics{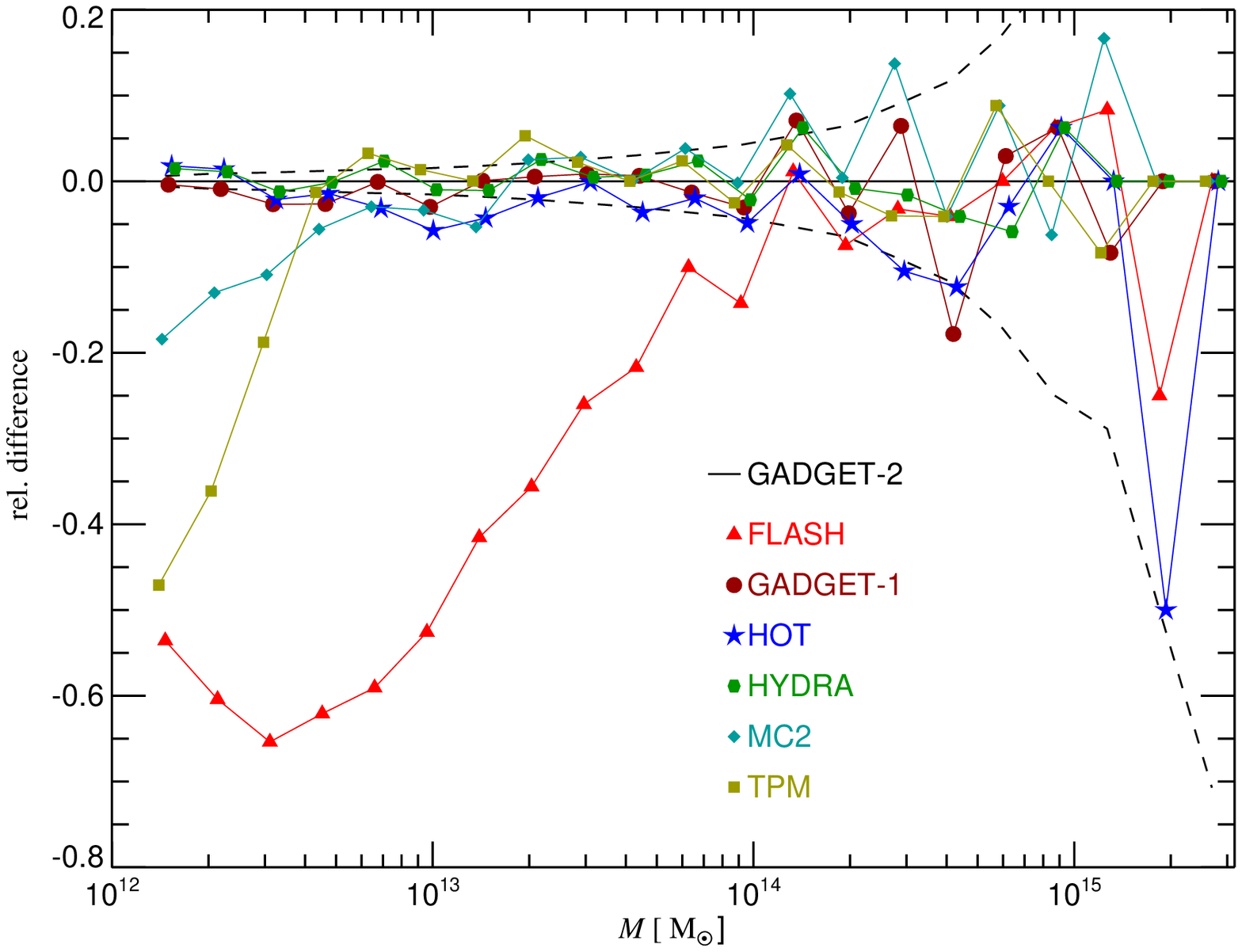}}
\caption{Comparison of the differential halo mass function obtained
  with different simulation codes for a $256^3$ $\Lambda$CDM simulation in a
  $256\,h^{-1}{\rm Mpc}$ box. The top panel compares the results from {\small
    GADGET-2} (red symbols with Poisson error bars) with those obtained with
  the old version {\small GADGET-1} (blue circles). The bottom panel shows the
  relative differences with respect to {\small GADGET-2} for a larger pool of
  6 codes. The evolved density fields for the latter have been taken from
  \citet{Heitmann2004}. The dashed lines indicate the size of the expected
  1$\sigma$ scatter due to counting statistics.
\label{figMassFuncComp}}
\ec
\end{figure}

\begin{figure}
\bc
\resizebox{8.0cm}{!}{\includegraphics{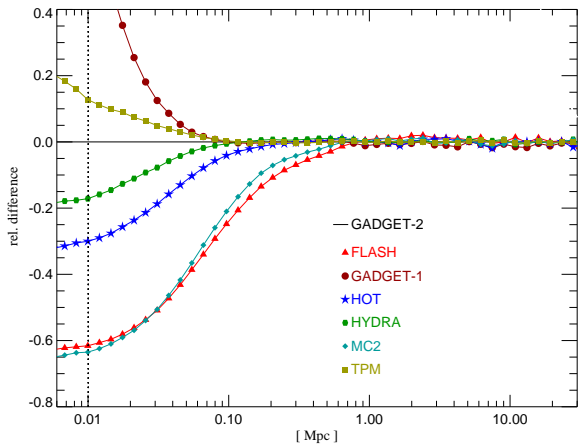}}
\caption{Comparison of different codes with respect to the two-point
  correlation of the evolved density of a $256^3$ $\Lambda$CDM
  simulation in a $64\,h^{-1}{\rm Mpc}$ box. We show the relative
  differences with respect to {\small GADGET-2} for the group of 6
  codes considered by \citet{Heitmann2004}. The vertical dotted line
  marks the gravitational softening length of $\epsilon = 10\,{\rm
  kpc}$ we used for our {\small GADGET-2} calculation. We explicitly
  checked that the latter is fully converged with respect to the time
  integration and force accuracy settings.
\label{figCorrelFuncComp}}
\ec
\end{figure}

\subsection{Dark matter halo mass function and clustering} \label{SecHeitmann}

Cosmological simulations of structure formation are the primary target
of {\small GADGET-2}. Because the dominant mass component is dark
matter, the accuracy and performance of the collisionless N-body
algorithms in periodic cosmological boxes is of tantamount importance
for this science application.  To compare results of {\small GADGET-2}
to other codes, we make use of a recent extensive study by
\cite{Heitmann2004}, who systematically compared the dark matter
results obtained with a number of different simulation codes and
techniques. Among the codes tested was also the old public version of
{\small GADGET-1} \citep{Springel2001}. As a useful service to
the community, \cite{Heitmann2004} have made their initial conditions
as well as the evolved results of their computations publicly
available. We here re-analyse the dark matter mass function and the
two-point autocorrelation function of their data using independent
measurement code and compare the results with those obtained by us
with {\small GADGET-2}.

The simulations considered are two runs with $256^3$ particles in
periodic boxes of sidelength $64\,h^{-1}{\rm Mpc}$ and
$256\,h^{-1}{\rm Mpc}$, respectively, in a $\Omega_m = 0.314$,
$\Omega_\Lambda=0.686$ universe with $h=0.71$. Further details about
the initial conditions are given in \citet{Heitmann2004}. We use a
comoving gravitational softening length equal to 1/35 of the mean
particle spacing.

Non-linear gravitational clustering leads to the formation of
gravitationally bound structures that over time build up ever more
massive halos. The abundance of halos as a function of mass and time
is arguably the most important basic result of structure formation
calculations.  In Figure~\ref{figMassFuncComp}, we show the
differential halo mass function, computed with the standard
friends-of-friends (FOF) algorithm using a linking length equal to
$0.2$ the mean particle spacing. The top panel compares our new
{\small GADGET-2} result for the large box at $z=0$ with the result
obtained by \citet{Heitmann2004} with {\small GADGET-1}. We obtain
very good agreement over the full mass range. The bottom panel of
Fig.~\ref{figMassFuncComp} extends the comparison to the five
additional codes examined by \cite{Heitmann2004}: The AMR code {\small
FLASH} \citep{Fryxell2000}, the parallel tree code {\small HOT}
\citep{Warren1995}, the adaptive P$^3$M code {\small HYDRA}
\citep{Couchman1995}, the parallel PM-code {\small MC$^2$} (Habib et
al., in preparation), and the tree-PM solver {\small TPM}
\citep{Bode2003}. We plot the relative halo abundance in each bin,
normalised to the {\small GADGET-2} result. While there is good
agreement for the abundance of massive halos within counting
statistics, systematic differences between the codes become apparent
on the low mass side. Particularly the codes based purely on
mesh-based gravity solvers, {\small MC$^2$} and {\small FLASH}, have
problems here and show a substantial deficit of small structures. It
is expected that some small halos are lost due to insufficient
resolution in fixed-mesh codes, an effect that can be alleviated by
using a sufficiently fine mesh, as {\small MC$^2$} demonstrates. It is
worrying however that current adaptive mesh refinement (AMR) codes
have particularly severe problems in this area as well. A similar
conclusion was also reached independently by \citet{OShea2003} in a
comparison of the AMR code {\small ENZO} \citep{OShea2004} with
{\small GADGET}. As gravity is the driving force of structure
formation, the novel AMR methods clearly need to keep an eye on this
issue and to improve their gravity solvers when needed, otherwise part of
the advantage gained by their more accurate treatment of
hydrodynamics in cosmological simulations may be lost.

In Figure~\ref{figCorrelFuncComp}, we show a similar comparison for
the two-point correlation function of the dark matter in the small
$64\,h^{-1}{\rm Mpc}$ box, again normalised to the {\small GADGET-2}
results. As discussed in more detail by \citet{Heitmann2004}, on
large-scales all codes agree reassuringly well, perhaps even better
than one might have expected. On small scales, the mesh-based codes
tend to show a deficit of clustering, consistent with the results for
the mass function. Interestingly, the result obtained by
\citet{Heitmann2004} for {\small GADGET-1} shows a noticeable excess
of clustering on very small scales compared to our computation with
{\small GADGET-2}. This happens on rather small scales, comparable to
the gravitational softening scale. This could simply be the result of
a different choice of gravitational softening length, but we also
believe that the {\small GADGET-2} result is the more accurate here.
As shown by \citet{Power2003}, the time integrator of {\small
GADGET-1} has the property that insufficient time integration settings
can lead to an increase of the central density in halos due to secular
integration errors, while for very poor timestepping the halo density
is eventually suppressed. The numerical steepening of the central
density profile caused by this effect could then show up as a
signature of enhanced clustering at very small scales, just like what
is seen here in the {\small GADGET-1} result.

\begin{figure*}
  \bc
\resizebox{8.0cm}{!}{\includegraphics{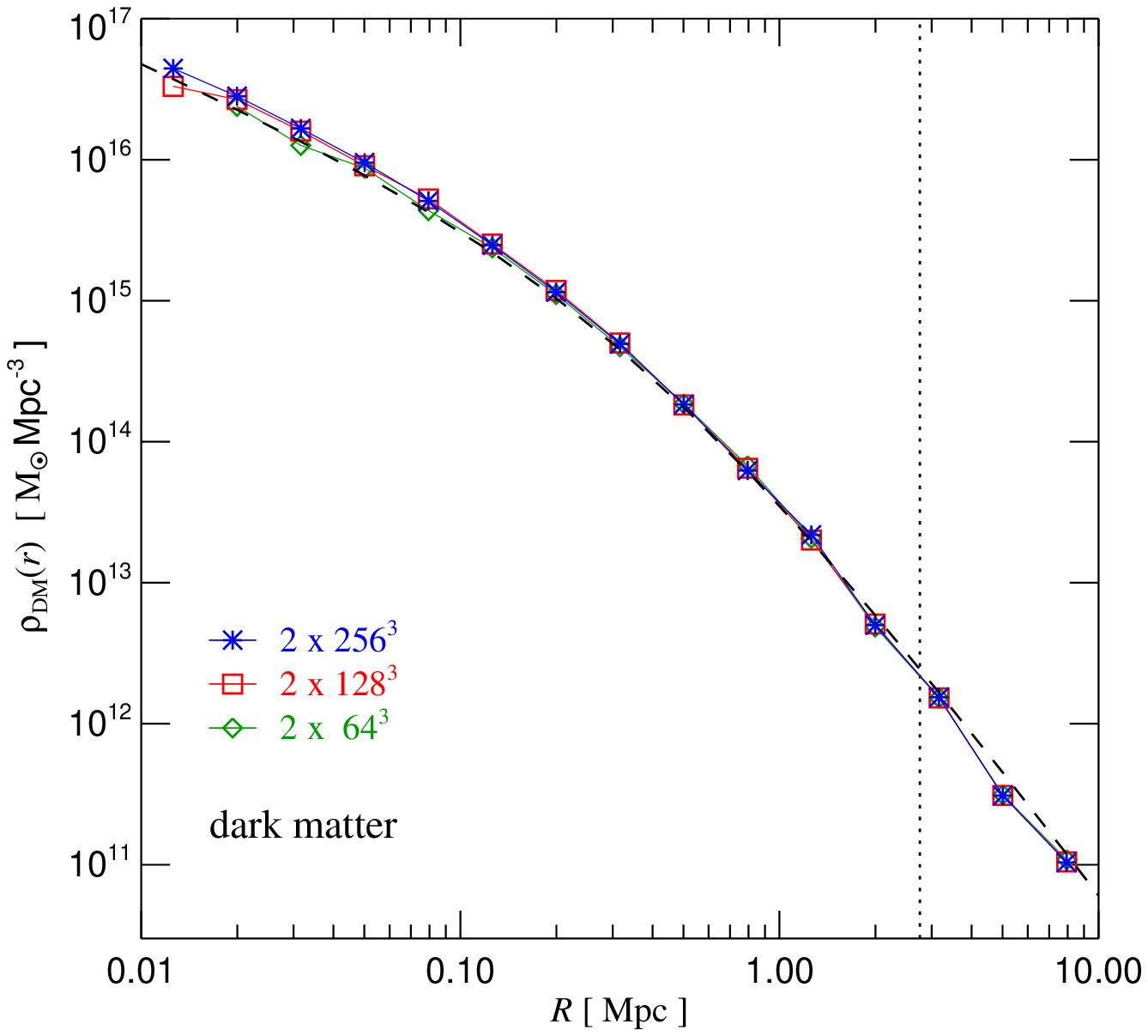}}%
\resizebox{8.0cm}{!}{\includegraphics{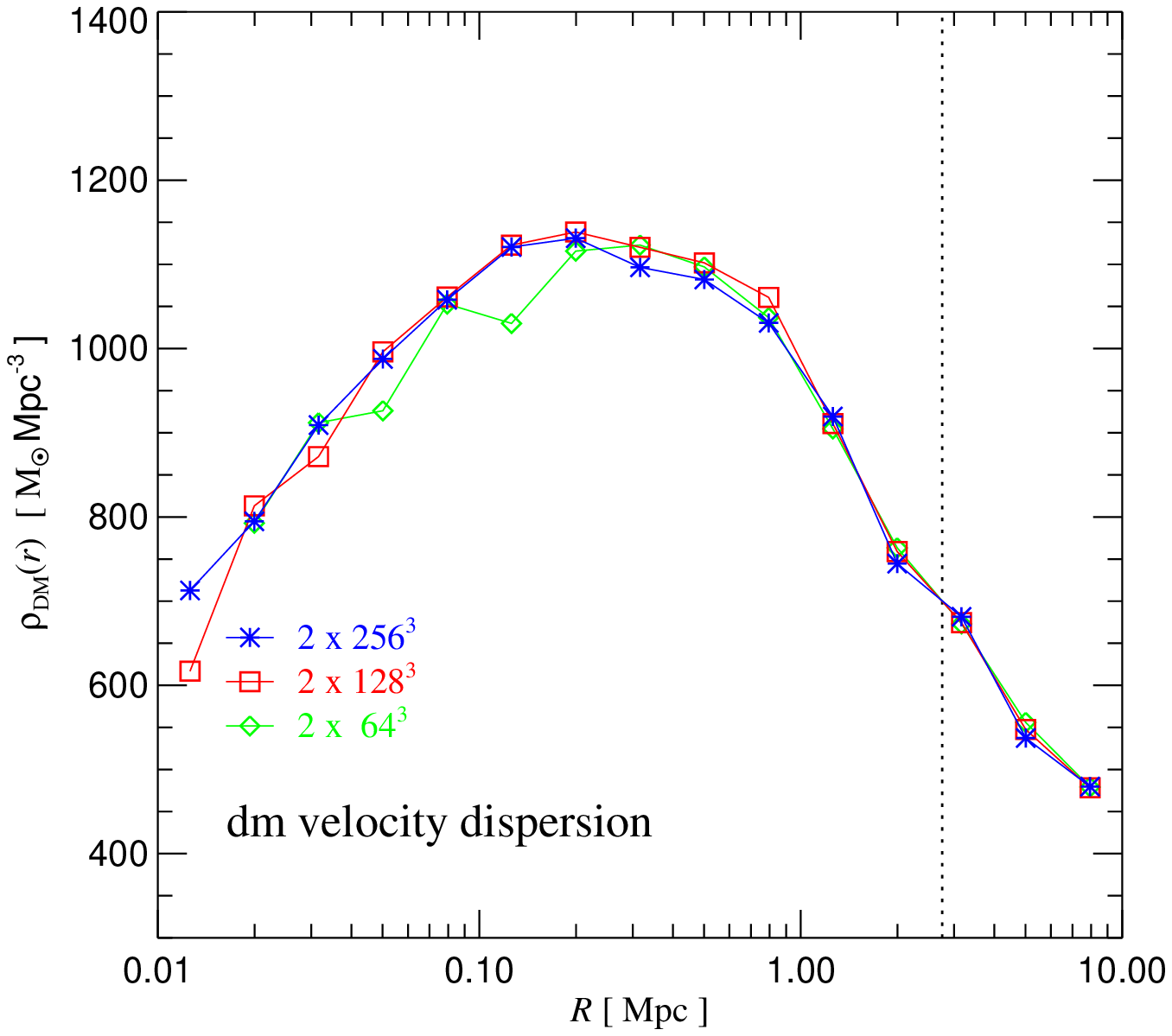}}\vspace*{-0.3cm}\\%
\resizebox{8.0cm}{!}{\includegraphics{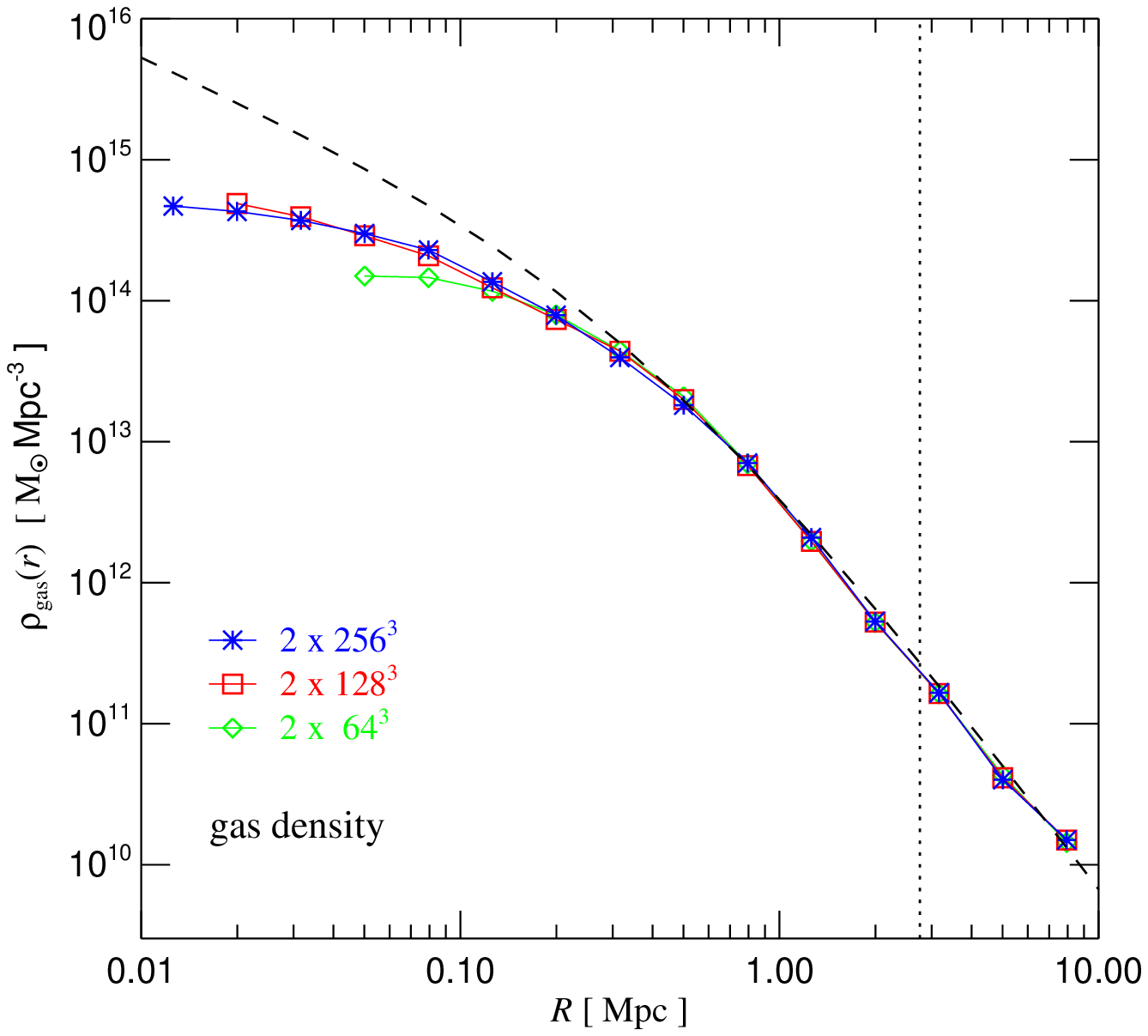}}%
\resizebox{8.0cm}{!}{\includegraphics{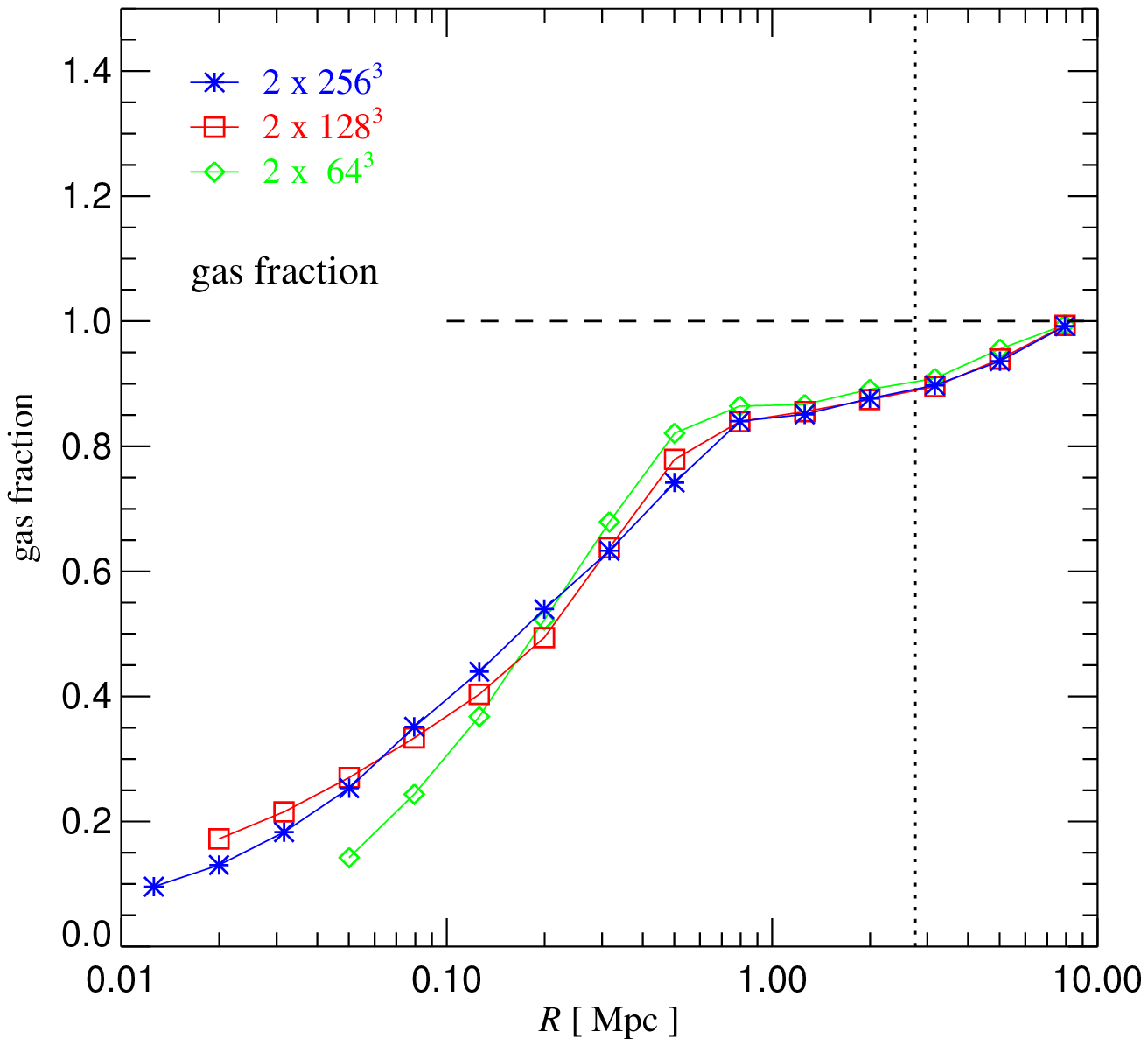}}\vspace*{-0.3cm}\\%
\resizebox{8.0cm}{!}{\includegraphics{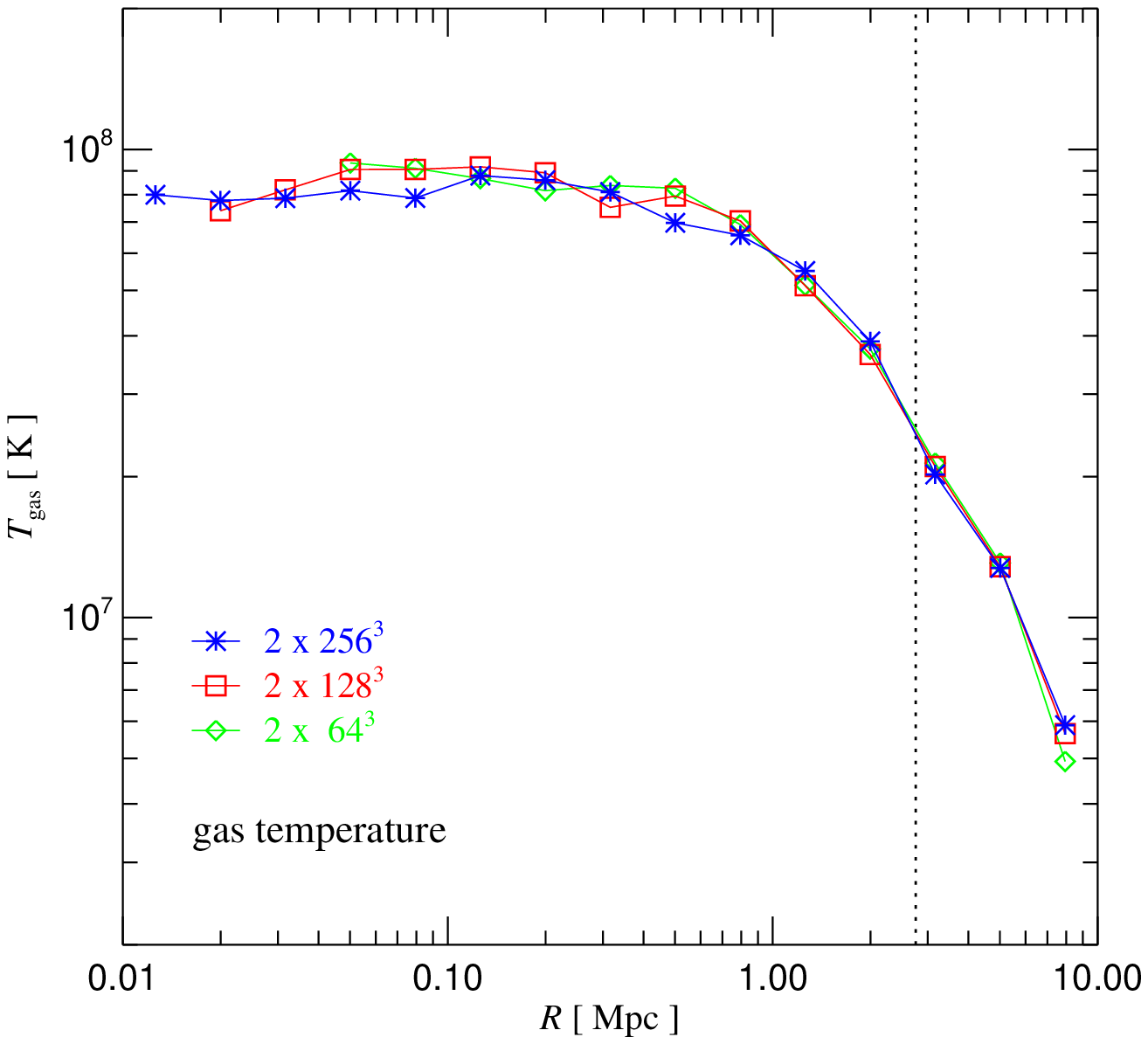}}%
\resizebox{8.0cm}{!}{\includegraphics{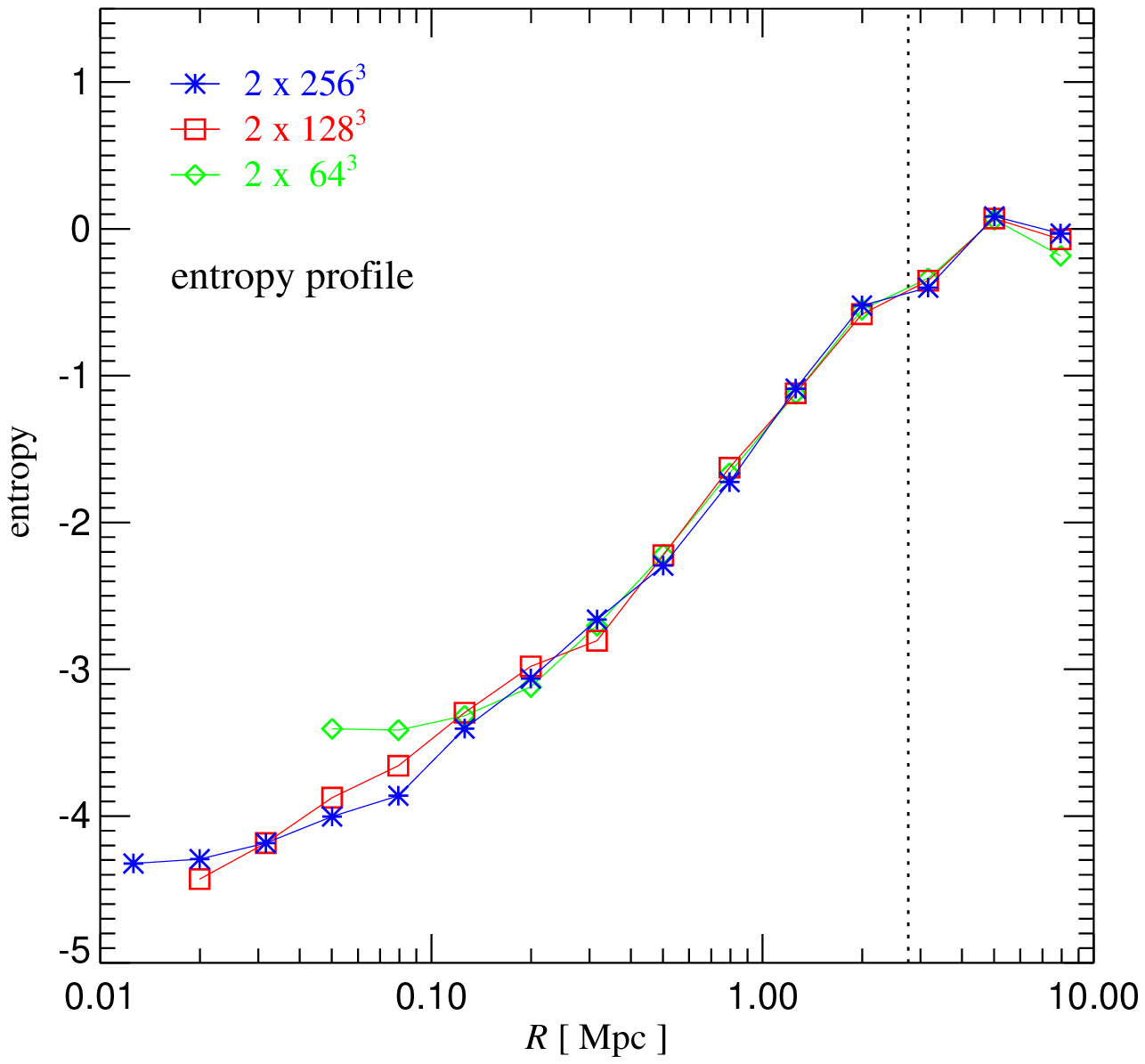}}\vspace*{-0.3cm}\\
\caption{Radial profiles of the Santa Barbara Cluster. From top left
  to bottom right, we show spherically averaged profiles of dark
  matter density, gas density, temperature, dark matter velocity
  dispersion, enclosed gas fraction, and specific entropy. In each
  case we compare results for three different resolutions. The
  vertical line marks the virial radius of the cluster. The dashed
  line in the dark matter profile is a NFW profile with the parameters
  given by \citet{Frenk99}. The same profile is also shown in the gas
  density plot to guide the eye (scaled by the baryon to dark matter
  density ratio).
\label{figSantaBarbaraProfiles}}
\ec
\end{figure*}

\begin{figure*}
\bc
\vspace{0.25cm}\ \\%
\resizebox{17.5cm}{!}{\includegraphics{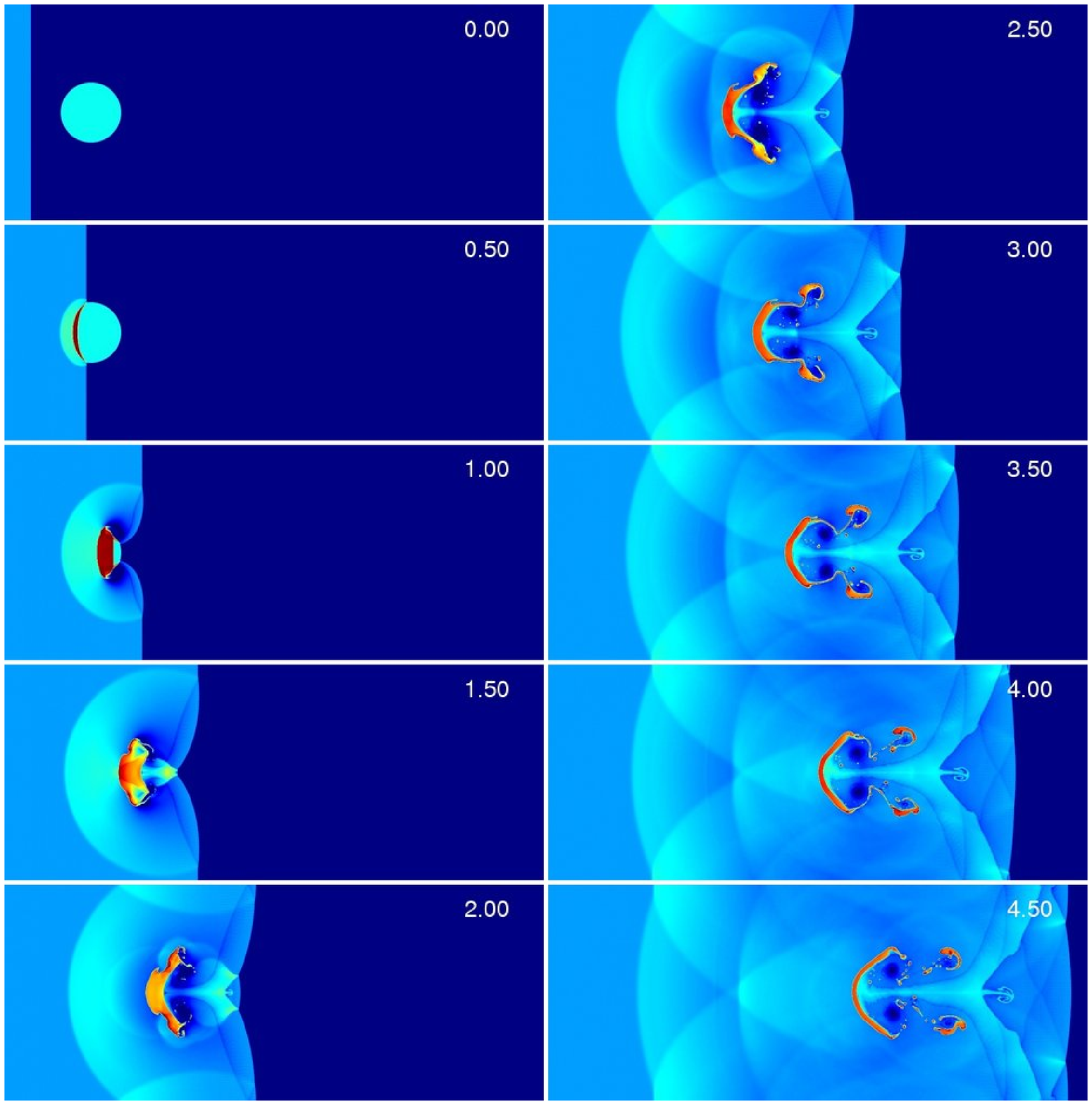}}
\vspace{0.35cm}\\%
\resizebox{10.0cm}{!}{\includegraphics{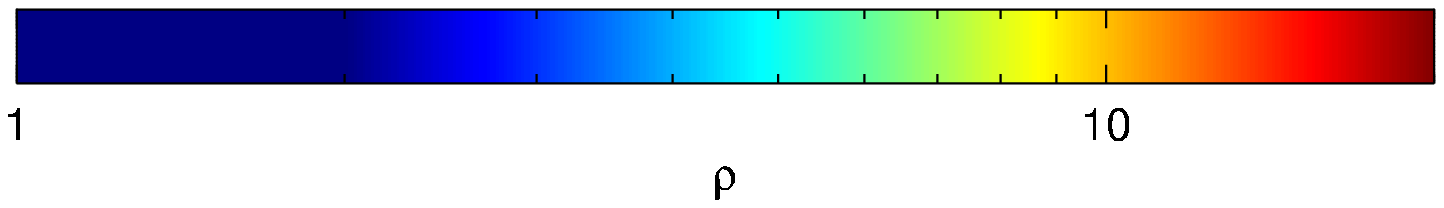}}\\%
\caption{Time evolution of the interaction of a strong shock wave with
  a dense cloud of gas. The cloud of radius $r=3.5$ has an initial
  relative overdensity of 5, and is embedded at pressure equilibrium
  in ambient gas of unit density and unit pressure.  From the left, a
  shock wave with Mach-number $M=10.0$ approaches and strikes the
  cloud. The gas has $\gamma = 5/3$, giving the shock an incident
  velocity of $v=9.586$ and a compression factor of $3.884$ with
  respect to the background gas. Each panel shows the gas density in a
  region of size $62.5\times 25.0$, with the time indicated in the top
  right corner. The computation assumed periodic boundaries at the top
  and bottom.
\label{figShockCloud}}
\ec
\end{figure*}

\subsection{Santa Barbara cluster} \label{SecSB}

In the `Santa Barbara cluster comparison project' \citep{Frenk99}, a
large number of hydrodynamic cosmological simulation codes were applied
to the same initial conditions, which were set up to give rise to the
formation of a rich cluster of galaxies in a critical density CDM
universe, simulated using adiabatic gas physics. In total 12 codes
were compared in this study, including SPH and Eulerian codes, both
with fixed and adaptive meshes. Each simulation group was allowed to
downsample the initial conditions in a way they considered reasonable,
given also their computational resources and code abilities, so that
the final comparison involved computations of different effective
resolutions.

The overall results of this comparison were encouraging in the sense
that bulk properties of the cluster agreed to within $\sim 10\%$ and
the gas properties were similar in most codes, although with large
scatter in the inner parts of the cluster.  However, there have also
been some systematic differences in the results, most notably between
mesh-based and SPH codes. The former showed higher temperatures and
entropies in the cluster centre than the SPH codes. Also, the enclosed
gas fraction within the virial radius was systematically higher for
mesh codes and closer to the universal baryonic fraction, while the
SPH codes only found about 90\% of the universal fraction in the
virial radius.  Since then, the Santa Barbara cluster has been
repeatedly used as a test problem for cosmological codes, but the
question which is the `correct' entropy profile and gas fraction in
adiabatic cluster has not been settled conclusively so far.

We have simulated the Santa Barbara cluster at three different
numerical resolutions ($2\times 256^3$, $2\times 128^3$, and $2\times
64^3$) with {\small GADGET-2}, in each case using a homogeneous
sampling for the periodic box. \citet{Frenk99} supplied displacement
fields at a nominal resolution of $256^3$, which we directly used for
our high resolution $2\times 256^3$ run. The initial conditions for
the lower resolution runs were constructed by applying a filter
in Fourier space to eliminate modes above the corresponding Nyquist
frequencies, in order to avoid aliasing of power.

In Figure~\ref{figSantaBarbaraProfiles}, we compare our results in
terms of spherically averaged profiles for dark matter density, dark
matter velocity dispersion, gas density, enclosed gas fraction,
temperature, and specific entropy.  We use the original binning
prescriptions of \cite{Frenk99}, and the same axis ranges for easier
comparison. Our simulations converge quite well in all their
properties, apart from the innermost bins (we have plotted bins if
they contained at least 10 dark matter or gas particles). However, we
note that we confirm the finding of \cite{Wadsley2004} that there is a
merger happening right at $z=0$; in fact, in our high-res $2\times
256^3$ run, the infalling clump is just passing the centre at $z=0$,
while this happens with a slight time offset in the other two runs. We
have therefore actually plotted the results at expansion factor
$a=1.02$ in Figure~\ref{figSantaBarbaraProfiles}, where the cluster
has relaxed again. The results at $z=0$ look very similar, only the
temperature, gas entropy, and dark matter velocity dispersion at
$r<0.1$ show larger differences between the simulations.  As
\citet{Wadsley2004} point out, the effects of this unfortunate timing
of the merger presumably also contribute to the scatter found in the
results of \citet{Frenk99}.

Our results agree very well with the mean profiles reported in the
Santa Barbara cluster comparison project. Our resolution study also
suggests that {\small GADGET-2} produces quite stable convergence for
a clean set of initial conditions of different resolution. The mass
resolution has been varied by a factor 64 and the spatial resolution
per dimension by a factor 4 in this series; this is already a
significant dynamic range for 3D simulations, thereby helping to build
up trust in the robustness of the results of the code.

The entropy profile  our results at small radii ($R\sim 0.1$)
appears to lie somewhat above the SPH results reported in
\citet{Frenk99} for other SPH codes. This is in line with the findings
of \citet{Ascasibar2003}, and perhaps a consequence of the
entropy-conserving formulation of SPH that we have adopted in {\small
GADGET-2}. Also, the entropy profile appears to become slightly
shallower at small radii, which suggests a small difference from the
near power-law behaviour seen in other SPH codes \citep[see for example
the high-res result of][]{Wadsley2004}.  However, this effect appears
to be too small to produce the large isentropic cores seen in the mesh
simulations of \citet{Frenk99}. Such a core has also been found in the
new AMR code by \citet{Quilis2004}. The systematic difference between
the different simulation methods therefore continues to persist. We
suggest that it may be  caused by entropy production due to
mixing; this channel is absent in the SPH code by construction while
it operates efficiently in the mesh codes, perhaps even too
efficiently.

Another interesting point to observe is that our SPH simulations
clearly predict that the enclosed baryon fraction is well below the
universal baryon fraction at the virial radius of the adiabatic
cluster. It seems a solid result that our results converge at values
of around 90\%, in clear difference with results near $\sim 100\%$
predicted by the majority of mesh codes in the study by
\citet{Frenk99}. However, we note that the new AMR code {\small ART}
of \citet{Kravtsov2005} also gives values below the universal baryon
fraction, although not quite as low as the SPH codes.  We can also
observe a clear break in the profile at $\sim 0.6\,{\rm Mpc}$, which
could not be discerned as easily in the results of
\citet{Frenk99}. At this radius, the gas profile begins to notably
flatten compared with the dark matter profile.

\begin{figure}
\bc
\mbox{\resizebox{8.4cm}{!}{\includegraphics{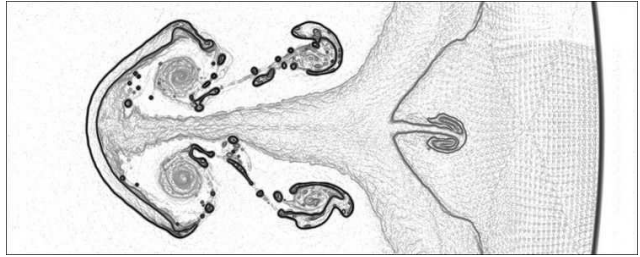}}}
\caption{Local gradients in the gas density field at time $t=4.5$,
visualised by a grey-scale image with intensity proportional to
$\log(|\vec{\nabla}\rho| /\rho)$. Clearly visible are the two pairs of
primary and secondary vortices, as well as the stem of the backflow.
The region shown has a size of $31.25\times 12.5$.
\label{figShockCloudGrad}}
\ec
\end{figure}

\subsection{Interaction of a strong shock with a dense gas cloud}

As a final hydrodynamical test problem we consider the interaction of
a strong shock wave with an overdense cloud embedded at pressure
equilibrium in a background gas.  This can be viewed as a simple model
for the interaction of a supernova blast wave with a dense cloud in
the interstellar medium.  When the shock strikes the cloud, a
complicated structure of multiple shocks is formed, and vortices are
generated in the flow around the cloud which lead to its (partial)
destruction. Aside from its physical relevance for simple models of
the interstellar medium, this makes it an interesting hydrodynamical
test problem.  The situation has first been studied numerically in a
classic paper by \citet{Klein1994}.  Recently, \citet{Poludnenko2002}
have readdressed this problem with a high-resolution AMR code; they
also extended their study to cases of multiple clouds and different
density ratios and shock strengths.

As initial conditions, we adopt a planar shock wave of Mach number
$M=10$ which enters gas of unit density and unit pressure from the
negative $x$-direction. In the frame of the ambient background gas,
the shock approaches with velocity $v=9.586$, leading to a post-shock
density of $\rho'=3.884$. We adopt a two-dimensional computational
domain with periodic boundaries in the $y$-direction, and formally
infinite extension in the $x$-direction. The boxsize in the
$y$-direction is 25 length units, and the radius of the spherical
cloud of overdensity 5 is $r=3.5$. The set-up of SPH particles was
realized with a glass like particle distribution using equal-mass
particles. We have first evolved the incident shock-wave independently
in order to eliminate transients that typically arise if it is set-up
as a sharp discontinuity, i.e.~our incident shock is consistent with
the SPH smoothing scheme.

In Figure~\ref{figShockCloud}, we show density maps of the system at
different times of its evolution. When the shock strikes the cloud, a
complicated structure of forward and reverse shocks develops. A
detailed description of the various hydrodynamical features of the
flow is given by \citet{Poludnenko2002}. Two pairs of primary vortices
develop in the flow around the cloud and start shredding the
cloud. This can be seen particularly well in the `Schlieren' image of
Figure~\ref{figShockCloudGrad}, where we show a grey-scale map of the
local density gradient. Overall, our SPH results look similar to the
AMR results of \citet{Poludnenko2002}, but there are also clearly some
differences in detail. For example, the small `droplets' of gas
chopped of from the cloud still survive in the SPH calculation for a
comparatively long time and are not mixed efficiently with the
background material, a clear difference with the mesh-based
calculations.

\section{Performance and scalability}  \label{SecPerformance}

The performance of a parallel simulation code is a complex function of
many factors, including the type of physical problem studied, the
particle and processor numbers employed, the choices made for various
numerical parameters of the code (e.g.~time integration settings,
maximum allowed memory consumption, etc.), and finally of hardware and
compiler characteristics. This makes it hard to objectively compare
the performance of different codes, which should ideally be done at
comparable integration accuracy for the same physical system. Given
these difficulties, we restrict ourselves to a basic characterisation of the
performance and scaling properties of {\small GADGET-2} without
attempting to compare them in detail with other simulation codes.

\begin{table}
\begin{center}
\begin{tabular}{lrr}
\hline
Simulation Boxsize ($256^3$)& $256\,h^{-1}{\rm Mpc}$ & $64\,h^{-1}{\rm Mpc}$ \\
\hline
Timesteps                  & 2648  & 5794 \\
Total wallclock time [sec] & 60600 & 173700 \\
Tree walk                  &52.8 \% & 41.0 \%\\
Tree construction          & 4.6 \% & 6.4 \% \\
Tree walk communication    & 0.9 \% & 1.6 \% \\
Work-load imbalance        & 6.7 \% & 14.4 \% \\
Domain decomposition       &13.0 \% & 15.2 \% \\
PM force                   & 4.4 \% & 4.9 \% \\
Particle and tree drifts   & 5.3 \% & 4.9 \% \\
Kicks and timestepping     & 1.4 \% & 1.1 \%\\  
Peano keys and ordering    & 8.0 \% & 7.8 \% \\ 
Misc (I/O, etc.)           & 2.9 \% & 2.6 \% \\
\hline
\end{tabular}
\end{center}
\caption{CPU-Time consumption in different parts of the code for two
typical $256^3$ dark matter simulations. The initial conditions for
the two simulations are those of \citet{Heitmann2004}. We first give
the total number of timesteps and the elapsed wallclock time to evolve
the simulation to $z=0$ on 8 CPUs of a Pentium-IV cluster. The total
consumed time is then broken up in time spent in different parts of
code, as measured by the timing routines built into {\small
GADGET-2}.\label{TabDMRunTime} }
\end{table}

\subsection{Timing measurements for cosmological simulations}

In Table~\ref{TabDMRunTime}, we list the total wallclock time elapsed
when running the two $256^3$ dark matter simulations discussed in
Section~\ref{SecHeitmann}, based on the initial conditions of
\citet{Heitmann2004}. The measured times are for all tasks of the
code, including force computations, tree construction, domain
decomposition, particle drifts, etc. A detailed break-down of the
relative contributions is given in the table as well. The hardware
used was a 8 CPU partition on a small cluster of Pentium-IV PCs (2.4
GHz clock speed, 2 CPUs per machine), using the public MPICH library
for communication via gigabit ethernet.

We can see that the CPU consumption is dominated by the short-range
tree computation, while the PM force is subdominant overall. The raw
force speed in the short-range tree walk of these TreePM simulations
(using a $384^3$ mesh) reaches about 21000 forces/sec/processor. This
is a high number, significantly in excess of what is reached with pure
tree algorithms. In fact, the latter tend to be significantly slower
for this type of simulation, typically by a factor 4-10.

Most of the auxiliary tasks of the simulation code, for example
particle drifting, I/O, and so on, typically require a few percent of
the total CPU. Some of these tasks are due to the parallelisation
strategy, namely the domain decomposition, the wait times due to
work-load imbalance, and the time needed for communication
itself. However, provided these contributions stay subdominant, we can
still expect a significantly faster time to solution as a result of
parallelisation, besides the possibility to carry out larger
simulations because of the availability of the combined memory of all
processors.

\begin{figure}
\bc
\fbox{\resizebox{7.6cm}{!}{\includegraphics{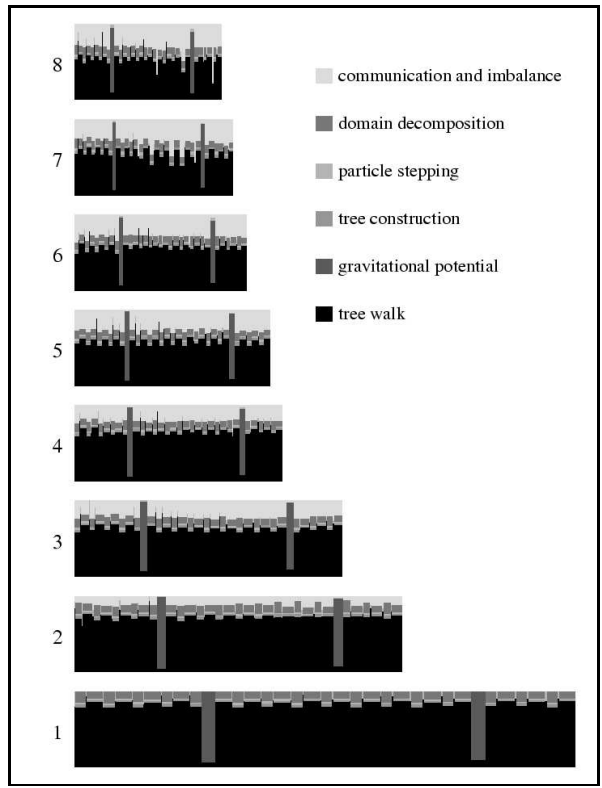}}}
\ec
\caption{Diagram for the time consumption of a rather small galaxy
  collision simulation evolved with a different number of processors
  between 1 and 8. We show a sample of 64 timesteps in each case, each
  represented by a vertical bar with a width proportional to the
  elapsed wall-clock time during this step. Each step is additionally
  subdivided into different constituent parts, drawn in different shades
  of grey as indicated in the legend.
\label{FigScalingGalaxy}}
\end{figure}

In cosmological hydrodynamical TreePM simulations, we find that the
CPU time required for the SPH computations is roughly equal to that
consumed for the short-range gravitational tree-forces. This is for
example the case in the simulations of the Santa Barbara cluster
discussed in Section~\ref{SecSB}. The cost of self-gravity is hence
comparable to or larger than the cost of the hydrodynamical
computations in {\small GADGET-2}. Even in simulations with
dissipation this ratio shifts only moderately towards a higher
relative cost of the hydrodynamics, but of course here the total cost
of a simulation increases substantially because of the much shorter
dynamical times that need to be resolved.

\subsection{Scalability}

The problem size is an important characteristic when assessing the
performance of a massively parallel simulation code. Due to the tight
coupling of gravitational problems, it is in general not possible to
obtain a nearly linear speed-up when a small problem is distributed
onto many processors. There are several reasons that make this
impossible in practice: (1) There is always some irreducible serial
part of the code that does not parallelise; this overhead is fixed and
hence its relative contribution to the total cost keeps getting larger
when the parallel parts are accelerated by using more processors. (2)
The more processors are used, the less work each of them has to do,
making it harder to balance the work equally among them, such that
more and more time is lost to idle waiting of processors. (3) When
more processors are used, a smaller particle-load per processor
results, which in turn leads to a larger communication-to-compute
ratio in tightly coupled problems.

\begin{figure}
\bc
\mbox{\resizebox{8.0cm}{!}{\includegraphics{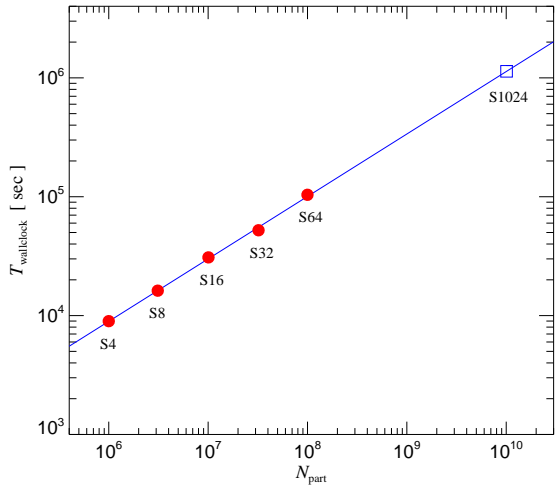}}}
\caption{Wall-clock times consumed by the test runs of the simulation
  series. An extrapolation to the size of a $2160^3$ simulation
  suggests that it should require about $1.1\times 10^6$ seconds on
  1024 processors.
\label{FigScalingTimings}}
\ec
\end{figure}

\begin{table}
\begin{center}
\begin{tabular}{ccccc}
\hline 
Name & $N_{\rm CPU}$ & $N_{\rm part}$ & $N_{\rm FFT}$ & $L_{\rm
  box}\;[h^{-1}{\rm Mpc}]$\\
\hline
S4  & 4 & $100^3$ & $128^3$ & 23.1\\
S8  & 8 & $146^3$ & $192^3$ & 33.8 \\
S16 & 16 & $216^3$ & $256^3$ & 50.0\\
S32 & 32 & $318^3$ & $384^3$ & 73.6 \\
S64 & 64 & $464^3$ & $576^3$ & 108.0\\
\hline
\end{tabular}
\end{center}
\caption{Simulations done for the scaling test. All runs used the same
  mass- and length resolution of $1.03\times 10^{9}\,h^{-1}{\rm
  M}_\odot$ and $5\,h^{-1}{\rm kpc}$, respectively, and were started
  at $z_{\rm init}=49$. The runs used equal settings for force
  accuracy and time integration parameters, and all were asked to
  produce the same number of outputs, at which point they also did
  group finding, power spectrum estimation and two-point correlation
  function computation. \label{TabScalTest}}
\end{table}

For all of these reasons, perfect scalability at fixed problem size
can in general not be expected. In Figure~\ref{FigScalingGalaxy}, we
illustrate this with a rather small galaxy collision simulation,
consisting of two galaxies with 30000 collisionless particles each,
distributed into a stellar disk and an extended dark matter halo. We
have evolved this simulation with {\small GADGET-2} using different
processor numbers, from 1 to 8. The diagram in
Figure~\ref{FigScalingGalaxy} shows the time consumption in different
parts of the code, during 64 typical steps taken from the
simulation. Each step is shown with an area proportional to the
elapsed wall-clock time, and different shades of grey are used for
different parts of the code within each step. In particular, black is
used for the actual tree walk, while light grey marks losses of some
sort or the other (primarily wait times due to work-load imbalance,
and communication times). We see that the relative fraction of this
light grey area (at the top) relative to the total keeps growing when
the number of processors is increased. In fact, the scaling is
disappointing in this example, falling significantly short of perfect
scaling where the total area for the 64 steps would decline as the
inverse of the processor number. However, this result is not really
surprising for such a small problem; when typical timesteps last only
fractions of a second and the particle-load per processor is very low,
the problem size is simply too small to allow good scaling with
{\small GADGET-2}'s massively parallel algorithms.  We also see that
the widths of the different steps follow a particular pattern,
stemming from the individual timestep integration scheme, where the
occupancy of certain steps with `active' particles constantly
changes. The two large grey bars represent the computation of the
gravitational potential for all particles, which was here carried out
in regular intervals to monitor energy conservation of the code.

However, arguably of more practical relevance for assessing the
scaling of the code is to consider its performance when not only the
processor number but {\em at the same time} also the problem size is
increased. This is of immediate relevance for practical application of
a simulation code, where one typically wants to employ large numbers
of processors only for challengingly large problems, while small
problem sizes are dealt with correspondingly fewer processors. A
simultaneous variation of problem size and processor number can
alleviate all three of the scaling obstacles listed above. However,
changing the problem size really means to change the physics of the
problem, and this aspect can be easily confused with bad scaling when
analysed superficially. For example, increasing the problem size of a
simulation of cosmological structure formation either improves the
mass resolution or the volume covered. In both cases, typically more
simulation timesteps will be required to integrate the dynamics,
either because of better spatial resolution, or because more massive
systems of lower space-density can form. The intrinsic computational
cost of a simulation therefore typically scales (sometimes
considerably) faster than linear with the problem size.

With these caveats in mind, we show in Figure~\ref{FigScalingTimings} the
required run times for a scaling experiment with cosmological $\Lambda$CDM
dark matter simulations, carried out with {\small GADGET-2} on a cluster of
IBM p690 systems. In our series of five simulations, we have increased the
particle number from $10^6$ to $10^8$, in each step by roughly a factor of
$\sqrt{10}$. At the same time we also doubled the number of processors in each
step.  We kept the mass and spatial resolutions fixed at values of
$10^9\,h^{-1}{\rm M}_{\odot}$ and $\epsilon = 5 \,h^{-1}{\rm kpc}$,
respectively, i.e.~the volume of the simulations was growing in this series.
We also increased the size of the FFT mesh in lock step with the particle
number.  In Table~\ref{TabScalTest}, we list the most important simulation
parameters, while in Figure~\ref{FigScalingTimings}, we show the total
wall-clock times measured for evolving each of the simulations from high
redshift to $z=0$, as a function of particle number. We note that the
measurements include time spent for computing on-the-fly friends-of-friends
group catalogues, two-point correlation functions, and power spectra for 64
outputs generated by the runs. However, this amounts only to a few per cent of
the total CPU time.

We see that the simulation series in Figure~\ref{FigScalingTimings}
follows a power-law. For a perfect scaling, we would expect $T_{\rm
wallclock} \propto N_{\rm part}/N_{\rm cpu}$, which would correspond
to a power-law with slope $n=1-\log(4)\simeq 0.4$ for the
series. Instead, the actually measured slope (fitted blue line) is
$n=0.52$, slightly steeper. However, the perfect scaling estimate
neglects factors of $\log(N_{\rm part})$ present in various parts of
the simulation algorithms (e.g. in the tree construction), and also
the fact that the larger simulations do need more timesteps than the
smaller ones. In the series, the number of timesteps in fact increases
by 23\% from S4 to S64. Overall, the scaling of the code is therefore
actually quite good in this test.  In fact, an extrapolation of the
series to $2160^3 \simeq 1.0078\times 10^{10}$ particles in a
$500\,h^{-1}{\rm Mpc}$ box suggests that such a simulation should be
possible on 1024 processors in about $1.1\times 10^6\,{\rm sec}$.
This simulation has in fact been realized with {\small GADGET-2} in
the first half of 2004, finishing on June 14. This `Millennium'
simulation by the Virgo consortium \citep{SpringelMS2005} is the
largest calculation carried out with {\small GADGET-2} thus far, and
it is also the largest high-resolution cosmological structure
formation simulation at present, reaching a dynamic range of $10^5$
everywhere in the periodic simulation box. The total wall-clock time
required for the simulation on the 512 processors actually used was
slightly below 350,000 hours, only about 10\% more than expected from
the above extrapolation over two orders of magnitude. This shows that
{\small GADGET-2} can scale quite well even to very large processor
partitions if the problem size is sufficiently large as well.

\begin{figure}
\bc
\mbox{\resizebox{8.0cm}{!}{\includegraphics{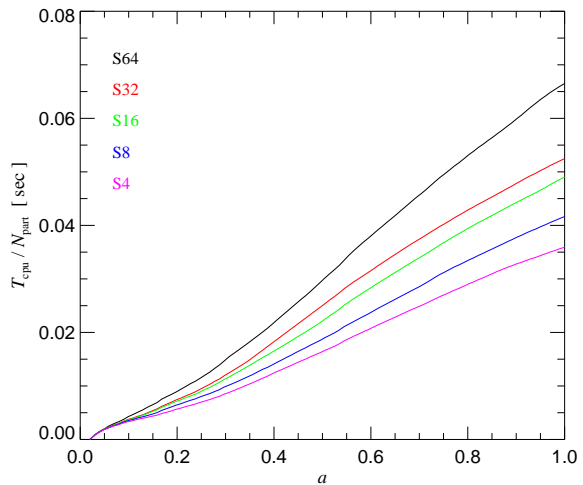}}}
\caption{Total elapsed wall-clock time per particle of each test run
  as a function of cosmological scale factor. The elapsed run-times of
  each simulation have been multiplied by the processor number, and
  normalised by the particle number.
\label{FigTimingsVsScaleFac}}
\ec
\end{figure}

Finally, in Figure~~\ref{FigTimingsVsScaleFac} we show the cumulative
CPU time consumed for the five simulations of the series as a function
of cosmological scale factor. We have normalised the total CPU time
consumption, $T_{\rm cpu} = T_{\rm wallclock}\times N_{\rm cpu}$, to
the number of particles simulated, such that a measure for the
computational cost per particle emerges. To first order the required
CPU times scales roughly linearly with the scale factor, and grows to
order of a few dozen milliseconds per particle. At the time of the
test run, the p690 cluster was not yet equipped with its fast
interconnection network, which led to the comparatively poorer
performance of the S64 simulation as a result of the communication
intensive PM part taking its toll.  On current high-end hardware
(which is already faster than the p690 machine), {\small GADGET-2}
reaches a total CPU cost of about $10\,{\rm ms}$ per dark matter
simulation particle in realistic simulations of cosmological structure
formation evolved from high redshift to the present.

\subsection{Memory consumption}

The standard version of {\small GADGET-2} in TreePM mode uses 20
variables for storing each dark matter particle, i.e.~80 bytes per
particle if single-precision is used. For each SPH particle, an
additional 21 variables (84 bytes) are occupied. For the tree, the
code uses 12 variables per node, and for a secondary data structure
that holds the centre-of-mass velocity and maximum SPH smoothing
lengths of nodes, another 4 variables. For a typical clustered
particle distribution on average about $\sim0.65$ nodes per particle
are needed, so that the memory requirement amounts to about 42 bytes
per particle. Finally, for the FFTs in the PM component, {\small
GADGET-2} needs 3 variables per mesh-cell, but the ghost-cells
required around local patches increase this requirement
slightly. Taking 4 variables per mesh-cell as a conservative upper
limit, we therefore need up to 16 bytes (or 32 bytes for double
precision) per mesh-cell for the PM computation. This can increase
substantially for two-level PM computations, since we here not only
have to do zero padding but also store the Greens function for the
high-resolution region.

While being already reasonably memory-efficient, the standard version
of {\small GADGET-2} is not yet heavily optimised towards a lean
memory footprint. This has been changed however in a special lean
version of the code, where some of the code's flexibility was
sacrificed in favour of very low memory consumption. This version of
the code was used for the Millennium simulation described above.  The
memory optimisations were necessary to fit the simulation size into
the aggregated memory of 1 TB available on the supercomputer partition
used. By removing explicit storage for long- and short-range
accelerations, particle mass and particle type, the memory requirement
per particle could be dropped to 40 bytes, despite the need to use
34-bit numbers for labelling each particle with a unique number. The
tree storage could also be condensed further to about 40 bytes per
particle. Since the memory for PM and tree parts of the gravitational
force computation are not needed concurrently, one can hence run a
simulation with a peak memory consumption of about 80 bytes per
particle, provided the Fourier mesh is not chosen too large. In
practice, one has to add to this some additional space for a
communication buffer. Also, note that particle-load imbalance as a
result of attempting to equalise the work-load among processors can
lead to larger than average memory usage on individual processors.

\section{Discussion} \label{SecDiscussion}

In this paper, we have detailed the numerical algorithms used in the
new cosmological simulation code {\small GADGET-2}, and we presented
test problems carried out with it. We have emphasised the changes made
with respect to the previous public version of the code. We hope that
the improvements made in speed, accuracy and flexibility will help
future research with this code by allowing novel types of simulations
at higher numerical resolution than accessible previously.

In terms of accuracy, the most important change of the code lies in an
improved time-integration scheme, which is more accurate for
Hamiltonian systems at a comparable number of integration steps, and
in an `entropy-conserving' formulation of SPH, which especially in
simulations with radiative cooling has clear accuracy benefits. Also,
large-scale gravitational forces are more accurate when the TreePM
method is used and offer reduced computational cost compared to a pure
tree code.

In terms of speed, the new code has improved in essentially all of its
parts thanks to a redesign of core algorithms, and a complete rewrite
of essentially all parts of the simulation code. For example, the
domain decomposition and tree construction have been accelerated by
factors of several each. Likewise, the SPH neighbour search has been
sped up, as well as the basic tree-walk, despite the fact that it now
has to visit many more nodes than before due to the lower order of the
multipole expansion.

In terms of flexibility, the code can now be applied to more types of
systems, for example to zoom simulations with a 2-level TreePM
approach, or to gas-dynamical simulations in two dimensions. {\small
GADGET-2} also uses considerably less memory than before which makes
it more versatile. The code can now be run on an arbitrary number of
processors, and has more options for convenient I/O. Also, the code
has become more modular and can be more easily extended, as evidenced
by the array of advanced physical modelling already implemented in it,
as discussed in Section~\ref{SecAddPhys}.

In summary, we think {\small GADGET-2} is a useful tool for simulation work
that will hopefully stimulate further development of numerical codes. To
promote this goal, we release {\small GADGET-2} to the public\footnote{The
  web-site for download of GADGET-2 is {\tt
    http://www.mpa-garching.mpg.de/gadget}}. In a time of exponentially
growing computer power, it remains an ongoing challenge to develop numerical
codes that fully exploit this technological progress for the study of
interesting astrophysical questions.

\section*{Acknowledgements}

The author acknowledges many helpful discussions with Simon D. M. White, Lars
Hernquist, Naoki Yoshida, Klaus Dolag, Liang Gao, Martin Jubelgas, Debora
Sijacki, Christoph Pfrommer, Stefano Borgani, Martin White, Adrian Jenkins,
Jasjeet Bagla, Matthias Steinmetz, and Julio Navarro, among others.  Special
thanks to Matthias Steinmetz for the 1D PPM results used in
Figure~\ref{figEvrard}, and to Katrin Heitmann for making the initial
conditions and results of her DM test simulations publicly available.

\bibliographystyle{mnras}
\bibliography{paper,citations2,add}

\end{document}